\documentclass[final,a4paper,english]{siamltex} 
\pdfoutput=1
\usepackage[T1]{fontenc} 
\usepackage[latin1]{inputenc} 
\usepackage{ulem}
\usepackage{amssymb} 
\usepackage{graphicx} 
\usepackage[usenames]{color}
\graphicspath{{.},{../../},{../../Fig/},{./fig/}}

\newcommand{\jacobe}[0]{\mathbf{J}} 
\newcommand{\mafig}[1]{\includegraphics[height=3.0cm,width=2.6cm]{#1}\hspace{-2mm}} 
\newcommand{\temps}[5]{\hspace{5mm}$#1$ \hfill $#2$\hfill $#3$\hfill $#4$\hfill $#5$\hspace*{10mm}}
\definecolor{magenta}{cmyk}{0.1,0.8,0,0.1}

\newcommand{\myfig}[1]{Fig.~(\ref{fig:#1})}
\newcommand{\myeq}[1]{Eq.~(\ref{eq:#1})}
\newcommand{\mysec}[1]{Section~\ref{sec:#1}}
\title{Time integration and steady-state continuation for 2d lubrication 
equations} 
 
\author{Philippe Beltrame\footnotemark[1]\ \footnotemark[2] 
	\and Uwe Thiele\footnotemark[3]}

\begin{document} 
\maketitle  
\renewcommand{\thefootnote}{\fnsymbol{footnote}} 
\footnotetext[1]{University of Avignon, 33 rue Pasteur, 84000 Avignon, France.
E-mail: philippe.beltrame@univ-avignon.fr}
 
\footnotetext[2]{Institut f\"ur Physik, Universit\"at Augsburg, D-86135 
Augsburg, Germany} 
\footnotetext[3]{Department of Mathematical Sciences, Loughborough University, 
Loughborough, Leicestershire, LE11 3TU, UK. 
E-mail: u.thiele@lboro.ac.uk} 
\renewcommand{\thefootnote}{\arabic{footnote}}

\begin{abstract} 
  Lubrication equations describe many structuring processes
  of thin liquid films. We develop and apply a numerical framework
  suitable for their analysis employing a dynamical systems
  approach. In particular, we present a time integration algorithm
  based on exponential propagation and an algorithm for steady-state
  continuation. Both algorithms employ a Cayley transform to overcome
  numerical problems resulting from scale separation in space and
  time.  An adaptive time-step allows one to study the dynamics close to
  hetero- or homoclinic connections. The developed framework is
  employed on the one hand to analyze different phases of the
  dewetting of a liquid film on a horizontal homogeneous substrate. On
  the other hand, we consider the depinning of drops pinned by a
  wettability defect.  Time-stepping and path-following are used in
  both cases to analyze steady-state solutions and their bifurcations
  as well as dynamic processes on short and long time-scales. Both
  examples are treated for two- and three-dimensional physical
  settings and prove that the developed algorithms are reliable and
  efficient for 1d and 2d lubrication equations.
\end{abstract} 
 
 

\thispagestyle{plain} 
 
\section{Introduction} 

The dynamics of structuring processes of thin liquid films, ridges and
drops on solid substrates is often described by thin film or
lubrication equations. They are obtained employing a long wave
approximation \cite{ODB97}. The notion 'thin' means that the thickness
of the film/drop is small as compared to all typical length scales
parallel to the substrate. Thin film equations model, for instance,
dewetting due to van der Waals forces
\cite{RuJa74,Mitl93,ShKh98,TVN01,Beck03}, the long-wave Marangoni
instability of a film heated from below \cite{OrRo92,BPT03,ThKn04},
and the evolution of a film of dielectric liquid in a capacitor
\cite{Lin01,VSKB05,MPBT05,JHT08}. Including driving forces parallel to
the substrate allows one to describe, e.g., droplets that slide down
an incline under gravity under isothermal \cite{PFL01,Thie01} and
non-isothermal conditions \cite{BPT03,ThKn04}, the evolution of
transverse front instabilities
\cite{SpHo96,BeBr97,ESR00,Kall00,ThKn03}, and the evolution of shocks
in films driven by a surface tension gradient against gravity
\cite{BMS99,SBB03}. Extensions describe, for instance,
two-layer films, films with soluble or non-soluble surfactants, films
of colloidal suspensions, and effects of evaporation,
complex rheology or slip at the substrate. For reviews see, e.g.\
references \cite{ODB97,KaTh07,Thie09,Bonn09}.

Thin film equations are related to other 'standard' equations employed
in studies of pattern formation out of equilibrium \cite{CrHo93}. For
instance, the (convective) Cahn-Hilliard 
and the Kuramoto-Sivashinsky equation 
may be obtained from a thin film equation for sliding drops and
flowing films as limiting cases for small and large lateral driving,
respectively \cite{ThKn04}.

Due to the strong non-linearity a typical thin film equation is
difficult to handle numerically, in particular, when describing
three-dimensional physical situations 
resulting in a partial differential equation (PDE) with two spatial
dimensions (2d).
We develop and apply a numerical framework
to study the time evolution and to follow steady
state solutions in parameter space for 1d and 2d equations.
For viscous fluids (small Reynolds number flows), for a surface
tension that dominates over viscosity (small capillary number), and a
small lateral driving force the long wave approximation
results in the following evolution equation for the film
thickness or drop profile $h(x,y,t)$ \cite{ODB97,KaTh07}
\begin{equation} 
\partial_t h \,=\, -\nabla\cdot\left[ m(h) \nabla \tilde{p}(h) +\vec{\mu}(h)\right],
\label{eq:film}
\end{equation} 
where $m(h)$ is a mobility function, $\vec{\mu}(h)$ represents the lateral driving 
force, and the pressure $\tilde{p}(h)$ may contain several terms.  A curvature or Laplace pressure results from capillary 
and stabilizes a flat film.  Its contribution results in a
Bilaplacian of the height $h$. That is the highest order derivative in the equation. Therefore it constitutes one of the main numerical difficulties.  Pressure contributions that  destabilize the flat film can result from various physical mechanisms 
\cite{ODB97}. Examples include an electrostatic pressure 
for dielectric liquids in a capacitor \cite{Lin01,MPBT05,VSKB05}, a 
disjoining or conjoining pressure for very thin films below 100 
nanometers thickness resulting from effective molecular interactions 
between the substrate and the free surface (wettability effects)
\cite{deGe85,DCM87,Isra92}, a 'thermal pressure' for a thin film 
on a heated plate (when the long-wave Marangoni mode is dominant)
\cite{Oron00c,BPT03} and a hydrostatic pressure due to
gravity, e.g., for a fluid film under a ceiling 
\cite{Ferm92,DeOr92,Burg01}. All the mentioned destabilizing effects 
result in a long-wave instability, i.e., an instability with wavenumber 
zero at onset. 
Here we use two variants of a disjoining pressure as examples for a
destabilizing mechanism because their particular thickness
dependencies are numerically demanding and extensive literature
results allow for detailed comparison.  However, the developed
algorithms can be readily applied to any other combination of
stabilizing and destabilizing pressure terms. 

In the absence of a lateral driving force ($\mu=0$), a film that is
linearly unstable evolves during a short-time evolution into a
structure of holes, drops or mazes with a typical structure length
determined by film thickness and other control parameters that reflect
the character of the destabilizing phenomenon
\cite{Reit92,ShKh98,Oron00c,KSR00,BeNe01,BPT03,Seem05}.  This process
is often called 'spinodal dewetting' \cite{Mitl93}.  However, the
resulting short-time structure is unstable (representing a saddle in
function space) with respect to coarsening and on a large time-scale
the structures coarsen until the system eventually approaches the
global energetic minimum, i.e., a single drop or hole
\cite{BGW01,GlWi03,TBBB03}. Indeed without a lateral driving force the
system follows a relaxation dynamics and the evolution equation can be
written using the variation of an underlying energy functional
\cite{OrRo92,Mitl93,Thie10}. The variational form of the (then
coupled) evolution equations can also be derived for multilayer films
in similar settings \cite{PBMT04,PBMT05}.  However, even in the
one-layer case, details of the different phases of the process are
still under investigation. Examples include the initial structuring
process that might occur via nucleation or a surface
instability \cite{TVN01,Beck03}, and the mechanisms of the transverse
instability of dewetting fronts \cite{ShRe96,ReSh01}.

A detailed understanding of dewetting films and the other processes
listed above is only possible if the pathways of time-evolution and
the steady states described by Eq.~(\ref{eq:film}) can be determined
in the one- and two-dimensional case using fast and versatile
algorithms. As the steady drop solutions in 1d can be seen as
periodic trajectories in a conservative dynamical system one can use
available continuation packages for ordinary differential equations
\cite{AUTO97} to map different solution families and
their linear stability (see, e.g., \cite{TVNP01,ThKn04,TVK06}).
Careful interpretation allows one, e.g., to predict for dewetting films
the dominance of different rupture mechanisms within the linear
unstable parameter range \cite{TVN01,TVNP01,Thie03}. No such
continuation tools are, however, available in the two-dimensional
case.

The situation is more involved when lateral driving forces are
present, i.e., gravity on an incline \cite{PFL01}, or
temperature gradients along the substrate
\cite{CHTC90,KaTr97,BMFC98}. There, new phenomena appear like
transverse instabilities of sliding liquid ridges, advancing and
receding fronts \cite{SpHo96,BeBr97,ESR00,Thie01,ThKn03}.
Another fascinating finding is related to sliding drops: Beyond a
critical driving force the drop forms a cusp at the back end and
'emits' smaller satellite drops \cite{PFL01,BCP01,LDL05}. For a
driven contact line, heterogeneities of the substrate can cause a
stick-slip motion \cite{deGe85}. In the setting of a sliding drop
this leads at a critical driving force to the depinning of drops
from such localized heterogeneities. In the vicinity of the
responsible sniper bifurcation the resulting motion resembles
stick-slip motion: The drop sticks a long time at a wettability defect
and then suddenly slips to the next defect.  This was studied in the
1d case in Refs.~\cite{ThKn06,ThKn06b}.  Differences in time scales
for the stick- and the slip-phase may be many orders of magnitude.  On
the homogeneous substrate one can, in the 1d case, regard stationary
periodic drop and surface wave solutions as periodic trajectories
of a dissipative dynamical system. They emerge from the trivial flat
film state via a Hopf bifurcation \cite{ThKn04}.  This allows one to
employ standard continuation packages \cite{AUTO97} to obtain solution
families and to track the various occurring bifurcations
\cite{Thie01,JBT05,TVK06,TGV09}. The same applies for fronts (shocks)
and drops that correspond to heteroclinic and homoclinic orbits,
respectively \cite{BMS99,Muen03,CBH08}.
Note, however, that at present little is known about the solution and
bifurcation structure in the 2d case.

In the previous decade many publications were devoted to the numerical
study of thin film dynamics. Studies focus, e.g., on drop spreading
for wetting liquids \cite{ZhBe00,GrRu01,DiKo02,WiBo03}, heated films
\cite{Oron00c,GrRu01,BPT03} pending drops \cite{GrRu01,Gruen03}, and
the dewetting of partially wetting films
\cite{ShKh98,Oron00,BeNe01,WiBo03}.  The analysis of different
numerical approaches \cite{ZhBe00,GrRu01,DiKo02,WiBo03,Gruen03} leads
to the conclusion that the positivity ($h\geq0$) and the convergence
of discrete solutions as well as the performance of the algorithm
depend on the way the mobility $m(h)$ is discretized. However, here,
the preservation of positivity is less of a problem because there
exists the precursor film. Moreover, such a positivity preserving
scheme \cite{ZhBe00} applied to our partial wetting model may lead to
stability problems (spatial oscillations) when the drop height is much
larger than the precursor film. Indeed, as pointed out by Gr\"un
\cite{Gruen03} the stability results only apply if the disjoining
pressure $\Pi(h)$ remains bounded from below (when $h\rightarrow
0$). Most of the disjoining pressures employed in the literature are
unbounded as the ones that we will use.  Therefore, here the
preservation of positivity is not a crucial stability criteria.
The classical way to overcome stability problems is to employ a
semi-implicit scheme (see, e.g.,~\cite{BPT03}). The implicit part
corresponds to the bilaplacian, i.e. the linear operator of highest
order which should be related to the eigenvalues of large modulus
\cite{TB00}. The approach normally works well if the ratio of drop
height and precursor film thickness is small. However, for larger
drops simulations often display spatial numerical oscillations.

We conclude that reliable algorithms for time integration and
  path-following for steady state solutions that are applicable
  equally well in most of the above introduced examples are not
  readily available.  Here, we develop and apply a time integration
  scheme with an adaptive time-step and tools for bifurcation analysis
  that are applicable in the 1d and (most importantly) 2d case.

Our starting point is the understanding that not only mobility is
  crucial for the numerical stability but also the disjoining
  pressure. Thus, in a semi-implicit scheme, a good choice of the
  linear part should contain contributions from these terms (see below
  section~\ref{sec:jacobe}). A natural candidate is the Jacobian
  matrix at each time-step.  Another efficient time integrator which
  involves the Jacobian matrix is the exponential propagation scheme.
  In general, it is more stable and converges better than
  semi-implicit methods \cite{Tokman06}.
  The exponential propagation scheme is based on the exact solution of
  the linearized equation at each time-step requiring the computation
  of the exponential of the Jacobian matrix. This operator is not
  directly computed: As commonly practiced for large and sparse
  matrices, only its action on vectors is estimated using projections
  on small Krylov subspaces of dimension $K\ll N$ \cite{Saad92}.  The
  standard algorithm to perform this task is the Arnoldi procedure
  possibly incorporating improvements as proposed in \cite{Tokman06}.
  The exponentiation can be performed at a negligible cost as long as
  $K$ is not too large.

  The application of such a scheme to lubrication equations proves to
  be reliable. However, the approximation in Krylov subspaces
  converges rather slowly.  We show that the necessary dimension $K$
  is about one hundred while in
  Refs.~\cite{FTDR89,Saad92,HoLu97,Tokman06} $K\approx10$ is
  sufficient. The difference results from the presence of the fourth
  order Bilaplacian in lubrication equations. Its effect is more
  disadvantageous than the one of a second order operator because the
  magnitude of the large negative eigenvalues increases with the order
  of the differentiation operator \cite{LiTi04}.  To our knowledge, no
  literature study analyzes the efficiency of the Krylov subspace
  approximation to a matrix exponential operator that contains a
  fourth order operator.  We improve this step by coupling existing
  methods for the determination of the rightmost spectrum with the
  classical Arnoldi procedure \cite{Meerbergen1996}. In particular,
  the application of a Cayley transform proves to be most
  powerful. The resulting scheme is well suited to efficiently
  adapt the time-step to the corresponding time-scale of the
  dynamics. This allows for very large time-steps as, for instance, in
  problems involving coarsening and stick-slip drop motion.
 
The outlined scheme can not only be used for time-stepping. We
show that the Krylov reduction associated with the Cayley transform
can as well be applied to track steady states in parameter space
employing a continuation scheme that consists of the determination of
a tangent predictor, the application of Newton's algorithm along the
secant direction, the detection of bifurcation points and finally the
determination of the direction of bifurcating branches of steady
solutions \cite{Seydel}.
Note, that in both algorithms, time-stepping and continuation,
  the Cayley transform is used in a original and to our knowledge
  novel way.

  The paper is structured as follows.  Section~\ref{sec:mod} presents
  the lubrication equation and its spatial discretization.  Then we
  describe in Section~\ref{sec:time} the exponential propagation
  method and discuss different ways to exponentiate the Jacobian. In
  Section~\ref{sec:cont} we adapt the developed algorithms to employ
  them for the continuation of steady states.  Two appendices give
  details for the Krylov reductions (\ref{sec:krylov}) and convergence
  of the algorithms (\ref{sec:concalgo}).  In the remaining part, we
  apply the algorithms to two typical situations. The first one
  (Section~\ref{sec:dew}) is the dewetting process of a thin film on a
  horizontal substrate.  The slow coarsening process that follows the
  initial fast patterning provides an excellent test for the adaption
  of the time-step. We do as well study steady state solutions in 2d,
  in particular, we discuss several solution branches corresponding to
  quadratic and hexagonal arrays of drops.  Second, in
  Section~\ref{sec:het} we study the pinning/depinning of drops on an
  inclined heterogeneous substrate. The path-following algorithm is
  applied to determine branches of steady state solutions and, in
  consequence, the onset of depinning in the 2d case. The stick-slip
  motion of drops beyond depinning is investigated using our
  time-stepping algorithm.  For comparison with the literature we do
  as well provide selected results for the 1d case.  Our conclusions
  are found in Section~\ref{sec:conc}.
%
\section{Modeling and Spatial Discretization} 
\label{sec:mod} 
\subsection{Lubrication equation} 
\begin{figure} 
\includegraphics[width=12cm]{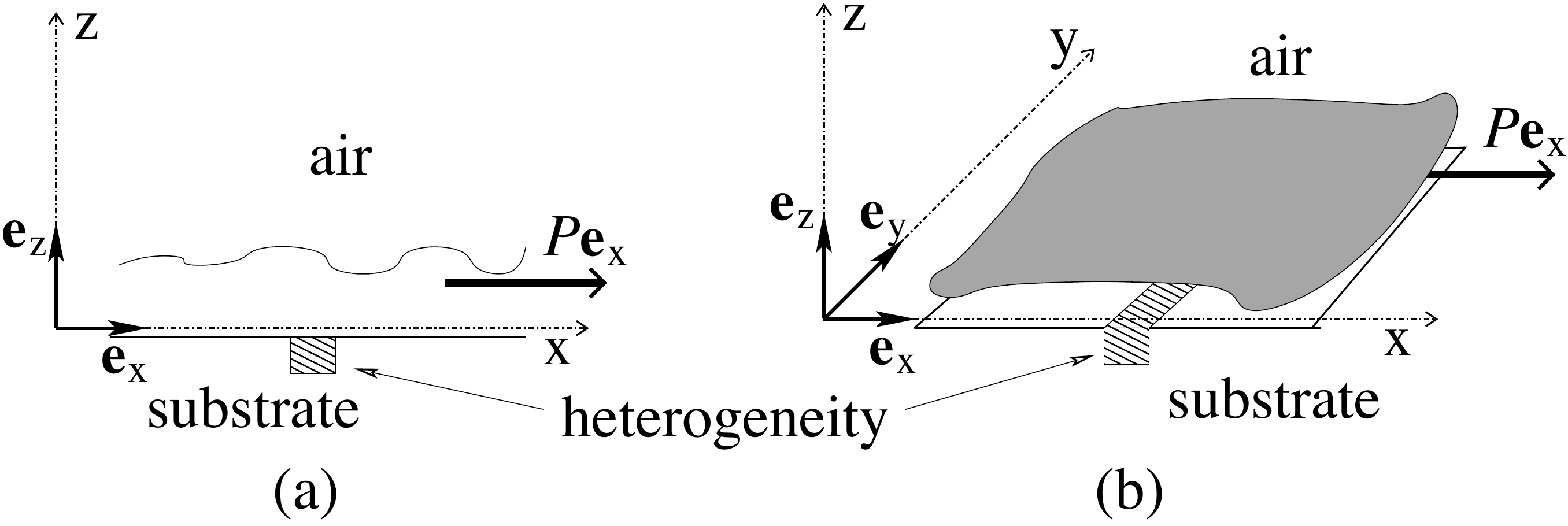} 
\caption{Sketch of the (a) two-dimensional and (b) three-dimensional 
  physical setting of the problem: A free-surface liquid film rests or 
  flows on a possibly heterogeneous substrate. Thereby the 
  heterogeneous wettability is assumed to depend on the location in the $x$-direction only. 
A driving force $P$ might act along the $x$-direction. Note, that the resulting 
film evolution equations have one (a) and two (b) spatial dimensions
and are referred to as the 1d and 2d case, respectively.
\label{fig:sketch}
}
\end{figure} 
Consider a liquid layer on an (inhomogeneous) one- or two-dimensional 
solid substrate (Fig.~\ref{fig:sketch}). The liquid partially wets 
the substrate and might be subject to a constant lateral force $P$. 
Using the long-wave approximation, the dimensionless evolution 
equation for the film thickness profile $h(x,y,t)$ derived from the
Navier-Stokes equations, continuity and boundary conditions is 
\cite{ODB97,KaTh07} 
\begin{equation} 
\partial_{t}h=F(h,x) = -\nabla\cdot\left\{ m(h)\left[ 
\nabla\left(\Delta h+\Pi(h,x)\right)+P\mathbf{e_{x}} 
\right]\right\}
\label{eq:lub} 
\end{equation} 
where $\nabla=(\partial_{x},\partial_{y})$ is the planar gradient
operator and $\Delta=\partial_{xx}^{2}+\partial_{yy}^{2}$ is the
planar Laplacian.  Note, that Eq.~(\ref{eq:lub}) has one and two
spatial dimensions for two- and three-dimensional physical settings,
respectively. In the following they are referred to as the 1d and 2d
case, respectively.  The mobility function $m(h)=h^{3}$ corresponds to
Poiseuille flow without slip at the substrate. The term $\Delta h$
represents the Laplace pressure. Wettability is modeled by the
disjoining pressure $\Pi(h,x)$ that for a striped heterogeneous
substrate depends on film thickness and on the position $x$. Here, the
lateral force acts as well in the $x$-direction
(Fig.~\ref{fig:sketch}).  Many particular forms of the disjoining
pressure are discussed \cite{deGe85,KaTh07,Bonn09}. The most common
ones allow for the presence of an ultra-thin wetting layer (a
so-called 'precursor film') of about 1-10 nm thickness.  The
short-time dewetting dynamics of an unstable film results in large
amplitude structures, either drops or holes. It is often called
initial 'film rupture' even if a stable precursor film remains present
-- a convention we follow.

To facilitate comparison to the literature
we employ the disjoining pressures
\begin{equation}  
\Pi(h)=-2e^{-h}(1-e^{-h})-Gh\label{eq:dis1} 
\end{equation} 
(as used in \cite{Thie01,TVNP01}), and 
\begin{equation} 
\Pi(h)=-\frac{b}{h^{3}}+e^{-h}\label{eq:dis3} 
\end{equation} 
(as used in \cite{Shar93,TVN01}). We call them (I) and (II), 
respectively. For case (II) we incorporate varying wettability properties 
due to a heterogeneous coating as
\begin{equation} 
  \Pi(h,x)=\frac{b}{h^{3}}-\left[1+\epsilon\xi(x)\right]e^{-h},\label{eq:dispin}\end{equation} 
where $\xi(x)$ is the heterogeneity profile and $\epsilon$ the 
amplitude of the heterogeneity (as in \cite{ThKn06,ThKn06b}). 
%
\subsection{Functional space} 
\label{sec:space}

The considered domain is $D=[0;L]$ (1d case) or 
$D=[0;L_{x}]\times[0;L_{y}]$ (2d case). Eq.~(\ref{eq:lub}) 
defines a partial differential equation (PDE) in $H^{4}(D)$ 
\begin{equation} 
\partial_{t}h=F(h,x),h\in H^{4}(D)\label{eq:edp} 
\end{equation}  
To obtain a well posed PDE system, we introduce periodic boundary
conditions. In the studied case of non-volatile liquids the mass
$M=\int_{D}h(x,y)dxdy$ is conserved. If $S_D$ denotes the surface
  of the domain $D$, the measure $H=M/S_D$ represents the mean height
and $u=h-H$ is the perturbation (sometimes denoted $\delta
  h$). The PDE (\ref{eq:edp}) defined in the space
\begin{eqnarray} 
E_0=&&\left\{ u\in H^{4}(D) : \int_{D} u(x,y)dxdy=0\right.\nonumber\\
&&\left.\mbox{and periodicity of}\quad\partial_{x}^{i}u\quad\mbox{and}\quad\partial_{y}^{j}u,\mbox{}0\leq i,j\leq 3\right\} 
 \end{eqnarray} 
is well defined: (i) $F$ operates on the Euclidian space $\{ H+E_0\}$,
and (ii) linear operators operate on the linear space $E_0$.
In this space, we define the $L^2$ norm (denoted $||.||$) of $u$:
\begin{eqnarray}
||u||=||\delta h||&=&\frac{1}{S_D}\left( \int_D |u(x,y)|^2 dx dy \right)^{1/2}, \label{l2norm}
\end{eqnarray}
where $|u|$ is the Euclidean. The notation $||\delta h||$ is used for
the presentation of the numerical results while $||u||$ is reserved
for the description of the algorithm. 
\subsection{Spatial discretization}

Here, we devote much effort to the time integration but chose the
  spatial discretization as simple and generic as possible.
  A finite difference scheme is used associated with a regular meshing of the domain. We use in $x$ and $y$ directions, $(N_{x}+1)$ and $(N_y+1)$ mesh points at distances $\delta x=N_{x}/L_{x}$ and $\delta y=N_{y}/L_{y}$,
respectively. The number of discretization points is
$N=(N_{x}+1)\times (N_{y}+1)$ in the 2d case and $N=N_{x}+1$ in the 1d
case.
The differentiation operators are approximated by a centered
five-point stencil, i.e., for the Bilaplacian $\Delta^2$ and
the third order operator $\nabla\Delta$ it is a second-order
approximation while for the gradient or divergence operator $\nabla$
and the Laplacian $\Delta$ it is a fourth-order approximation. Thus,
these operators are sparse band matrices with maximally
5 (1d) or 25 (2d) non-zero elements in each row.
The map $F(h,x)$ defined in \myeq{edp} is discretized according to
\myeq{lub}, i.e., in the mass conserving form. The Jacobian of the map
$F$ is discretized using \myeq{jacobe} below, resulting in a
second-order approximation. The discretized Jacobian matrix is not in
a mass conserving form to avoid a further increase of the number of
non-zero elements.
Mass conservation is, however, imposed during the Krylov projection (see appendix \ref{sec:krylov}).

\subsection{Jacobian matrix}
\label{sec:jacobe}
Next, we justify our choice to compute the Jacobian matrix at each time-step.
First, we explicitly determine the Jacobian matrix as linearization of the function $F(h,x)$ at $h=h_0$
\begin{equation}
	\jacobe=D_hF(h_0,x)
\end{equation}
where $D_h$ denotes the differentiation operator with respect to $h$.
The Jacobian matrix is a sum of differentiation operators up to fourth order:
\begin{eqnarray}  
\jacobe u&= & v_0(h_0,x)  u  + {\mathbf{v_1}}(h_0,x)\cdot\nabla u   \label{eq:jacobe}\\
& & +  m_0\Pi'_0  \Delta u-\nabla m_0  \cdot \nabla\Delta u-m_0  \Delta^{2}u. \nonumber
 \end{eqnarray}
with
\begin{eqnarray}  
v_0(h_0,x)&= & -m'_0\Delta^{2}h_0-P\partial_{x}m'_0\\
& & + m'_0\Delta\Pi_0+\nabla m'_0\cdot\nabla\Pi_0+\nabla m\cdot\nabla\Pi'_0+m\Delta\Pi'_0  \nonumber\\ 
\mathbf{v_1}(h_0,x) &= &- m'_0\nabla\Delta h_0+m'_0\nabla\Pi_0+2m_0\nabla\Pi'+\nabla m_0\Pi'+P\mathbf{e_x}
 \end{eqnarray}
and
\[
\begin{array}{lcrclcr}
m_0&=&m(h_0)& &m'_0&=&\frac{dm}{dh}(h_0)\\
\Pi_0&=&\Pi(h_0,x)& &\Pi'_0&=&\frac{d\Pi}{dh}(h_0,x)
\end{array}
\]
The Jacobian matrix $\jacobe$ has eigenvalues with large negative real
parts. They correspond to the fastest time scales related to the
differentiation operators.  As these eigenvalues are situated in the
left half of the complex plane they are called ``leftmost
eigenvalues''.  Their presence is the main cause for spatial numerical
oscillations in explicit time integration methods. The resulting
stability restriction on the time-step is overcome by treating the
linear term implicitly.  
The leftmost eigenvalues are due to the
differentiation operator of the highest degree. Therefore, on can
filter them out by treating only this operator
implicitly.  The implicit linear term normally
corresponds to the Laplacian for second order equations like, e.g.,
reaction-diffusion or Navier-Stokes equations and to the bilaplacian
for fourth order equations like, e.g., the Kuramoto-Sivashinsky
equation \cite{GBP07}.  Here, a natural candidate is the
bilaplacian.  However, it is multiplied by the Poiseuille flow
mobility $m_0=h_0^3$. This implies that, e.g., for a height variation
by a factor 10 like for a drop solution, a factor $10^3$ appears in
the linear operator. The resulting spatial scale separation
complicates the situation as compared, e.g., to the
Kuramoto-Sivashinsky equation.

Furthermore, here the lower order operators may contribute to the
leftmost eigenvalues because the vectorial factor in front of them can
have very large elements. For example, the Laplacian is scaled by
$m_0\Pi'_0$ related to the disjoining pressure that can be very large
close to the contact region of drops. Then the Laplacian gives a
non-negligible contribution to the leftmost spectrum of
$\jacobe$. This strongly localized prefactor is reminiscent of
  a point force at the contact line related to wettability as
  discussed, e.g., in \cite{deGe85}.  Thus the specific numerical
  problems of the thin film equation for partially wetting liquids are
  not only due to the Poiseuille flow mobility but also to phenomena
  related to wettability and contact angle.

In conclusion, it is not advisable to filter the leftmost
  eigenvalues employing a constant linear operator since flow maxima
  and front positions are time dependent. It is preferable to compute
  the Jacobian through a linearization at each time-step. It is then
  employed in the exponential propagation method as it is best suited
  to our purpose (cf.~introduction and appendix \ref{sec:concalgo}).

%
\section{Time integration} 
\label{sec:time}
\subsection{Exponential propagation} 
Starting from the known profile $h_0$ at $t_0$, the exponential
propagation scheme consists in solving the autonomous evolution
problem
\begin{eqnarray}
\frac{dh}{dt}&=&F(h,x),\\
h(t_0) &=& h_0
\end{eqnarray}
at each time-step.
This is done by expanding the operator $F(h,x)$ near the state $h_0$ in a Taylor series
\begin{equation}
F(h_0+u,x) = F(h_0,x) + D_hF(h_0,x) u + R(u),\label{eq:jtaylor}
\end{equation}
where $D_hF(h_0,x)$ is the Jacobian matrix at $h_0$ and $R(u)=O(||u||^2)$ contains the quadratic and higher order terms. To simplify the
notation we let
\begin{eqnarray}
\jacobe &=& D_hF(h_0,x)\\
b &=& F(h_0,x).
\end{eqnarray}
The height variation $u(\tau)=h(t_0+\tau)-h_0$ then solves the evolution problem
\begin{eqnarray}
\frac{du}{d\tau}&=& b + \jacobe u + R(u) ,\label{eq:sysevolu}\\
u(0) &=& 0 \nonumber
\end{eqnarray}
which admits as solution
 \begin{equation} 
u(\tau)=G(\jacobe\tau)b\tau+\int_{0}^{\tau}\exp((\tau-\tau')\mathbf{J}) R(u(\tau'))d\tau' ,\label{eq:convold} 
\end{equation}
with
\begin{equation} 
G(\jacobe\tau)=\frac{\exp(\mathbf{J}\tau)-\mathbf{I}}{\mathbf{J}\tau}.
\label{eq:g}
\end{equation}
Eq.~(\ref{eq:g}) is formally correct even if
$\jacobe$ is not invertible since the operator $G$ can be expressed as
the series $G(x)=\sum_{n=0}^{+\infty} x^{n}/(n+1)!$.  Since the
functions $G$ and $\exp$ tend to zero at $-\infty$, they filter the
leftmost spectrum of the operator $\jacobe$.  For a semi-implicit
scheme one has a similar situation: There is a rational function of
$\jacobe$ that plays the role of the filter.  Yet, a rational function
tends slower to zero than $G$ and $\exp$ when approaching $-\infty$.
Therefore, we expect the exponential propagation scheme to better
filter the leftmost eigenvalues.  See as well appendix~\ref{sec:concalgo}.

\subsection{Order of the scheme and error estimator}
If we only consider linear variations of $F(h,x)$,
i.e., $R(u) \equiv 0$ in \myeq{jtaylor}, the first term of \myeq{convold} is the exact solution of the evolution problem (\ref{eq:sysevolu}):
\begin{equation}
u^{(1)}(\tau) = G(\jacobe\tau)b\tau.\label{eq:ulin}
\end{equation}
The solution $h^{(1)}(\tau)=u^{(1)}(\tau)+h_0$ is then an
approximation of second order with respect to time.  To employ
exponential schemes of higher order one has to take into account the
non-linear term $R(u)$ in \myeq{convold}. Then the perturbation $u$
can not be obtained explicitly from \myeq{convold} but can be estimated  by successive
approximations of the non-linear terms through the series
${u^{(\ell)}}$ \cite{FTDR89}. The resulting scheme of order $\ell+1$ is
 \begin{equation} \label{eq:uel}
u^{(\ell)}(\tau)=u^{(1)}(\tau)+\sum_{k=2}^{\ell} \int_{0}^{\tau}\exp((\tau-\tau')\mathbf{J}) c_{k}^{\ell}\left(\frac{\tau'}{\tau}\right)^{k}d\tau'  ,
\end{equation}
where the computation of the vectors $c_k^\ell$ is detailed in
\cite{FTDR89}.

Note that relation (\ref{eq:uel}) can be differentiated analytically
with respect to $\tau$. Moreover, the approximation $u^{(\ell)}$ is
close to the exact solution of the evolution problem \myeq{sysevolu}
if $du^{\ell}/d\tau\simeq F(u^\ell+h_0,x)$ and the difference
\[
u_{err}=\tau\left[\frac{du^\ell}{d\tau}-F(h_0+ u^\ell,x)\right],\]
represents an error vector of the time-step. Thus, a natural candidate
for a relative error is the ratio of the $L^2$ norms of $u_{err}$ and
$u+h$
\begin{eqnarray}\label{eq:err} 
  \epsilon_{r} & = & \frac{\left\Vert u_{err}\right\Vert }{\left\Vert u+h\right\Vert }.
\end{eqnarray} 
It may be used to efficiently control the numerical error.
%
\subsection{Krylov projection}
Since Ref.~\cite{FTDR89} was published a plethora of high order
approximations were developed \cite{Saad92,HoLu97,HLS98,Tokman06}.
However, the accuracy of the exponential scheme does not only depend
on the non-linear approximation of $R(u)$ but as well on the
approximation of the vectors $v_g=G(\jacobe\tau)b$ and
$v_e=\exp(\jacobe\tau)c$. 

They are commonly performed using a projection onto a small Krylov
subspace of dimension $m$ computed by the classical Arnoldi
algorithm. In the literature, this step does not constitute a
difficulty as a good approximation of the action of their highest
order operators is already obtained with a small Krylov subspace of
about $m= 10$. In contrast, here we need $m=100$ (see appendix
\ref{sec:krylov}).  The literature examples are normally second order
equations (reaction-diffusion equation, Schr\"odinger equation)
whereas the thin film equation contains a fourth order operator. This
indicates that the bad convergence of the Krylov Arnoldi algorithm
results mainly from the presence of the bilaplacian as it has a worse
conditioning than the Laplacian.  The large values that may be taken
by the mobility function and the disjoining pressure play as well an
important role.

In consequence, one has to improve the Krylov projection step to be
able to apply the exponential propagation method efficiently to thin
film equations. As our improvement can be applied to schemes of any
order, in the following we focus on a second order exponential scheme.

Appendix~\ref{sec:krylov} gives a detailed description of the two
methods used to estimate $G(\jacobe\tau)b$ and $\exp(\jacobe\tau)c$ --
the classical Krylov-Arnoldi projection and our variations of that
method. The key-idea of our improvement is to transform the spectrum
of the operator $\jacobe$ in order to accelerate the convergence of
the Krylov approximation.  It is well known that the Krylov-Arnoldi
algorithm first tends to the part of the spectrum that has the largest
modulus. However, the rightmost eigenvalues of $\jacobe$ are the ones
of primary interest for the time-stepping.  To reach fast convergence
we need to apply a transformation that allows these ''wanted''
eigenvalues to become the ones of largest modulus.
Such transformations are commonly used when estimating the rightmost
spectrum (see e.g.~\cite{WhBa06}). One can distinguish two main
methods: Chebyshev acceleration and Cayley transform
\cite{Meerbergen1996}. Here, only the latter is efficient. However, it
requires an Incomplete-LU (ILU) factorization which needs $O(N^{3/2})$
steps.  It is laid out in appendix~\ref{sec:krylov} that a
Cayley-Krylov method is most efficient for system sizes below
$N\approx O(10^5)$.

To summarize, the time integration of \myeq{lub} is performed using an
exponential propagation scheme that employs Krylov projection. The
scheme is stable independent of the particular Krylov approximation
used. For moderate system sizes of $N=O(10^5)$ the Cayley-Krylov
projection furthermore allows one to employ adaptive time-stepping. As
it provides as well the leading modes at each time-step it is a
valuable tool for studies of film and drop dynamics.
 
\section{Continuation of steady-state solutions} 
\label{sec:cont} 
\subsection{Introduction} 
Next, we develop an algorithm to follow branches of
steady-state solutions when varying a parameter $p$, i.e., we
seek the branch $(h,p)$ such that
\begin{equation} 
\forall x \in [0;L_x]: F(h,x,p)=0.\label{eq:stat} 
\end{equation} 
Note that the continuation parameter $p$ may be any parameter of
  the problem as, e.g., the lateral force $P$ or the heterogeneity
  strength $\epsilon$. We adopt the tangent predictor - secant
corrector scheme \cite{Seydel}. For both steps the Jacobian has to be
inverted.  Although, this operation is different from the
exponentiation required in the time-stepping algorithm, the employed
Krylov reduction using the Cayley transform is relevant for the
inversion that needs only a few eigen-directions associated to small
eigenvalues. Eigenvalues of large modulus --namely, the leftmost
eigenvalues-- are of negligible influence, their presence can even
lead to numerical oscillations.
Furthermore, the knowledge of the leading eigenvalues
facilitates the stability analysis and allows one to detect bifurcation points.
 
\subsection{Tangent-predictor/secant-corrector method} 

\begin{figure}\label{fig:schemecon} 
\centering 
\includegraphics[width=9cm,height=6cm]{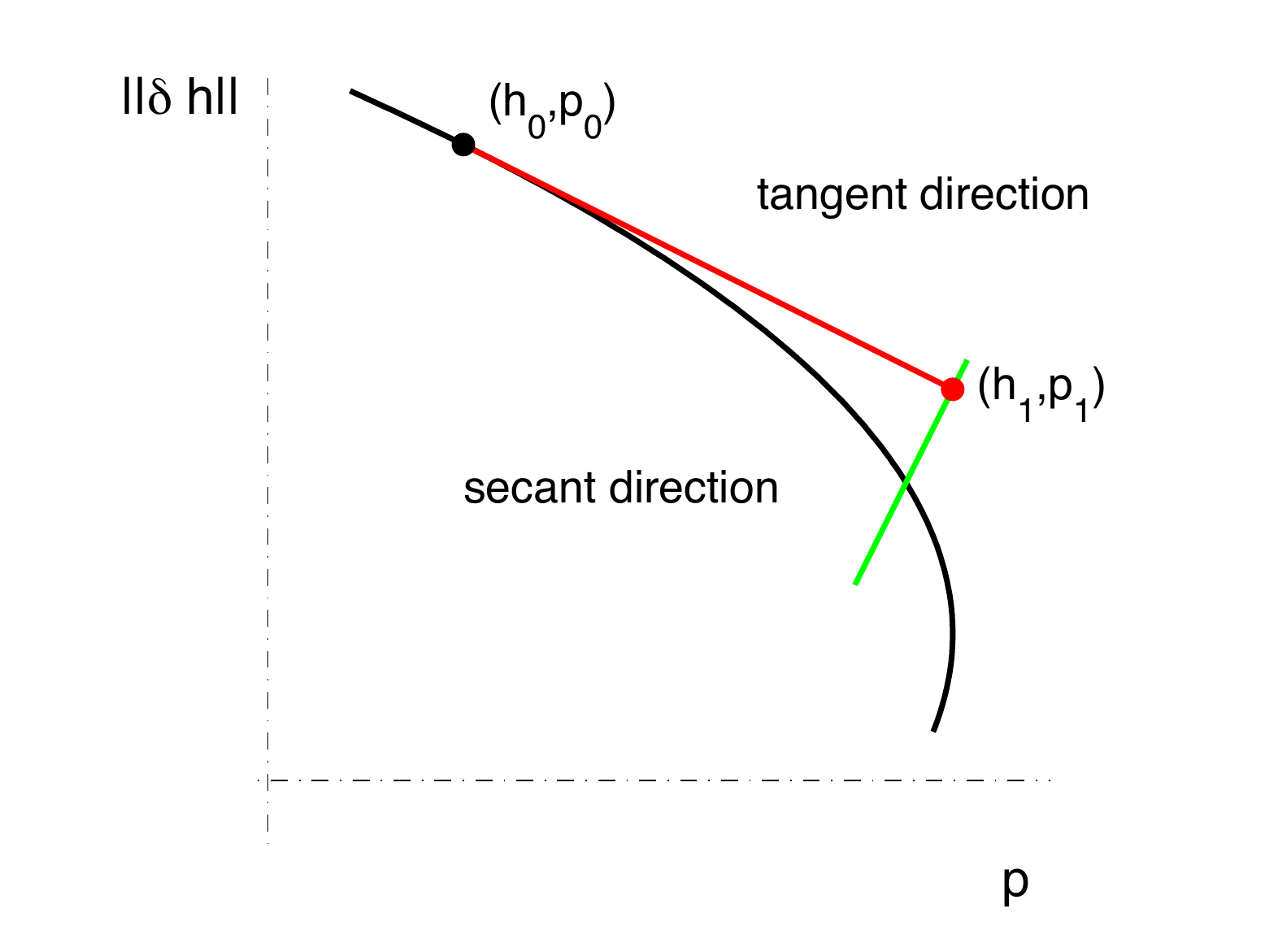} 
 \caption{Scheme of a continuation step using the tangent 
  predictor - secant corrector method.} 
\end{figure} 
First, we describe a continuation step as sketched in
Fig.~\ref{fig:schemecon}. One starts at point $(h_{0},p_{0})$
representing a steady state $h_{0}$ at parameter value
$p_{0}$. Differentiating (\ref{eq:stat}) one obtains the
tangent direction $(\delta u_{t},\delta p_{t})$ of the
continuing branch at the point $(h_{0},p_{0})$ as solution of the system
 \begin{equation} 
\mathbf{J_{0}}u_{t}=-D_{p}F(h_{0},x,p_{0})\delta p_{t}\label{eq:cont} 
\end{equation} 
where $\mathbf{J_{0}}=D_{h}F(h_{0},x,p_{0})$ is the Jacobian.  The
$(\delta u_{t},\delta p_{t})$ solution is entirely determined by fixing the
amplitude of $\delta p_t$.  This is done by finding the maximal amplitude of
$p_t$ such that
\begin{equation} 
||F(h_{0}+\delta u_{t},x,p_{0}+\delta p_{t})||<\epsilon_{t}||h_{0}||
\end{equation} 
to stay close to the steady state branch.  Typically
we take $10^{-3}<\epsilon_{t}<10^{-1}$.  We denote this intermediate point by $(h_{1},p_{1})$ with
\begin{eqnarray} 
h_{1} & = & h_{0}+\delta u_{t}\\ 
p_{1} & = & p_{0}+\delta p_{t}. 
\end{eqnarray} 
The sign of $\delta p_{t}$ remains to be chosen.  It only changes at
saddle-node bifurcations.  In the $(p,f(p))$ plane the passing of a
saddle-node bifurcation is characterized by a sign change of $f'(p)$
and an increase of $|f'(p)|$ before reaching the bifurcation. If both
conditions are fulfilled the sign of $\delta p_t$ has to be changed.

Next, one uses Newton's method to solve Eq.~(\ref{eq:stat}) close to
$(h_1,p_1)$.  The secant is taken orthogonal to the tangent $(\delta
u_t,\delta p_t)$ to be able to follow the branch even in the
neighborhood of a turning point.  For the Newton iteration step that
starts from $(h_k,p_k)$ the condition reads
\begin{eqnarray} 
\mathbf{J_{k}}u_{k+1} & = & -F(h_{k},x,p_{k})-D_{p}F(h_{k},x,p_{k})\delta p_{k}\label{eq:new}\\ 
\left<u_{t},\delta u_{k}\right> & = & 0\label{eq:neworth}\\ 
h_{k+1} & = & h_{k}+\delta u_{k}\\ 
p_{k+1} & = & p_{k}+\delta p_{k} 
\end{eqnarray} 
where $\mathbf{J_{k}}=D_{h}F(h_{k},x,p_{k})$ and
$\delta u_{k},\delta p_{k}$ are the unknowns at step $k+1$.
Eq.~(\ref{eq:new}) and the orthogonality condition (\ref{eq:neworth}) written as matrix equation are
\begin{eqnarray} 
\left[\begin{array}{cc} 
\mathbf{J_{k}} & D_{p}F(u_{k},x,p_{k})\\ 
u_{t} & 0\end{array}\right] 
\left[\begin{array}{c} 
		\delta u_{k}\\ 
		\delta p_{k} 
	\end{array}\right]  
 &=&  
\left[\begin{array}{c} -F(u_{k},x,p_{k})\\ 
0\end{array}\right] \nonumber\\ 
\mathbf{N} 
\left[\begin{array}{c} 
		\delta u_{k}\\ 
		\delta p_{k} 
	\end{array}\right]  
 &=&  
\left[\begin{array}{c} -F(u_{k},x,p_{k})\\ 
0\end{array}\right]. \label{eq:newmat} 
\end{eqnarray} 

One clearly sees, that the continuation step requires inversions of
the Jacobian matrix $\mathbf{J}$ (tangent predictor) and of the matrix
$\mathbf{N}$ (Newton corrector steps).  Except at bifurcation points,
systems are invertible in the space $E_{0}$.  As above, the
restriction to $E_0$ is ensured during the Krylov reduction (see
appendix \ref{sec:krylov}).

\subsection{Computation of the tangent predictor} 
The Cayley-Krylov reduction of (\ref{eq:cont}) leads to
\begin{equation} 
V_m\mathbf{J_{m}}V_m^{t}u_{t}=-b\delta p_{t}=-D_{p}F(h_{0},x,p_{0})\delta p_{t}. 
\end{equation} 
where ${V_m}$ is the Krylov basis consisting of $m$ vectors in
$E_{0}$.  It is constructed by letting the operator
$\mathbf{C}=(\mathbf{J}-c\mathbf{I})^{-1}$ act on the vector
$b=D_{p}F(u_{0},p_{0})$.  The choice of the scalar $c$
follows the same rules as in the time-stepping
(appendix~\ref{sec:krylov}).  Using the QR-method
the spectrum of the reduced $m\times m$ Jacobian is obtained:
\begin{equation} 
\jacobe_m=\mathbf{P_m}\mathbf{D_{m}}\mathbf{P_m^{-1}}. 
\end{equation} 
For the inversion we distinguish two cases: (i) if the kernel contains
a non-zero eigenvector $v_{0}$, then the pair $(v_{0},0)$ is the
solution of the problem; (ii) otherwise we perform the inversion and
the solution is given by
\begin{equation} 
u_{t}=-V_m\mathbf{{P_mD_m}^{-1}P_m^{-1}}V_m^{t}b\delta p_{t}. 
\end{equation} 
In this way one is able to detect bifurcation points
and ensures that the continuation works well at turning points.
\subsection{Computation of the secant corrector} 
To perform one Newton step (\ref{eq:newmat}), the same method
  as in the previous paragraph is applied to the matrix $\mathbf{N}$
  instead of $\jacobe_m$. Note that, because of the secant direction
  requirement (\ref{eq:neworth}) the matrix $\mathbf{N}$ is invertible
  in $E_0$ even close to a bifurcation point. 

As both, tangent predictor and secant corrector step, need an
  ILU-factorization, it is the most important numerical task in the
  continuation algorithm (as for the time-stepping algorithm).

\subsection{Bifurcation analysis and stability analysis} 
Because the rightmost spectrum of the Jacobian is known we are as well
able to assess the stability of solutions. Furthermore, during the
tangent predictor step one is able to detect the presence of
bifurcations. The direction of the bifurcating branch may be deducted
from the eigen-directions of the kernel.  Although, the rightmost
spectrum is normally well evaluated it may occur that the rightmost
eigenvalue $\lambda_{\rm max}$ is not in the Krylov space (see appendix
\ref{sec:krylov}).  Furthermore, the accuracy of the estimation of the
rightmost eigenvalues is only about $10^{-3}$. To
determine the bifurcation point more accurately one has to apply a
different algorithm as, e.g., a block
Arnoldi method \cite{Sadkane1993} adapted to the Cayley-Arnoldi algorithm.
In the following the developed algorithms are used to study important
questions related to the dewetting of a thin film (Section~\ref{sec:dew}) and the depinning of a drop (Section~\ref{sec:het}).

\section{Short-time dynamics and coarsening for dewetting thin films} 
\label{sec:dew} 
The dewetting of a thin film can be initiated by two different
mechanisms: Either via a surface instability (spinodal dewetting) or
by heterogeneous nucleation at finite size defects
\cite{deGe85,Thie03,Seem05,Thie07}.  Dewetting of metastable films can
only be initiated by nucleation, i.e., by finite disturbances larger
than a critical threshold given by an unstable steady state.  For
linearly unstable films, there exists a critical wavelength
$\lambda_c=2\pi/k_c$. Any disturbance associated to a wavelength
$\lambda>\lambda_c$ grows exponentially in time with a growth rate
$\beta=-m(h_0)k^2(k^2-k_c^2)$. The resulting short-time dewetting
structure consists of a drop, hole or labyrinthine pattern. Its
characteristic scale corresponds to the wave length
$\lambda_{m}=\sqrt{2}\lambda_c$ of the fastest growing linear mode.
However, whether this linear instability is the dominant mechanism
depends on the character of the primary bifurcation: Deep inside the
linearly unstable regime it is supercritical and the film necessarily
develops the surface instability.  However, closer to the metastable
region the primary bifurcation is subcritical. Then threshold
solutions are present. They offer as second pathway of evolution a
nucleation process as in the metastable regime
\cite{TVN01,TVNP01,Thie03,Beck03}. The latter normally dominates if
the growth rate of the threshold solution is much larger than the
largest growth rate of the linear instability $\beta_{m}$.  Details of
the dewetting process then strongly depend on experimental conditions
(amount of defects, roughness, noise).  When nucleation dominates one
expects larger drops or holes that are randomly distributed.  Here we
investigate (i) the dominance of either instability- or
nucleation-triggered dewetting in the linear unstable thickness region
in the 2d case; and (ii) quadratic and hexagonal steady state
solutions in the 2d case and the character of their primary
bifurcations.  The short-time dewetting dynamics is followed by a very
slow coarsening process resulting eventually in a single drop
coexisting with the precursor film. The coarsening advances via a
cascade of two- (and three-)drop mergers based on two mechanisms
related to a volume and a translation mode, respectively
\cite{GlWi03,KaTh07}.  In the volume mode all liquid flows through the
precursor film and the centers of mass of the drops do not move.  In
the translation mode the entire drops approach each other and merge.
The stabilization of the two modes by a substrate heterogeneity is
discussed in Ref.~\cite{TBBB03}. It implies that during coarsening the
major part of the dynamics occurs in the contact line regions. Because
the motion is related to the Goldstone mode of translational
invariance for a single liquid front \cite{KaTh07} the corresponding
eigenvalues are close to zero.  Note that a coarsening step can be
interpreted as a heteroclinic connection between unstable steady
states.
\subsection{The one-dimensional case}\label{sec:dew2d} 
We use the 1d case with disjoining pressure (I) to compare our
  results to the literature \cite{TVNP01}. The initial profile is a
  flat film with a small localized defect at the center.  The Cayley
  transform is used with a regular mesh with $N=1600$ for a domain
  size $\ell=16 \lambda_m$.
\begin{figure} 
{\centering 
\includegraphics[width=6.0cm,height=7cm]{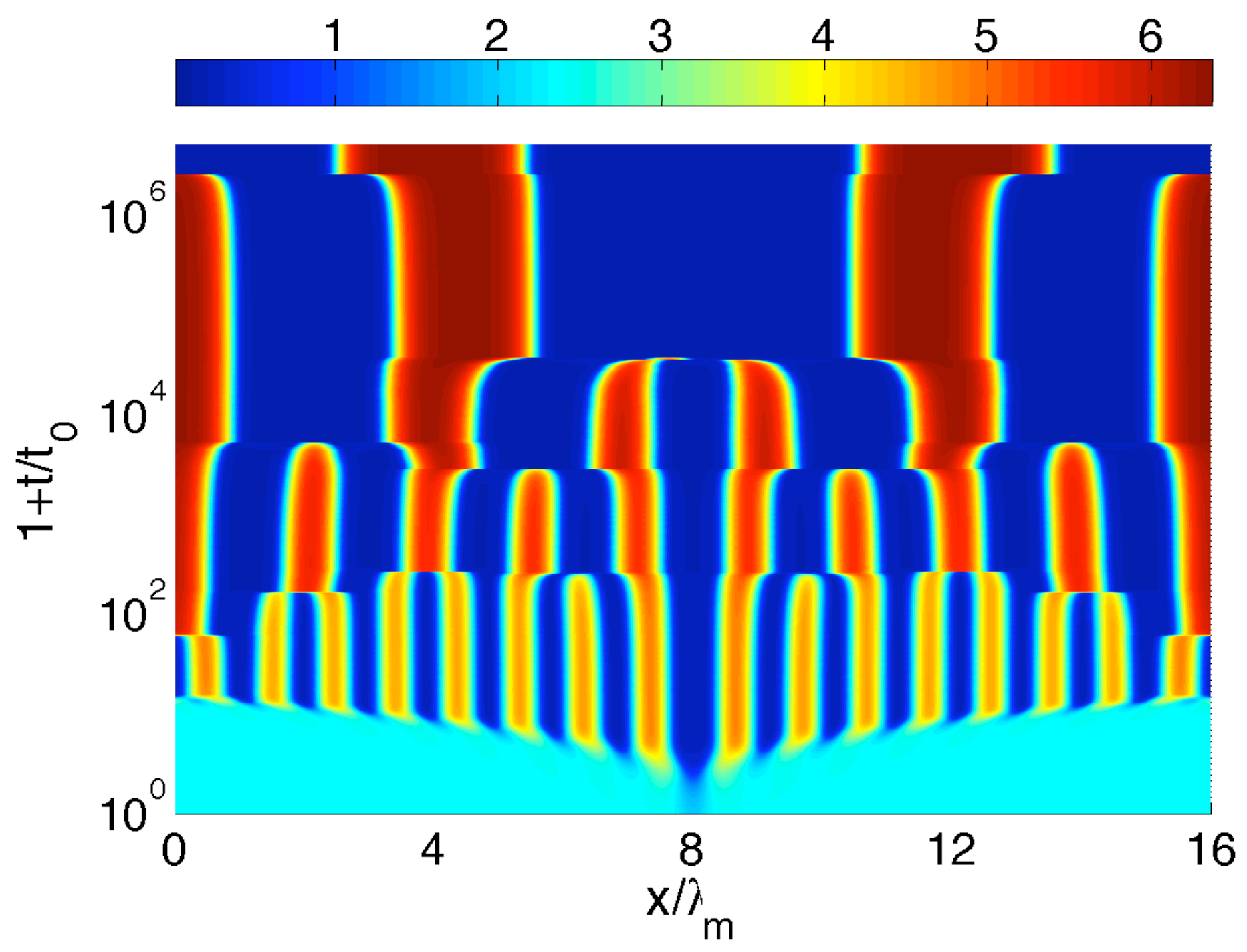} 
\includegraphics[width=6.0cm,height=7cm]{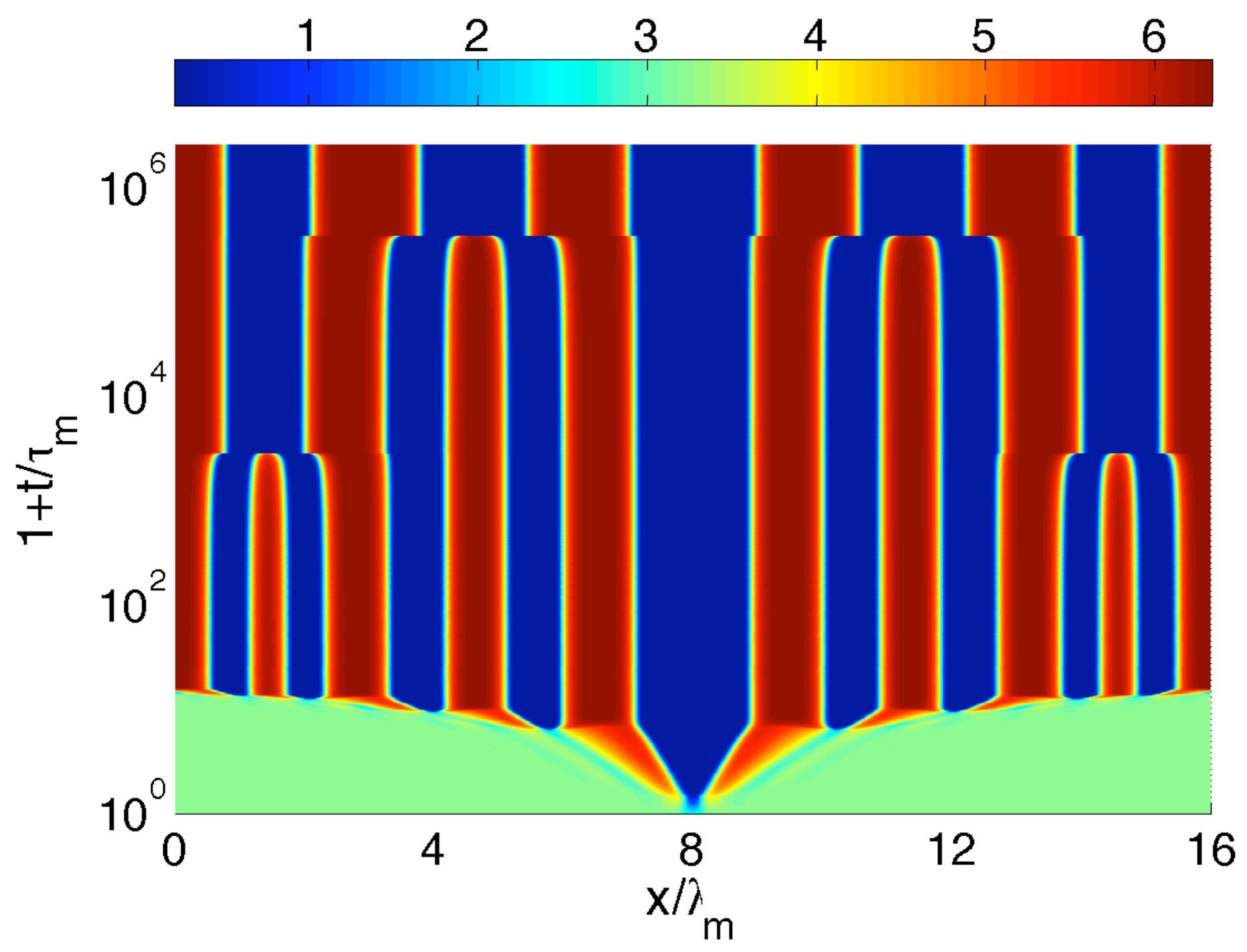} 
} 
\hspace*{3.0cm}(a) \hspace{6.3cm} (b) 
\caption{Space-time contour plots for the time evolution of the film
  thickness profile $h$ during dewetting in (a) the
  surface-instability-dominated regime ($H=2.4$) and (b) the
  nucleation-dominated regime ($H=3.2$). The initial film is linearly
  unstable in both cases. We consider the disjoining pressure (I) with
  $G=0.05$ and the mobility function $m(h)=(h^3-\ln(0.1))^3$ (for
  details see \cite{TVNP01}). The domain size is \(16 \lambda_m\) and
  the initial profile is a flat film with a single defect:
  $h_0=H\left( 1-0.1\cosh(5x/\lambda_m)^{-2}\right)$. The parameters
  agree with the ones used in Fig.~14 of
  {\cite{TVNP01}}. } \label{fig:dew2dp}
\end{figure} 
\begin{figure}\label{fig:dew2dt} 
 \centering 
 (a)\includegraphics[width=10cm,height=4cm]{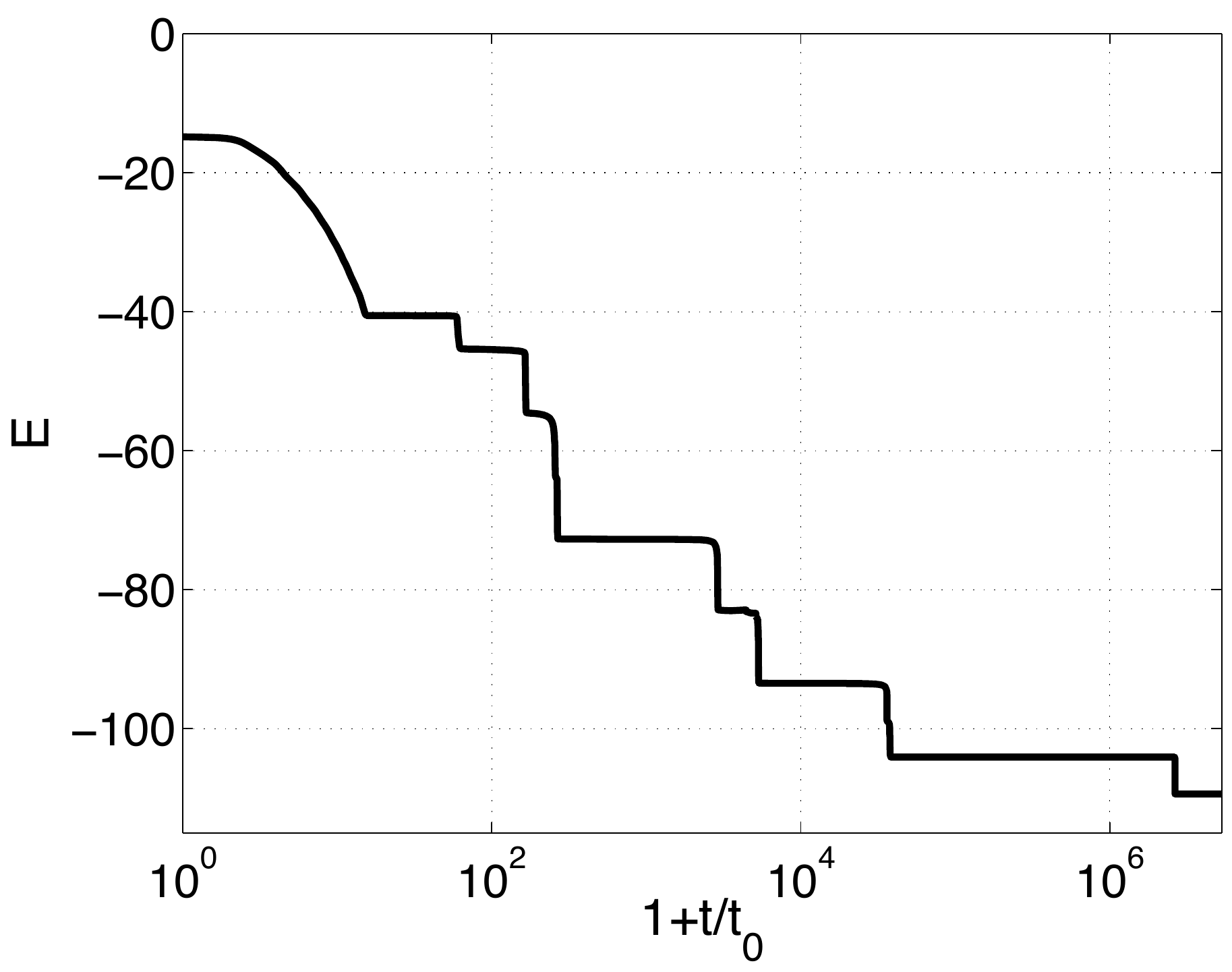}\\ 
 (b)\includegraphics[width=10cm,height=4cm]{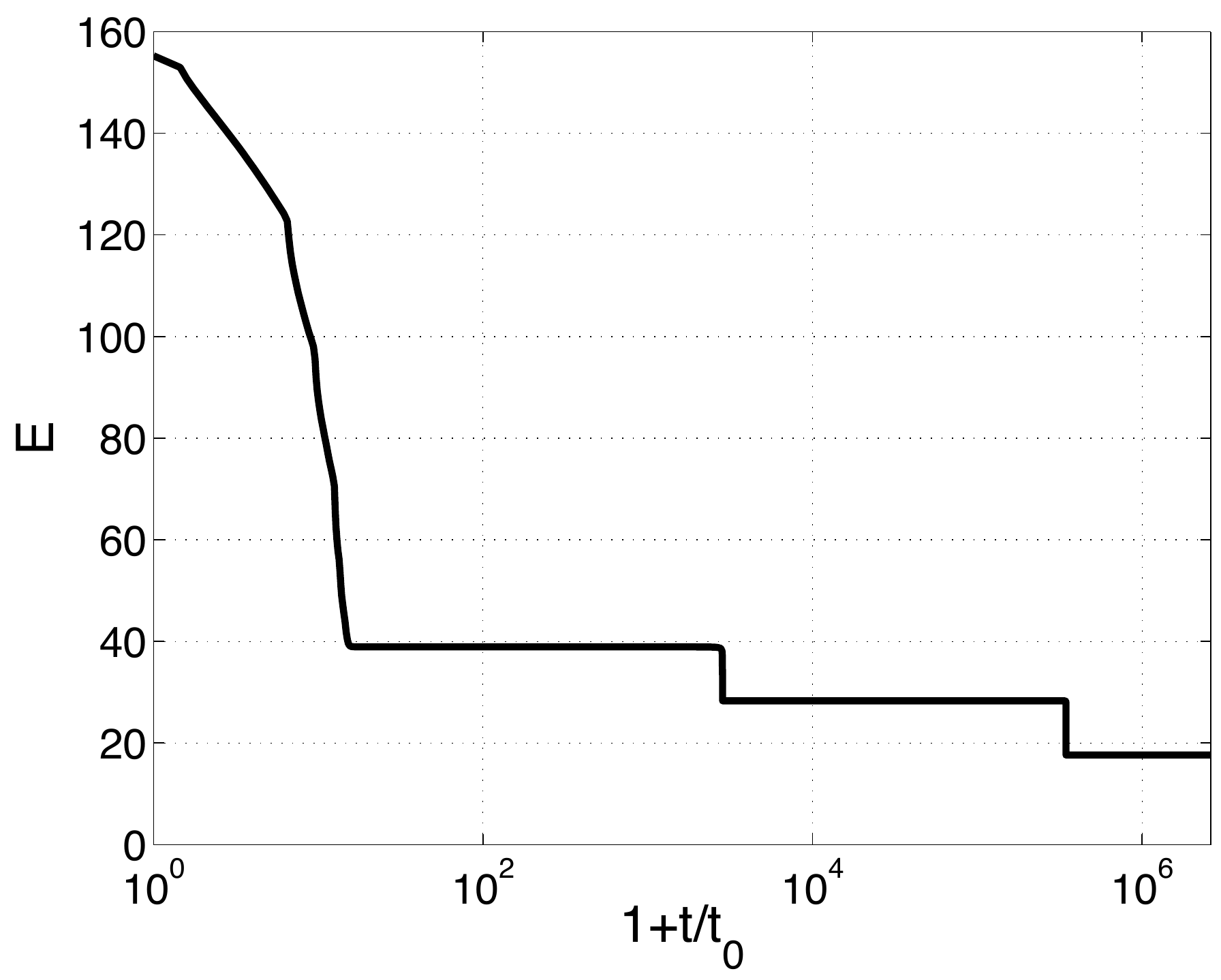}\\ 
 \caption{Change of the energy $E$ with time for dewetting at (a)
   $H=2.4$ and (b) $H=3.2$ corresponding to time evolutions given in
   Fig.~\ref{fig:dew2dp}. The energy $E$ is defined as in
   \cite{TVNP01}: $E=(1/L) \int [(\partial_x h)^2/2 + f(h)]dx$ with
   the local energy $f(h)=-\int\Pi(h)dh$.  }
\end{figure} 

Figures~\ref{fig:dew2dp}(a) and (b) give space-time plots of the
long-time evolution in the cases of dominating surface instability
($H=2.4$) and nucleation ($H=3.2$), respectively.  In
Fig.~\ref{fig:dew2dp}(a) the initial growth ($t\leq15\tau_{m})$
results in a regular array of 16 drops of distance $\lambda_{m}$. The
profile and the evolution of the norm and the relative energy
(Fig.~\ref{fig:dew2dt}) do well agree with \cite{TVNP01} (their
figures 14(a) and 16(a)).  In Fig.~\ref{fig:dew2dp}(b) the growth of
the hole resulting from the finite disturbance is faster than the
surface instability. Further holes are subsequently nucleated by
secondary nucleation events close to the primary hole.  The resulting
short-time structure consists of fewer larger drops than in the
surface instability regime.  The initial 'rupture' phase in
\ref{fig:dew2dt}(b) is in good agreement with Fig.~16(b) of
\cite{TVNP01}.

Our method allows us to study the long-time coarsening far beyond the
results of, e.g., \cite{Thie01,TBBB03}. In Fig.~\ref{fig:dew2dp} one
can identify both coarsening modes in agreement with
Refs.~\cite{GlWi03,KaTh07}.
The merging of drops does not occur continuously slow but starts
extremely slow and culminates very fast. In consequence, the evolution
of the energy (Fig.~\ref{fig:dew2dt}) shows long plateaus connected by
'jumps'.   Our adaptive time-step method copes very well with
this combination of slow and fast dynamics.
 
The coarsening process in the 
nucleation case (Fig.~\ref{fig:dew2dp}(b)) is slower than in the surface instability case and 
proceeds mainly via the mass transfer mode. 
Finally, two [five] drops remain in the surface instability [nucleation] case. In
principle, coarsening should continue, but the evolution becomes so slow
that we reach the limit of numerical accuracy, i.e., the eigenvalues
related to the coarsening modes become smaller than the numerical
accuracy. In particular, for leading eigenvalues smaller than
$10^{-7}$, the numerical noise is not negligible and we are not able
to observe the next coarsening step.
 
We conclude that the developed algorithm is well suited to study
the short- and long-time behavior in the 1d case. In particular, the
short-time evolution agrees well with results obtained using a semi-implicit
scheme with a constant time-step \cite{TVNP01}.  In contrast, here the
time-step varies by 6 orders of magnitude allowing us to study the
long-time coarsening. 
 
\subsection{The two-dimensional case} \label{sec:dew3d} 
\begin{figure}[h]
\centering 
\hspace*{-0.05\hsize}(a)\includegraphics[width=0.39\hsize]{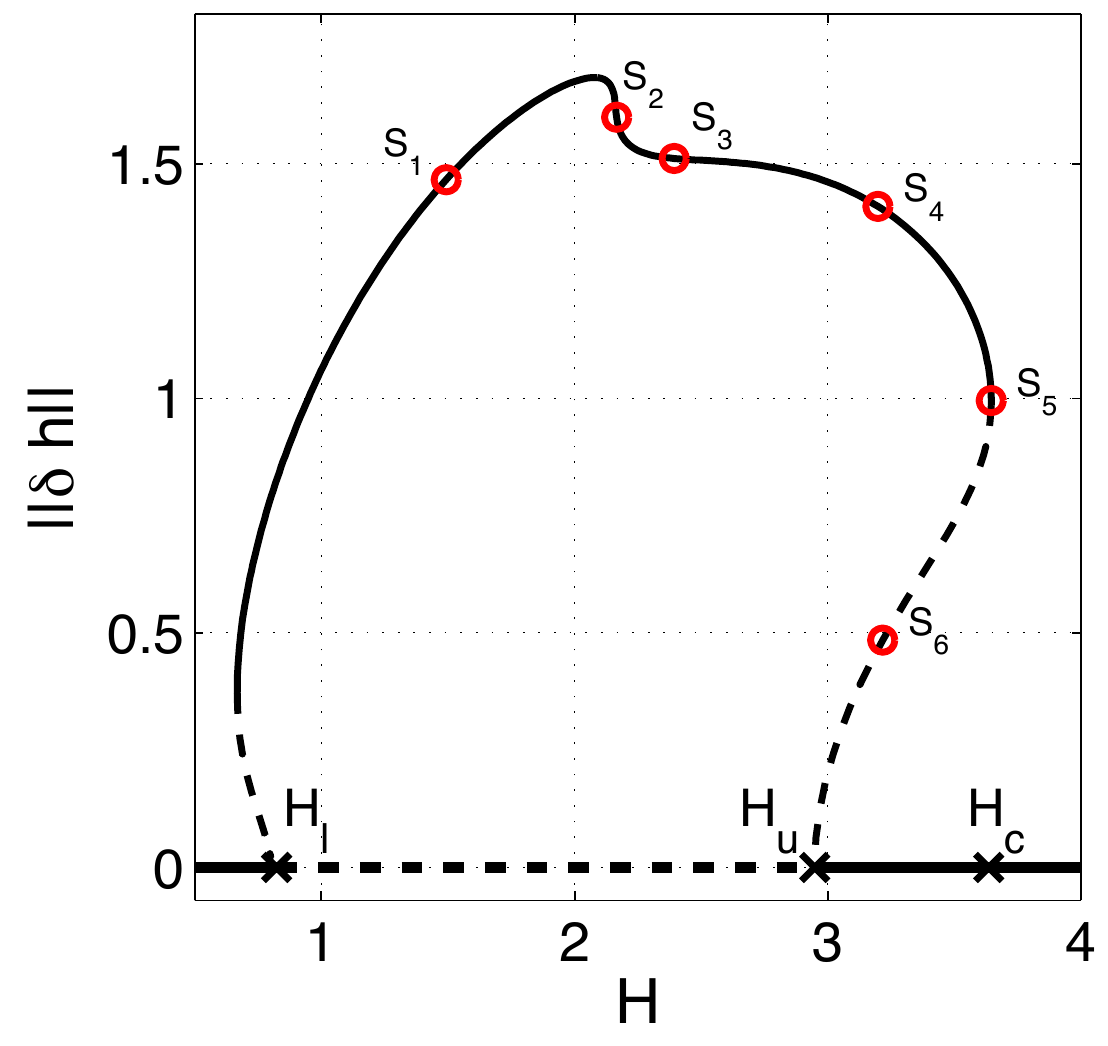} 
(b)\includegraphics[width=0.59\hsize]{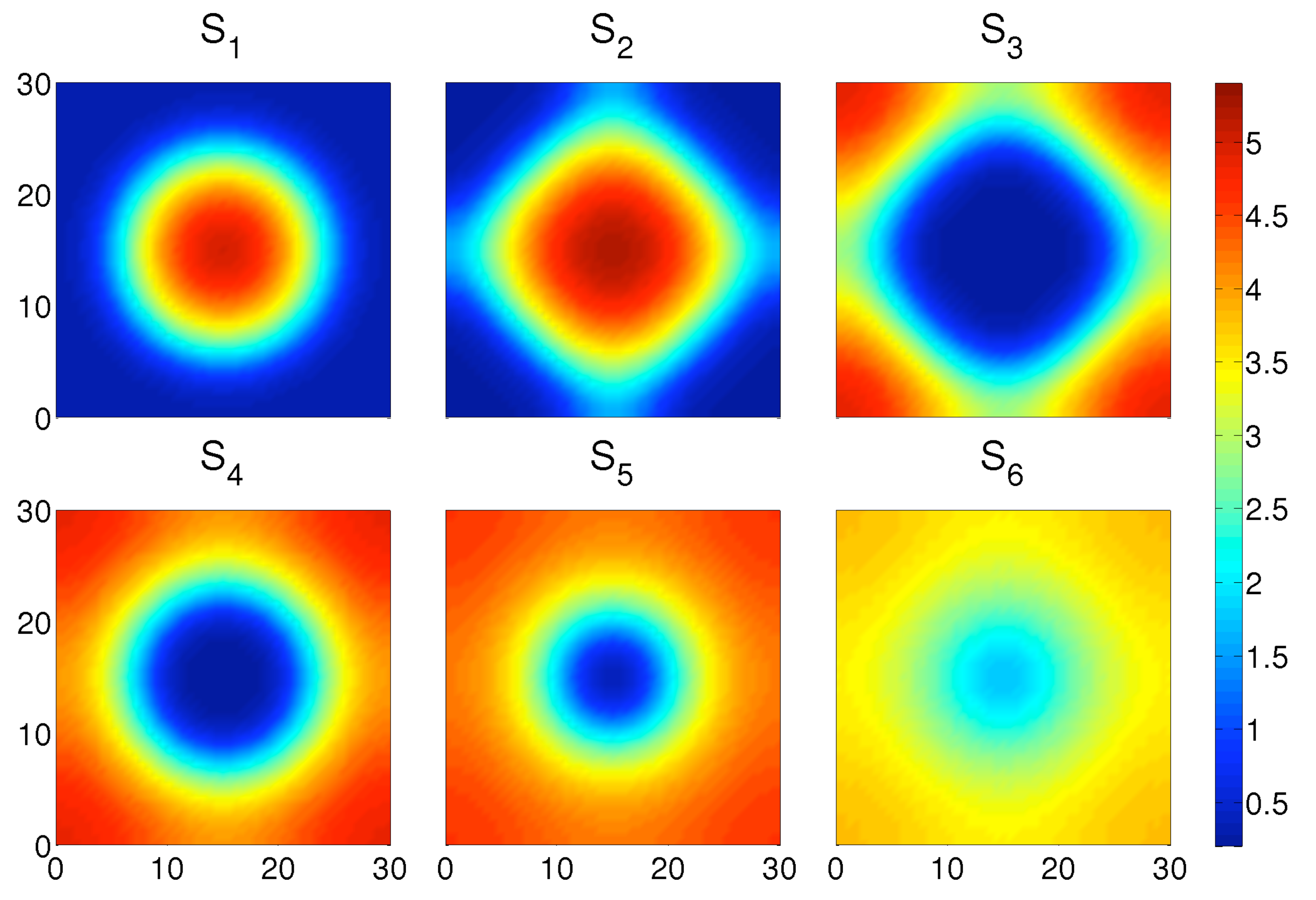}\hspace*{-0.05\hsize}
\caption{(a) Bifurcation diagram for a 2d square array of drops/holes
  on a horizontal homogeneous substrate. The domain size is
  $30\times30$. Shown is the $L^{2}$ norm $||\delta h||$ as a function
  of the mean height of liquid $H$. The steady state solutions may be
  linearly stable (solid line) or unstable (dashed line). (b) Contour
  plots of steady state solutions indicated by circles in the
  bifurcation diagram. With the exception of the control parameter
  $H$, all parameters are as given in Section~\ref{sec:dew3d}.
}\label{fig:conhsq}
\end{figure}

\begin{figure}[t]
\centering 
\includegraphics[width=0.7\hsize]{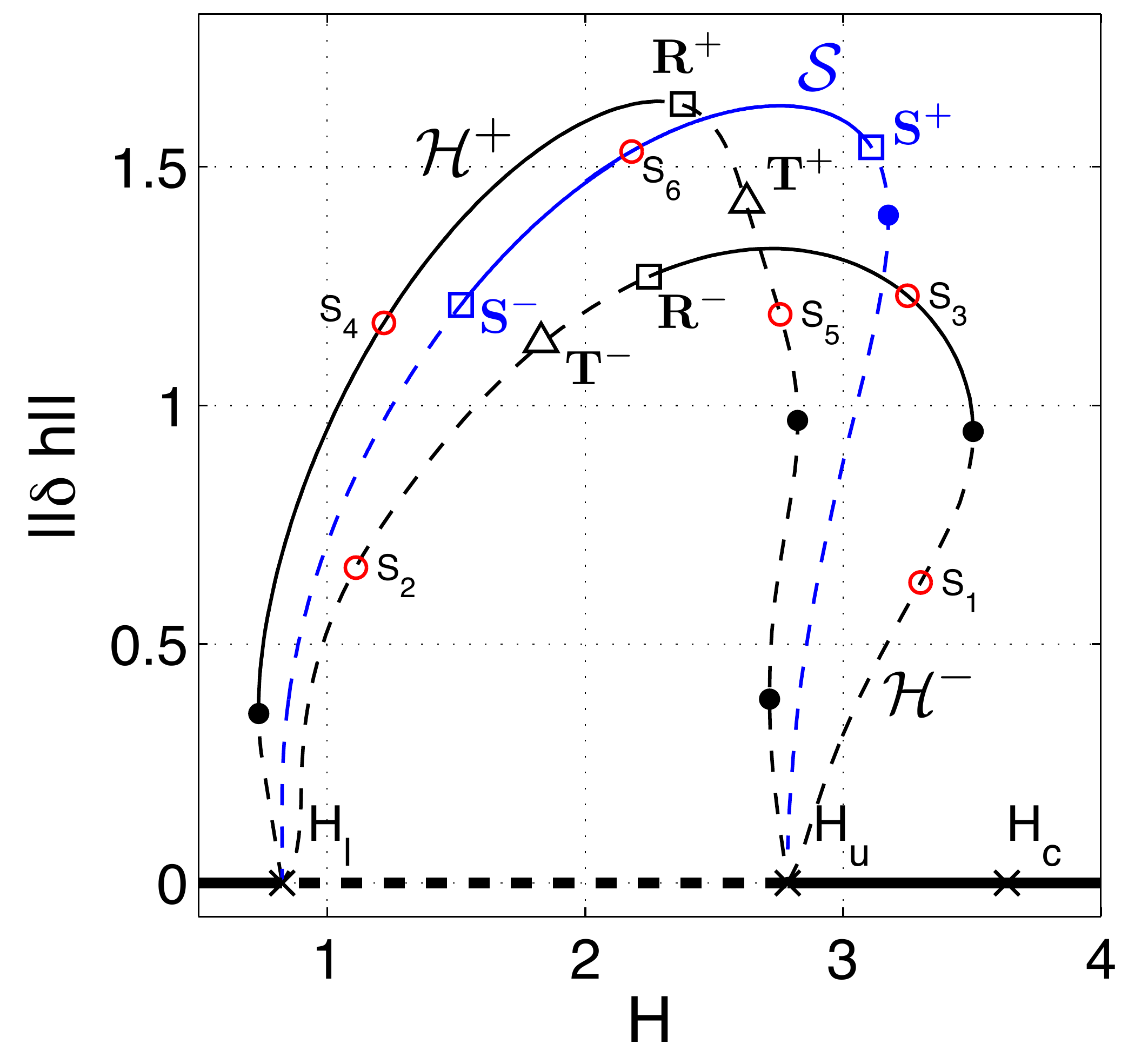} 
\caption{Bifurcation diagram for a 2d hexagonal array of drops/holes
on a horizontal homogeneous substrate. The domain size is
$30\times30\sqrt{3}$. Shown is the $L^{2}$ norm $||\delta h||$ as a
function of the mean height of liquid $H$. Black lines correspond to
the branches of hexagons ($\cal{H}^+$ and $\cal{H}^-$), whereas the
blue line corresponds to the branch of stripes ($\cal{S}$). The steady
state solutions may be linearly stable (solid line) or unstable
(dashed line).  The dots, squares and triangles represent secondary
bifurcations as explained in the main text. The hollow circles
indicate solutions that are displayed in Fig.~\ref{fig:snaphex}. The
remaining parameters are as given in Section~\ref{sec:dew3d}.}\label{fig:conhex}
\end{figure} 
\begin{figure}[th]
\centering 
\includegraphics[width=0.9\hsize]{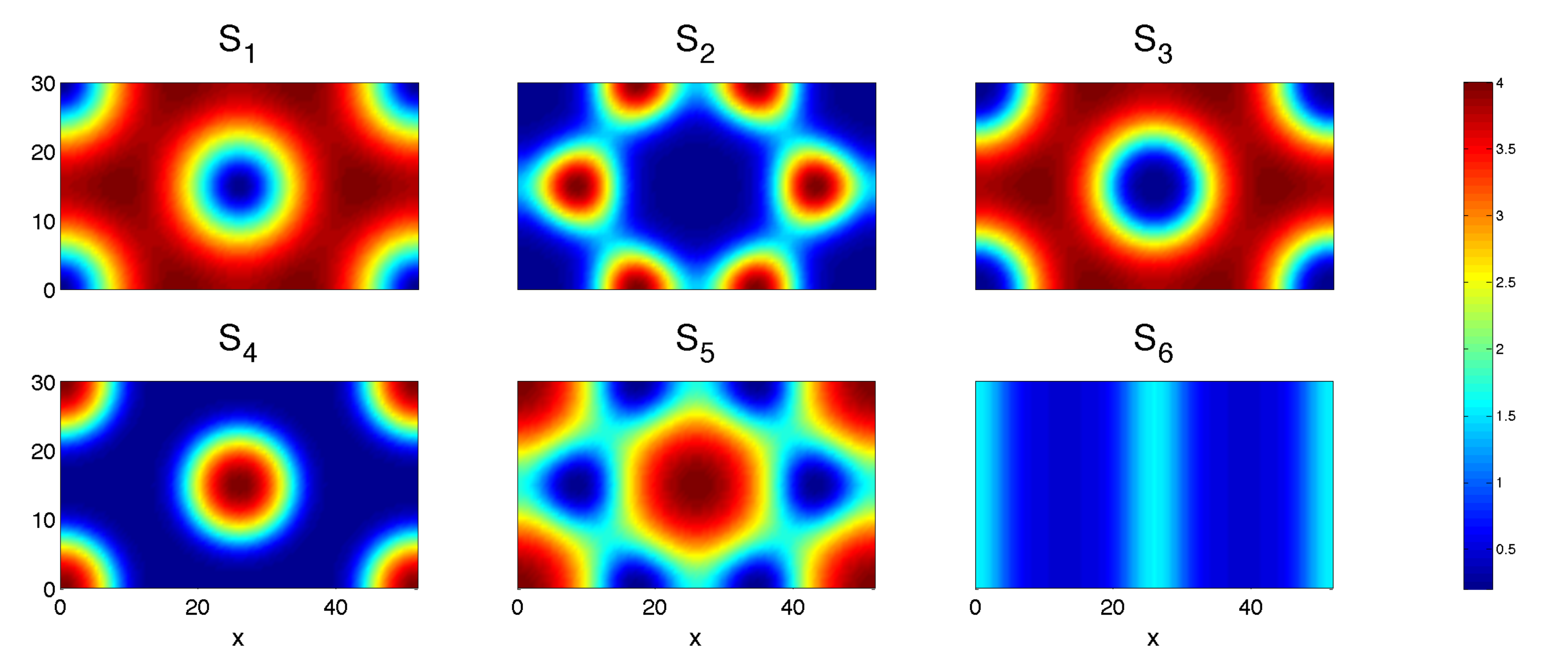}
\caption{Contour plots of steady state solutions indicated by
hollow circles in the bifurcation diagram of Fig.~\ref{fig:conhex}. In
particular, the upper row gives three hole solutions from the
$\cal{H}^-$ branch ($S_1$ to $S_3$), whereas the lower row gives two
drop solutions from the $\cal{H}^-$ branch ($S_4$ and $S_5$) and a
typical stripe solution ($S_6$). Of the shown solutions only $S_3$, 
$S_4$ and $S_6$ are linearly stable.}\label{fig:snaphex}
\end{figure} 
After having shown the reliability of our algorithm for 1d dewetting,
we next employ it to study the 2d case.  The above discussion of the
linear stability still applies. In particular, the fastest growing
wavelength $\lambda_m$ and corresponding growth rate $\beta_m$ are
identical.  However, in contrast to the 1d case, 2d patterns involve
two wave vectors $\mathbf{k_1}$ and $\mathbf{k_2}$ that can lead to a
variety of periodic steady state patterns \cite{Conway92}.  Here, we
track square and hexagonal patterns by imposing a periodic $30\times
30$ square and a $30\times30\sqrt{3}$ rectangular domain,
respectively. Choosing the mean film thickness $H$ as control
parameter we obtain the bifurcation diagrams in
Figs.~\ref{fig:conhsq}(a) and \ref{fig:conhex}(a), respectively. Note
that the finite domain size results in critical film thicknesses
different from the one for an infinite domain. In particular, one
finds $H_u=2.79$ for the upper limit instead of $H_c=3.63$ expected
for an infinite domain.  Fig.~\ref{fig:conhsq}(a) shows that for the
square pattern at $H_u$ an (unstable) subcritical branch bifurcates
from the trivial one. It stabilizes and turns towards smaller $H$ at a
saddle-node bifurcation at $H_{\mathrm{sn}}\simeq H_c$. In
consequence, for $H_u<H<H_{sn}$ the flat film is metastable and can
only dewet via nucleation that allows it to pass the unstable
threshold solution. In a similar way, the lower
  critical thickness $H_l=0.825$ differs slightly from the one for an
  infinite domain ($H_{cl} =0.747$, not shown in the bifurcation
  diagrams).%
All steady state profiles above $H\approx2.2$ correspond to hole
patterns (see Fig.~\ref{fig:conhsq}(b)).  Further decreasing $H$, a
morphological transition occurs (related to the steep variation in the
norm at $H\simeq2.2$) via a state of a rotated (by $\pi/4$, see states
$S_2$ in Fig.~\ref{fig:conhsq}(b)) checkerboard pattern to drop
patterns that always prevail at smaller $H$.  At another saddle-node,
the branch turns and becomes unstable again before subcritically
joining the flat film state at $H_1=0.84$.
  
Fig.~\ref{fig:conhex} shows branches of hexagonal symmetry that bifurcate
from the trivial solution. It does as well give the secondary
bifurcations.  Note, however, that here we do not show the secondary
branches although they are discussed in passing in the following.
In contrast to the square patterns in Fig.\ref{fig:conhsq}, the
  branches of hexagonal patterns do not display a transition between
  drops and holes at finite amplitude.  However, at each of the two
bifurcation points ($\bf H_l$ and $\bf H_u$) the hexagonal branch
crosses the trivial flat film solution branch in a transcritical
bifurcation. In the representation of Fig.~~\ref{fig:conhex} this
corresponds to two families of hexagons that seem to emerge at each
bifurcation point.  Ones of them consist of a hexagonal array of drops
(see states $S_4$ and $S_5$ in Fig.~\ref{fig:snaphex}) and is
equivalent to the $\cal{H}^+$ branch discussed in \cite{CrKn91}. The
other one represents a hexagonal array of holes (states $S_1$ to
$S_3$ in Fig.~\ref{fig:snaphex}) and corresponds to the $\cal{H}^-$
branch \cite{CrKn91}.

Each of these branches connects the two bifurcation points $\bf H_l$
and $\bf H_u$.  Additionally to the branches of hexagons, a branch of
stripe solutions $\cal{S}$ appears through a supercritical pitchfork
bifurcation at $\bf H_l$ and ends at a subcritical pitchfork
bifurcation at $\bf H_u$.  All these branches are unstable near the
primary bifurcation as expected for the generic case of bifurcations
with hexagonal symmetry \cite{CrKn91}.

The branches gain and loose stability through a number of secondary
bifurcations in a scenario that is similar to one described in
\cite{BuGo83}: The branch $\bf\cal{H}^+$ that emerges at $\bf H_l$ first
continues towards smaller $H$, then turns and gains stability at a
saddle-node bifurcation. It continues towards larger $H$ and loses its
stability again at a transcritical bifurcation (point $\bf R^+$ in
Fig.~\ref{fig:conhex}).  Note that it undergoes another two saddle-node
bifurcations before approaching $\bf H_u$.  The branch that crosses at
$\bf R^+$ has rectangular symmetry; one side connects to the bifurcation
point $\bf S^+$ on the branch of stripe solutions ($\cal{S}$); the other
one crosses the $\cal{H}^-$ branch transcritically at $\bf R^-$ before
ending at $\bf S^-$ on the branch of stripe solutions.

A similar scenario occurs for the branch $\cal{H}^-$ when starting at
the transcritical bifurcation at $\bf H_u$: It continues towards
larger $H$, then turns and gains stability at a saddle-node
bifurcation. It continues towards smaller $H$ and loses its stability
at $\bf R^-$.  Finally, the branch of stripe solutions $\cal{S}$ that
emerges at $H_l$ gains stability at $\bf S^+$ and loses it again at
$\bf S^-$. The saddle-node bifurcation at larger $H$ does not change
its stability.

This introduces all bifurcations necessary to understand the stability
of the primary branches. Note however, that there exist further
secondary bifurcations ($\bf T^+$ and $\bf T^-$ in
Fig.~\ref{fig:conhex}) involving a branch of triangular solutions. As
discussed in \cite{BuGo83}, this branch connects the two hexagonal
branches $\cal{H}^+$ and $\cal{H}^-$.

\begin{figure} 
\centering 
\mafig{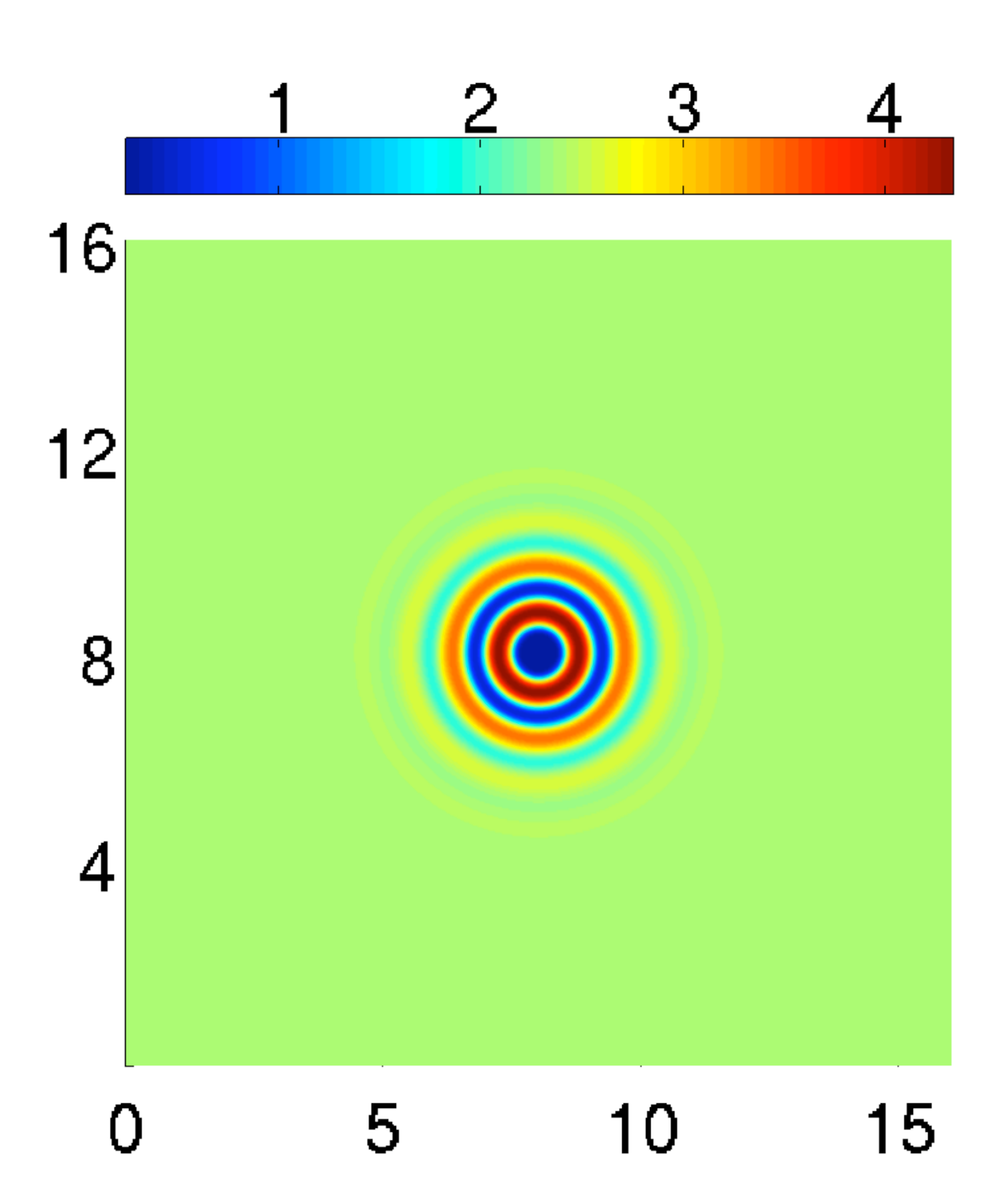} 
\mafig{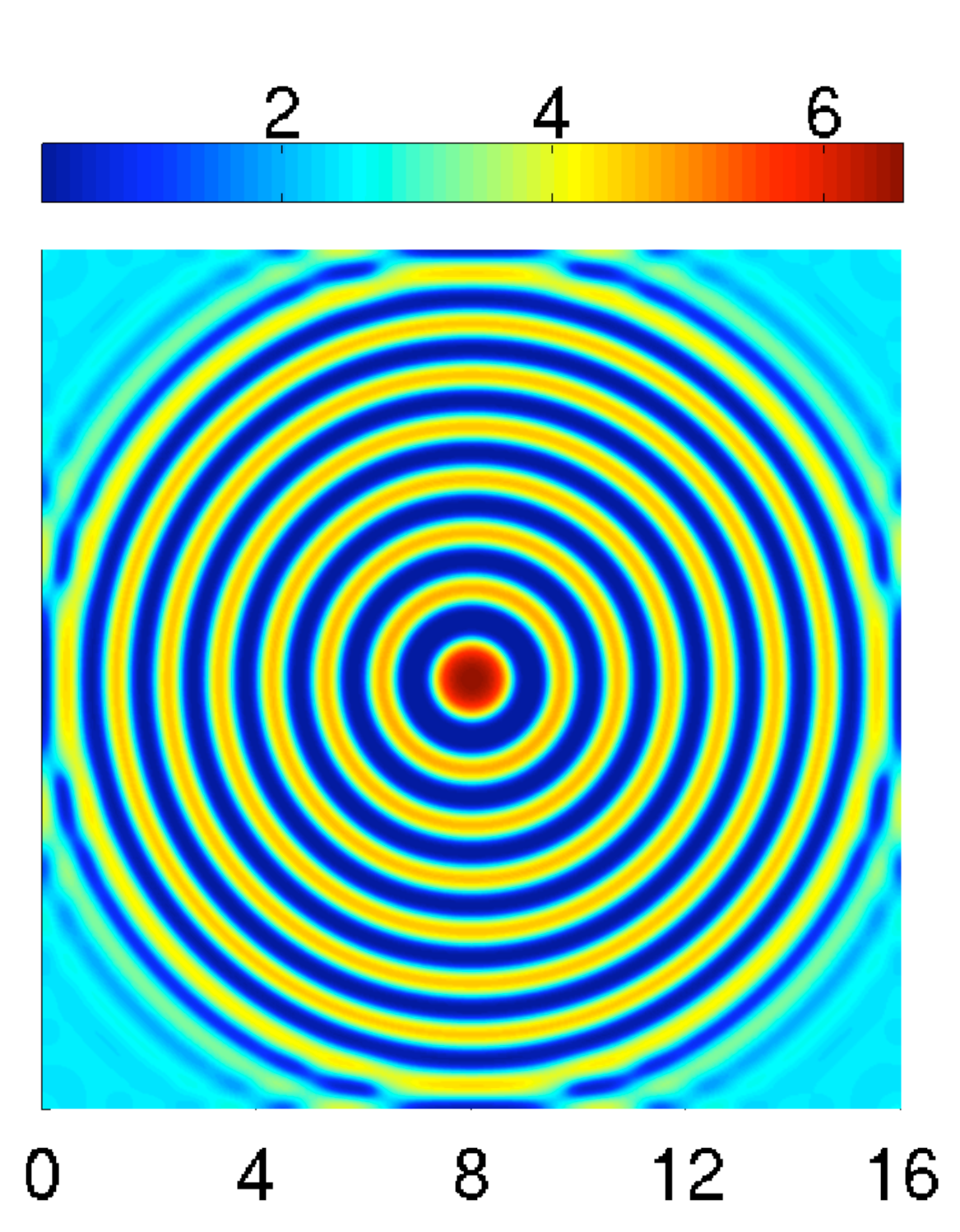} 
\mafig{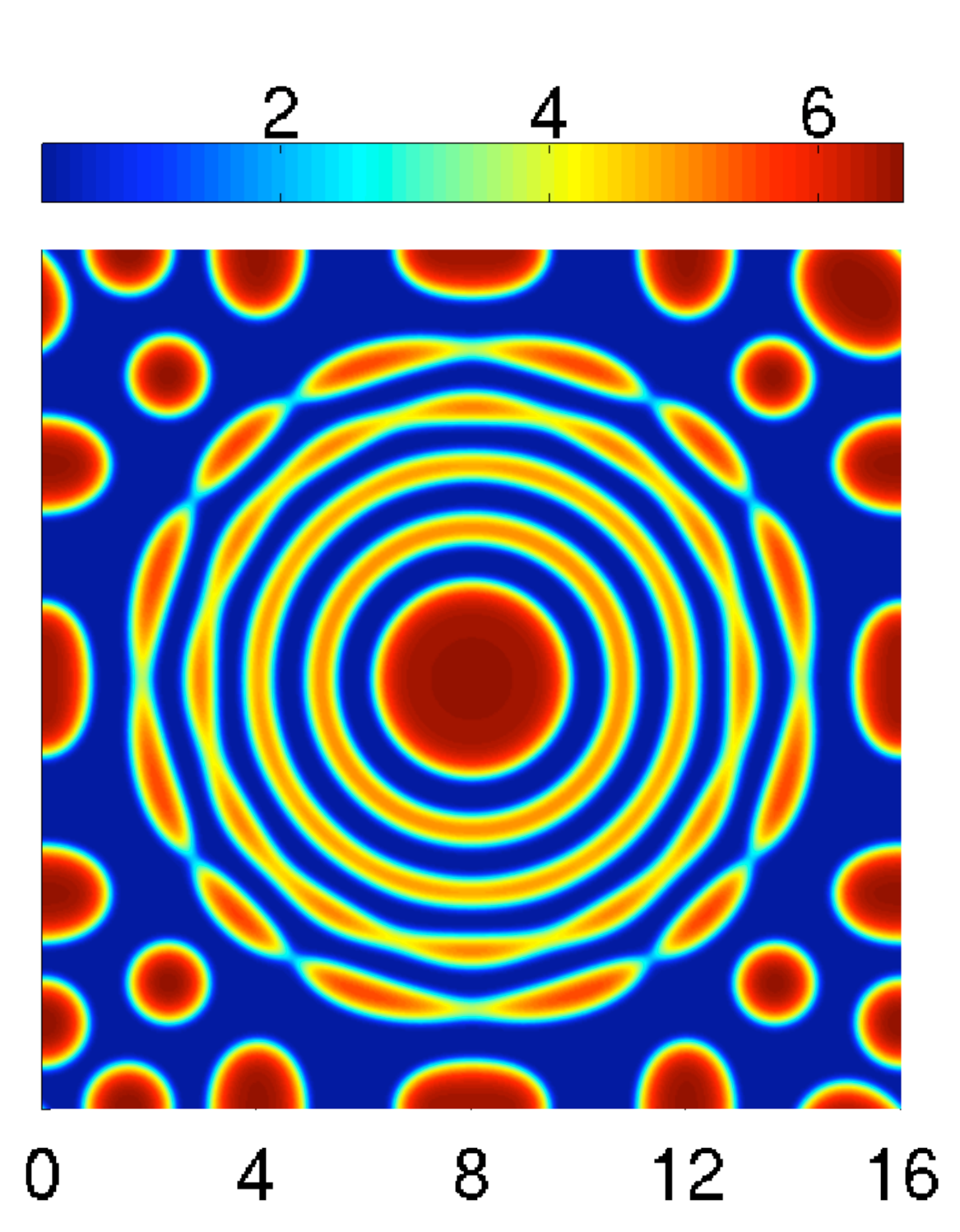} 
\mafig{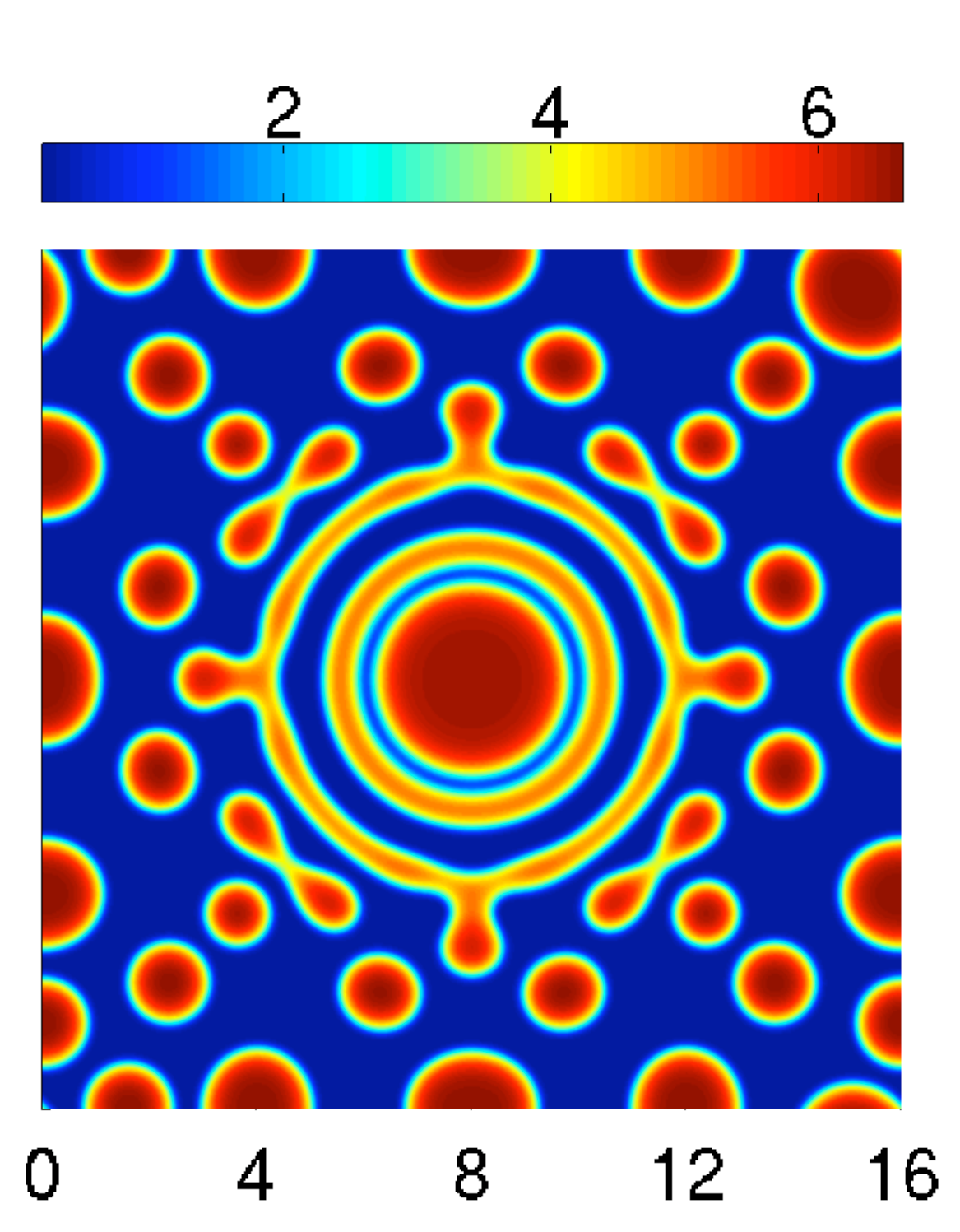} 
\mafig{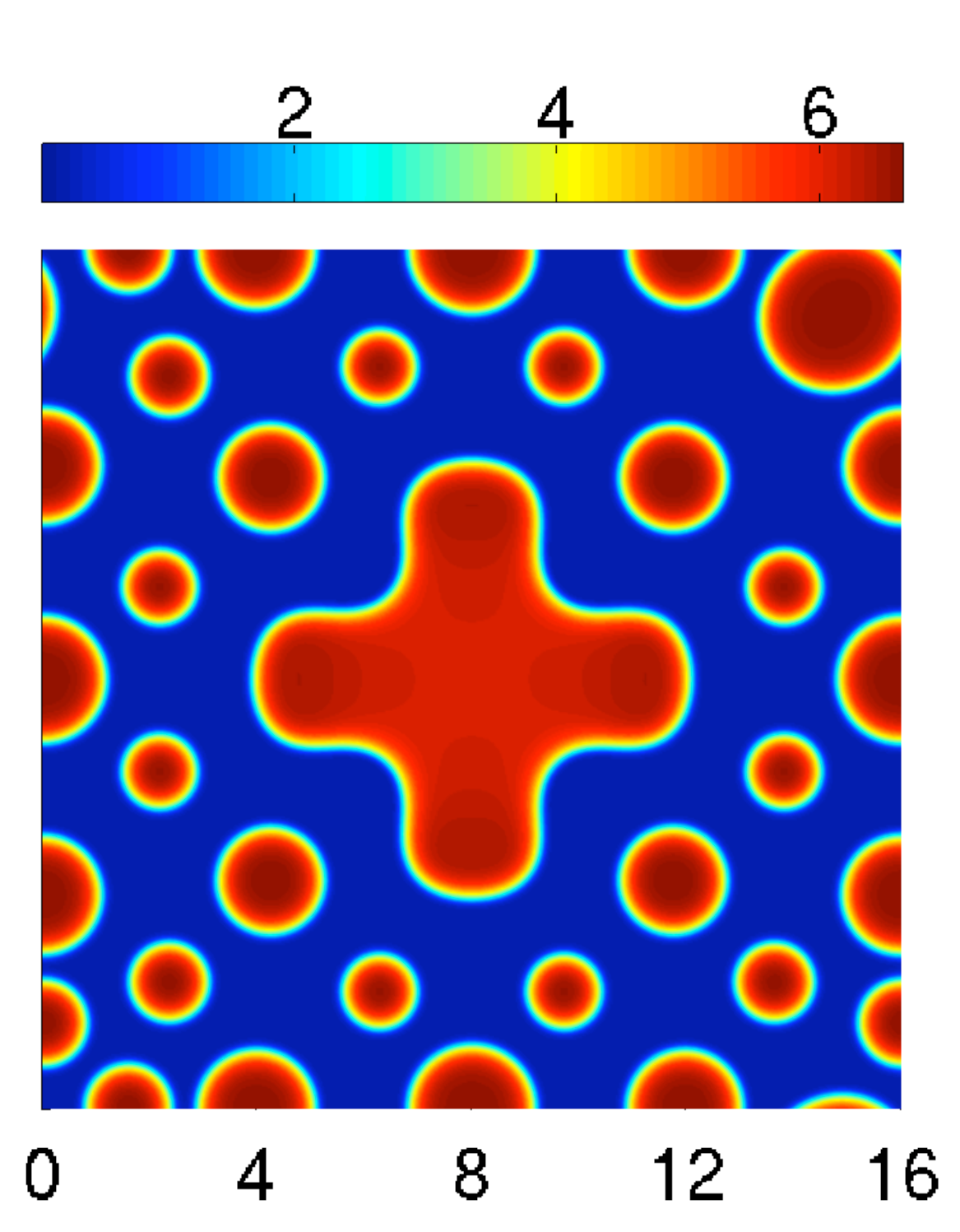}\\ 
(a)\temps{5.57}{15.4}{122}{259}{467}\\
\mafig{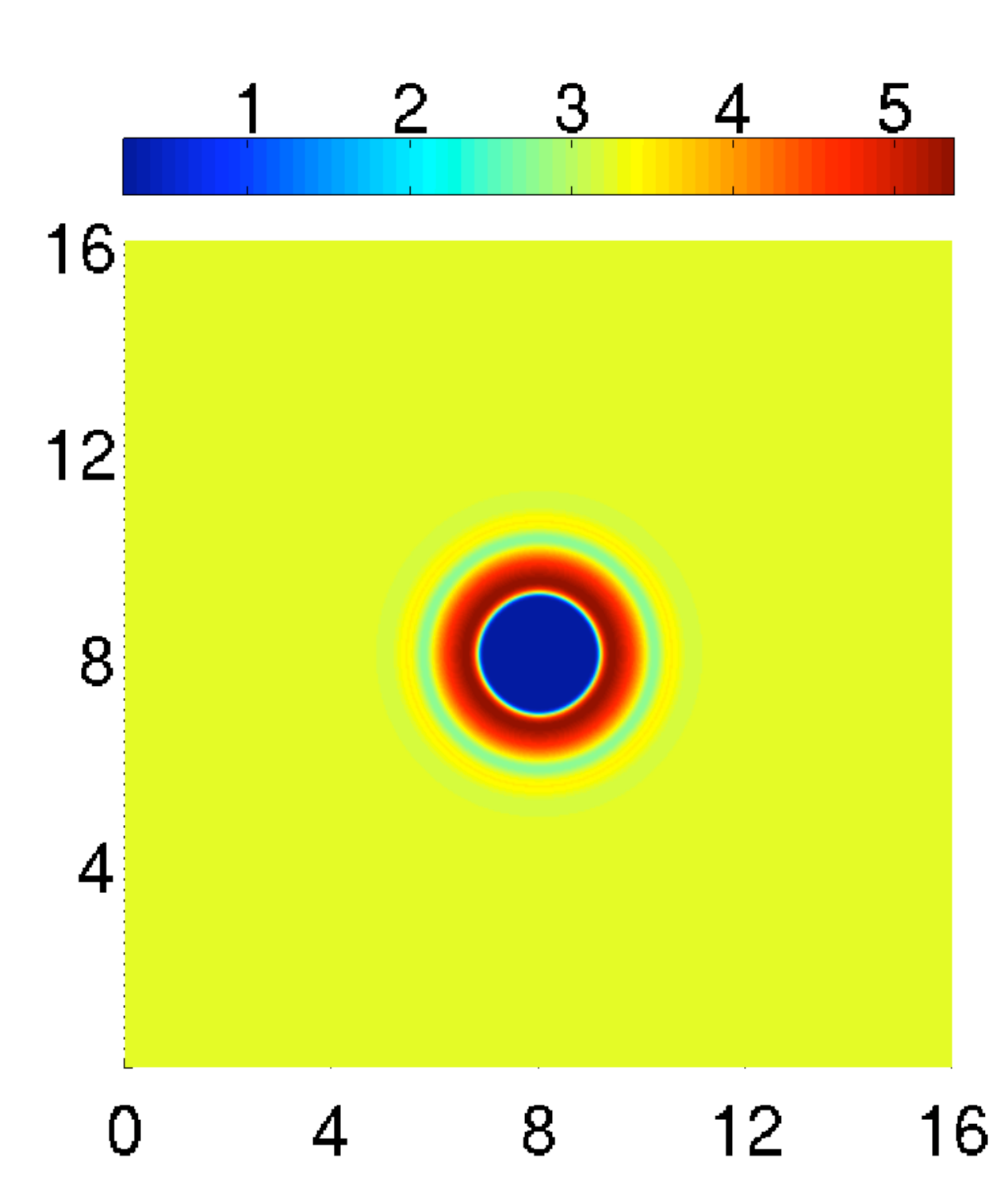} 
\mafig{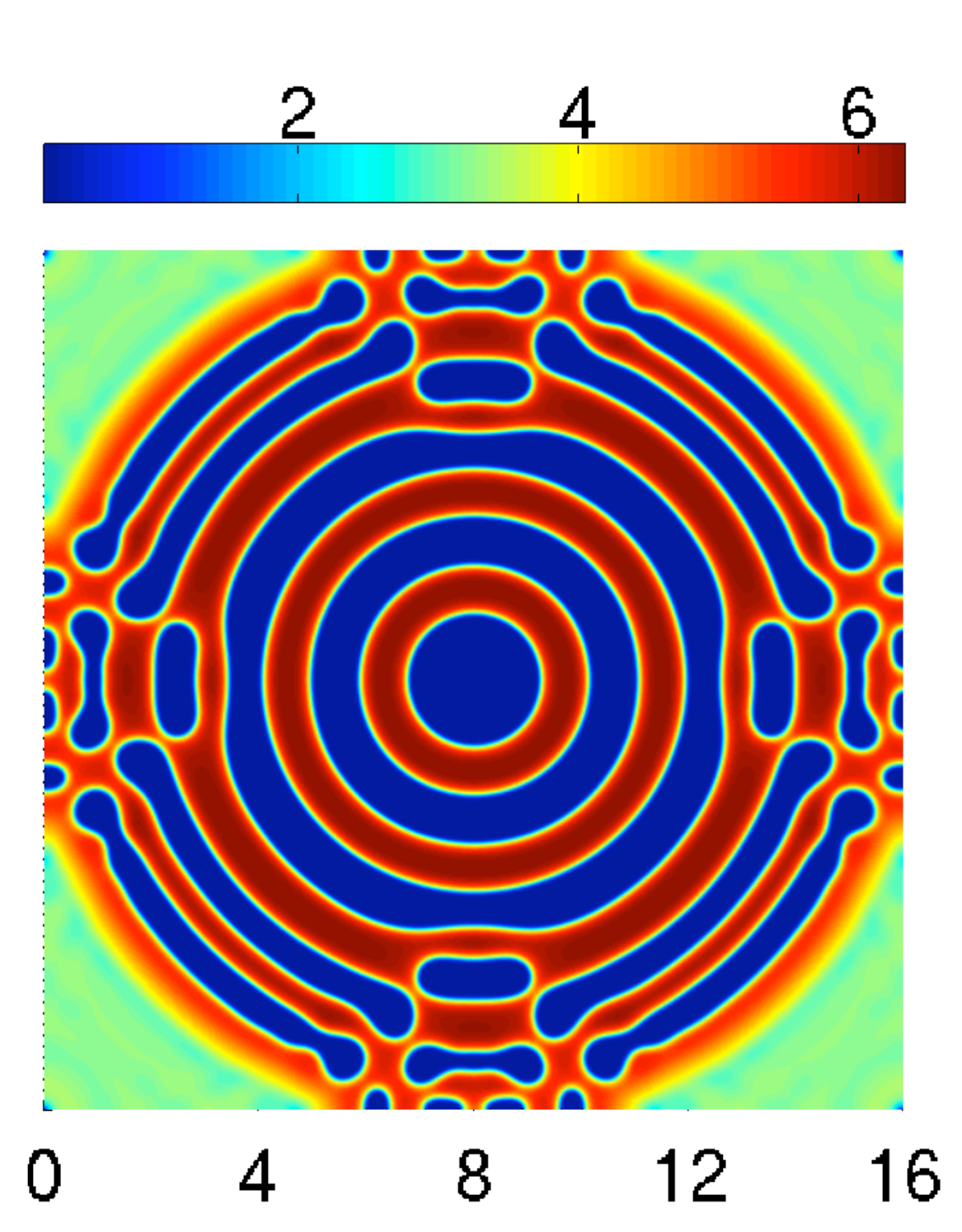} 
\mafig{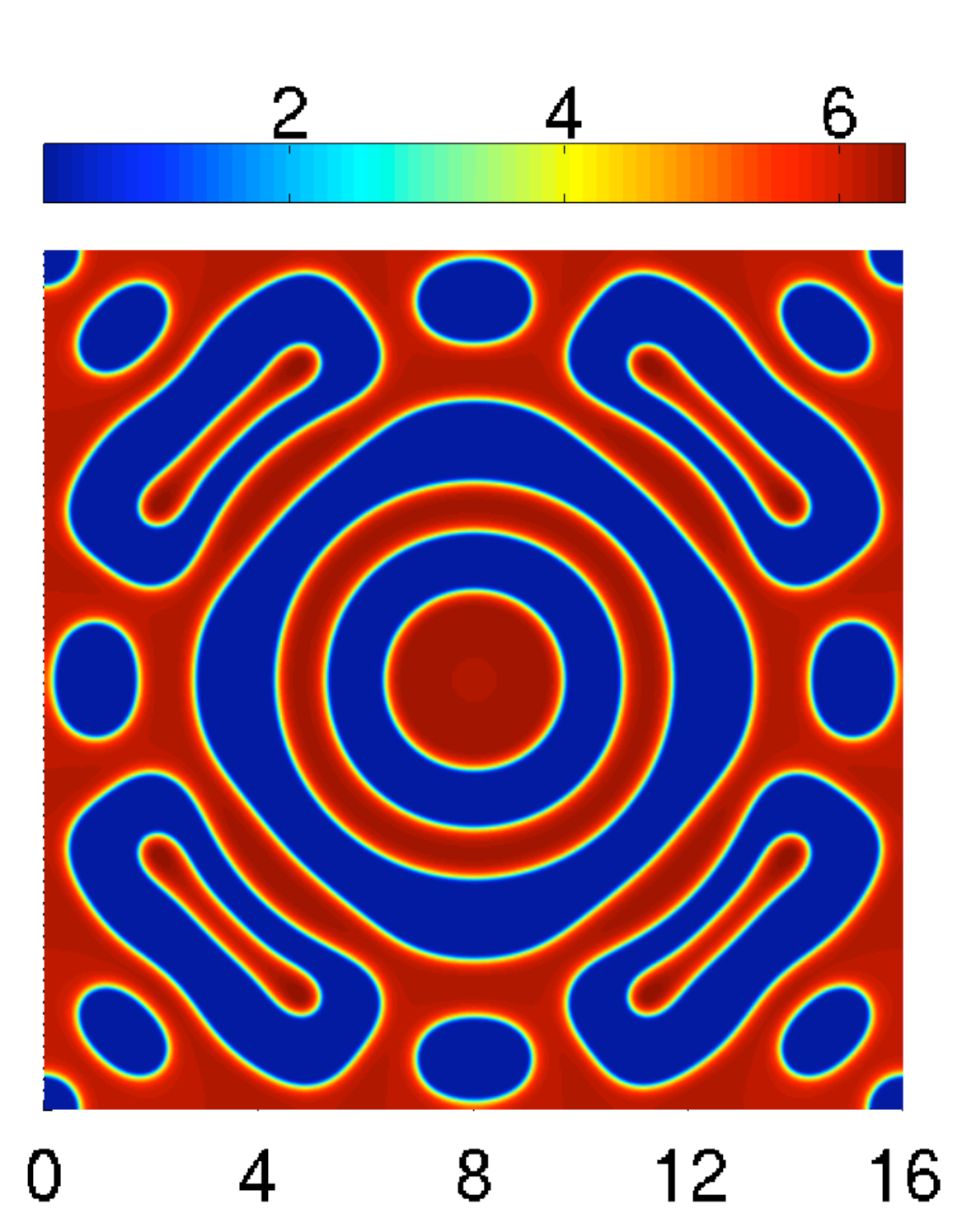} 
\mafig{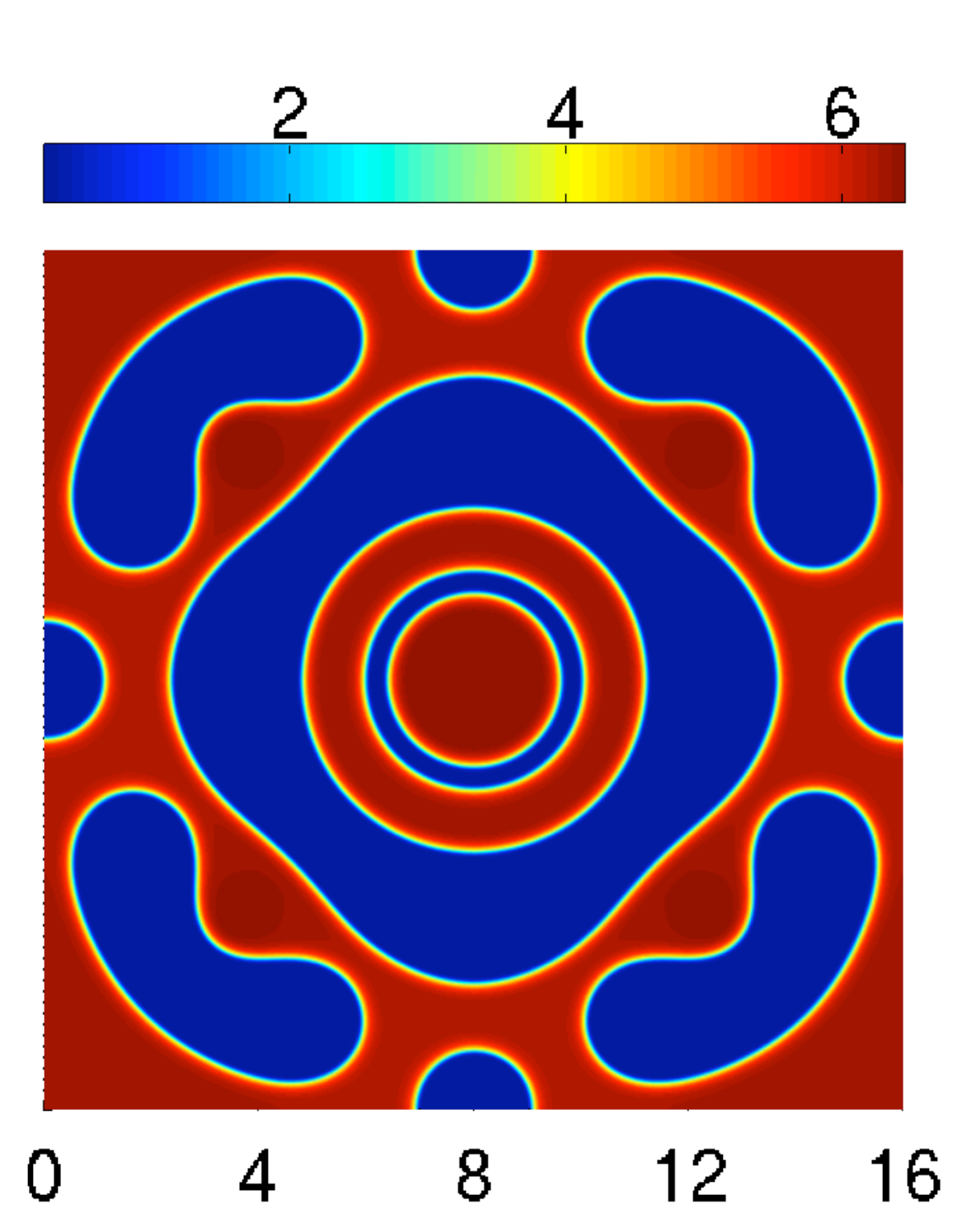} 
\mafig{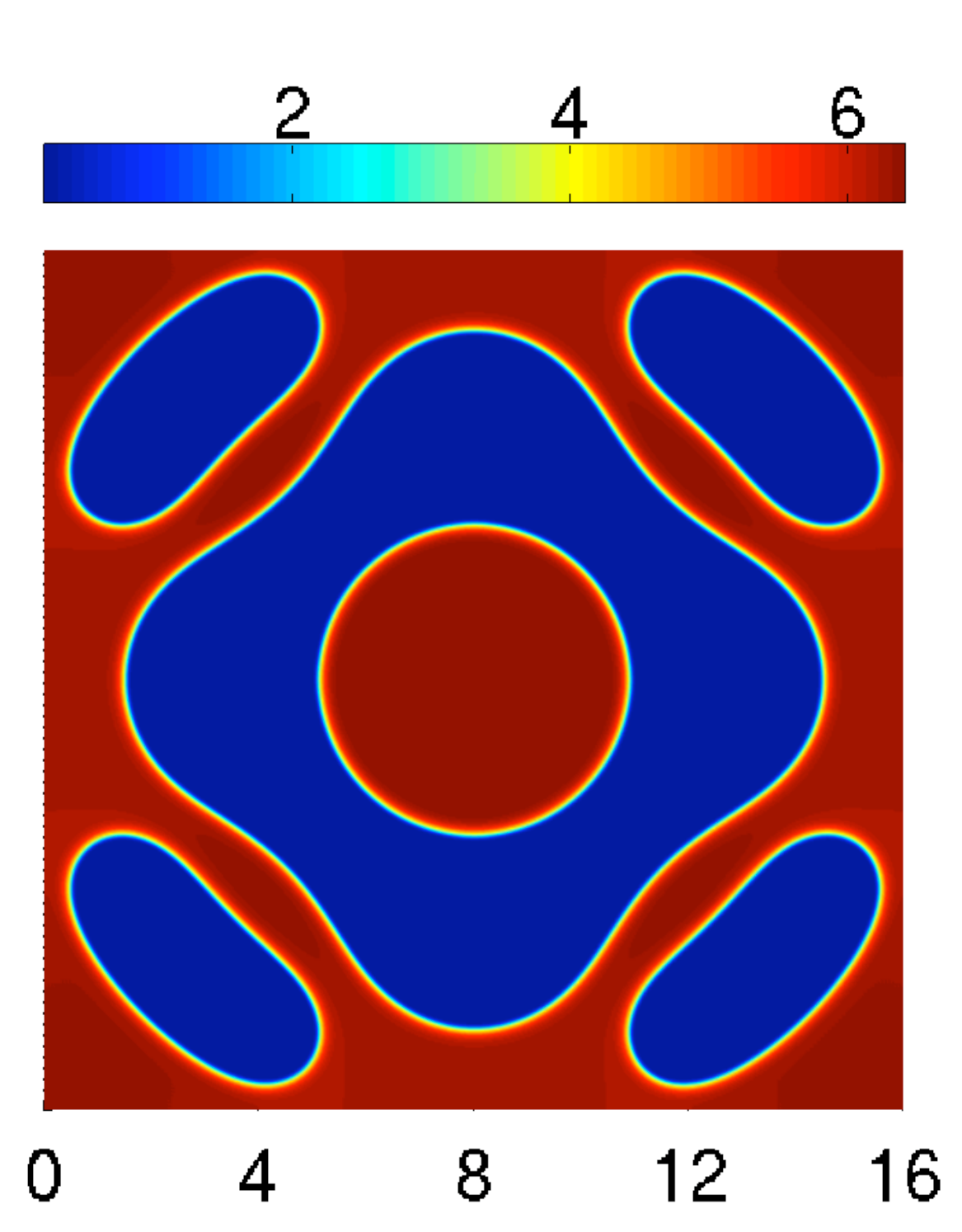}\\ 
(b)\temps{5.60}{14.5}{54.4}{140}{326} 
\caption{\label{fig:dew3dd} Snapshots from the evolution of dewetting
  thin films. The initial condition corresponds to a flat film with a
  central defect: $h_0=H\left( 1-0.1\cosh(5r/\lambda_m)^{-2}\right)$
  in the (a) surface-instability-dominated regime at $H=2.4$ and (b) in the
  nucleation-dominated regime at $H=3.2$.  The domain size is \(16
  \lambda_m\times 16 \lambda_m\) and we use periodic boundary
  conditions. The time is indicated below the individual panels in
  units of $\tau_m$. The remaining conditions are as in
  Fig.~\ref{fig:dew2dp}. }
\end{figure}

Although the bifurcation analysis is very instructive for a small
domain, it does not allow one to directly predict which mechanism
dominates 2d dewetting for larger system sizes. To study this question
we employ our time-stepping algorithm. The use of different
  initial disturbances allows one to discuss the prevalence of either
  dewetting by nucleation or by surface instability for linearly
  unstable films: We chose an radially symmetric defect of profile
  $h(r)$ (identical to the $h(x)$ in the 1d case) that shall compete
  with the surface instability that emerges from an initial film
  roughness.  Without
  initial roughness one finds that the short-time evolution conserves
  the radial symmetry until the boundaries of the square box are
  felt. In the instability dominated case ($H=2.4$,
  Fig.~\ref{fig:dew3dd}(a)) the distance between rings corresponds as
  expected to $\lambda_m$, whereas in the nucleation-dominated regime
  ($H=3.2$, Fig.~\ref{fig:dew3dd}(b)) it is roughly $2\lambda_m$ and
  the boundary is felt earlier. After the ring formation during the
  short-time evolution, the rings break and coarsening sets in.  In
  the radially symmetric part coarsening starts at the center and
  proceeds through a cascade of ring contractions. 

\begin{figure} 
\centering 
\mafig{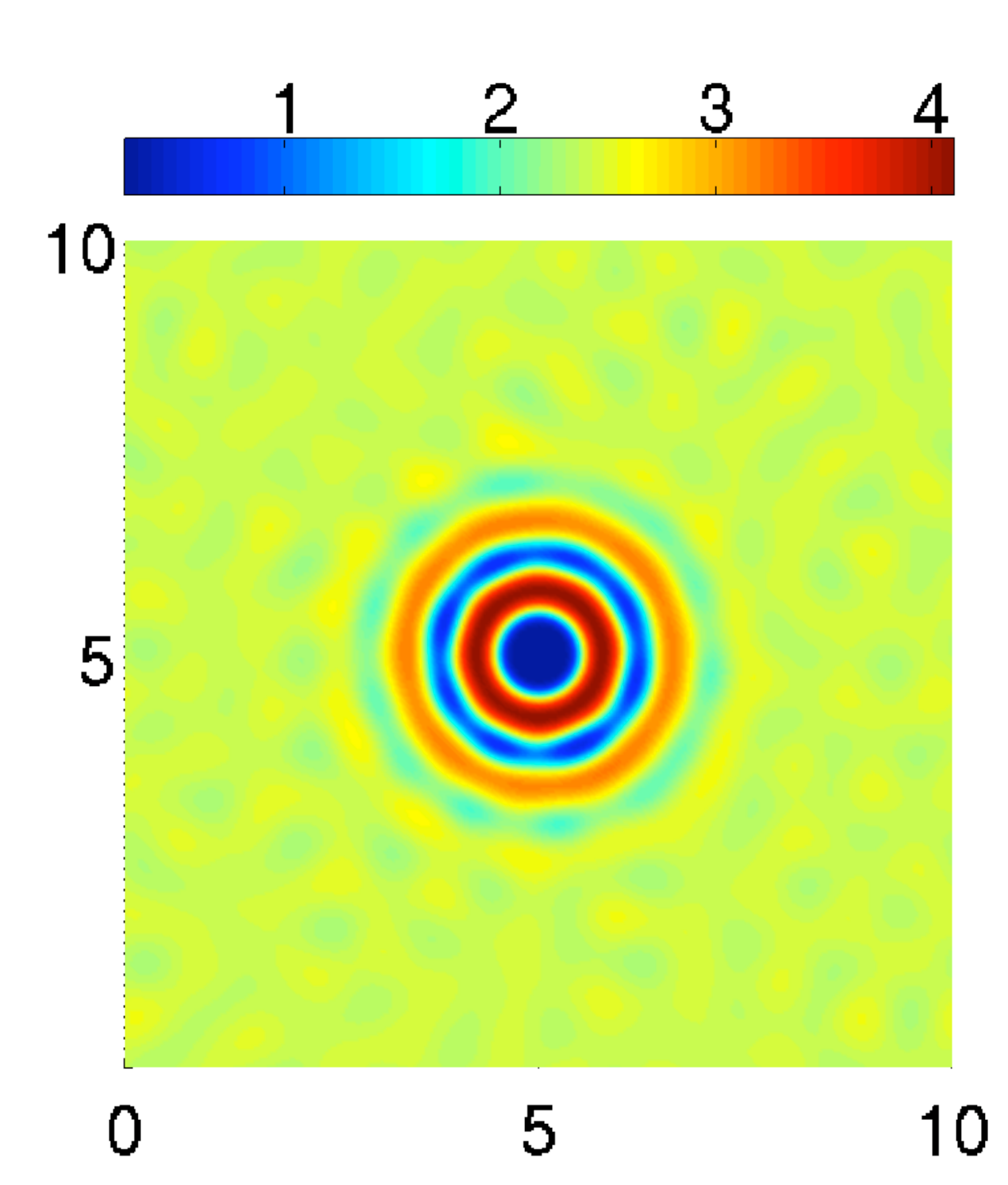} 
\mafig{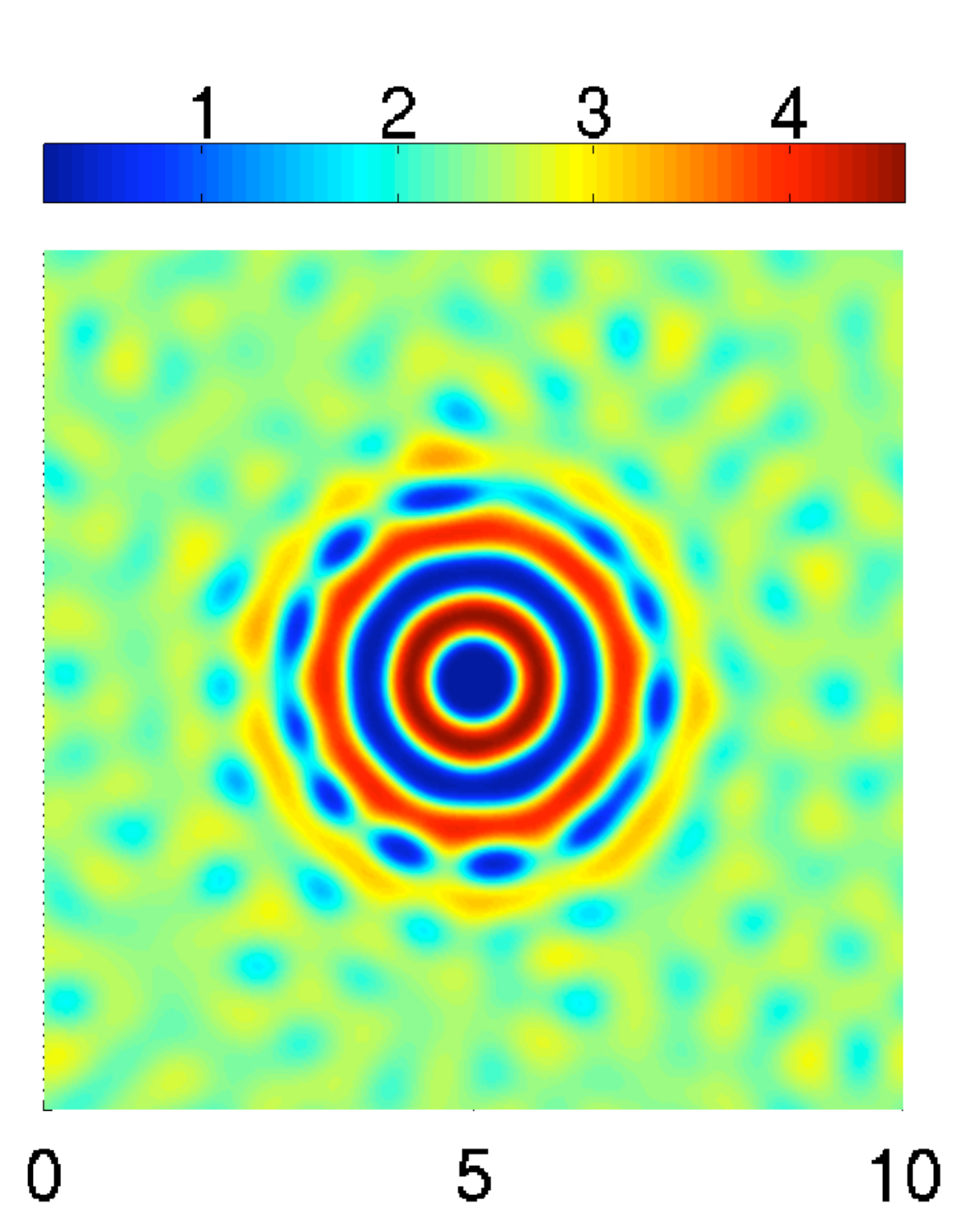} 
\mafig{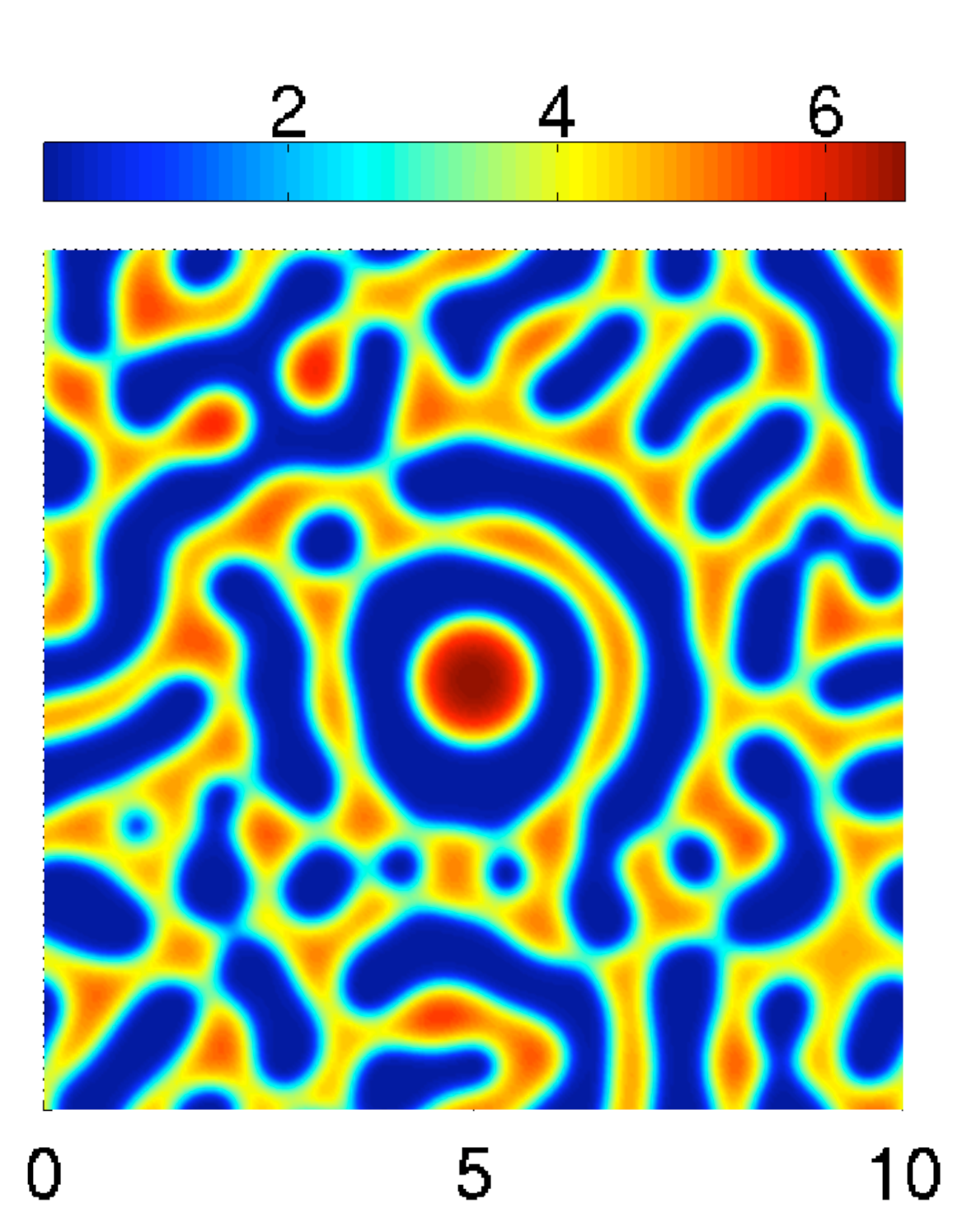} 
\mafig{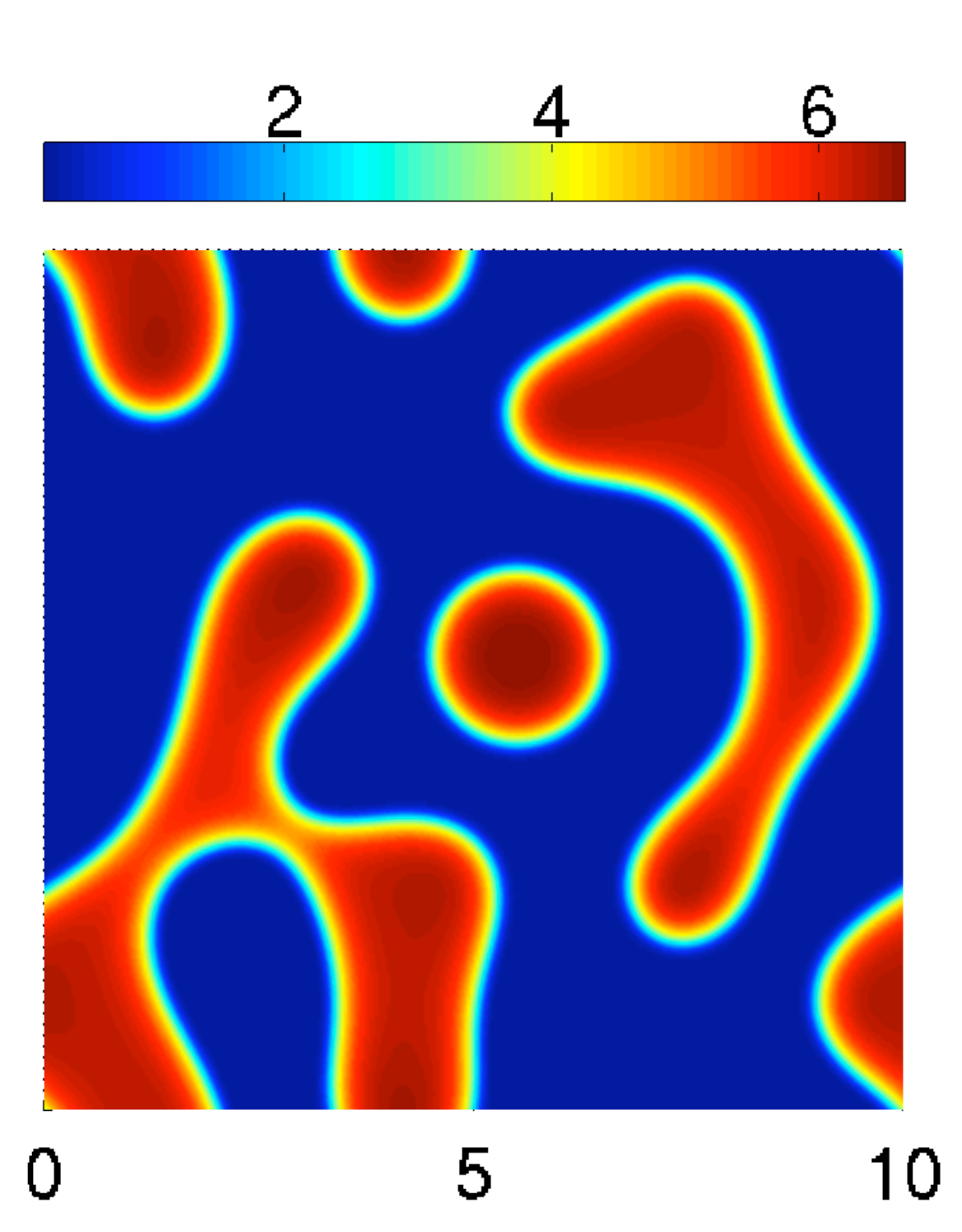} 
\mafig{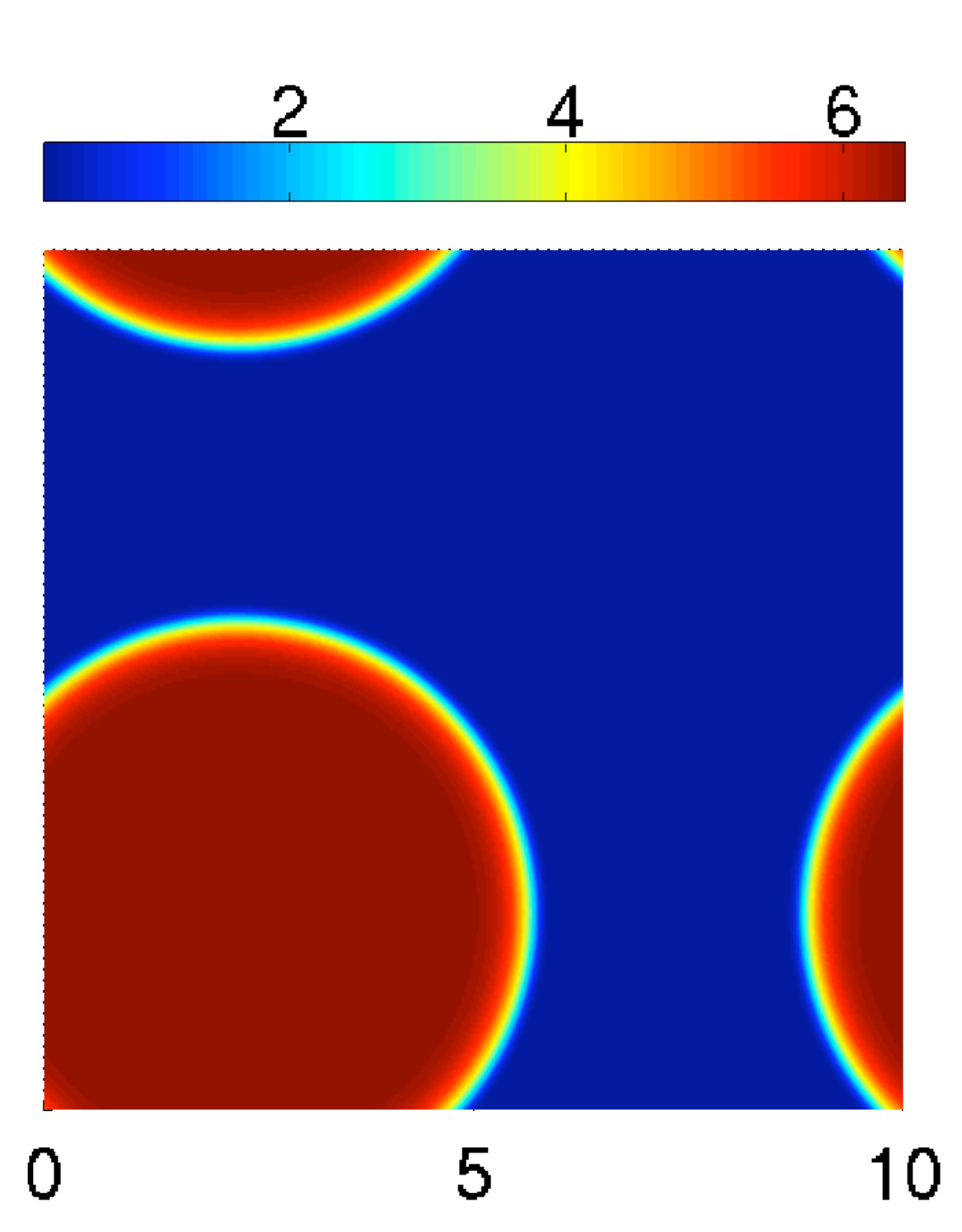}\\ 
(a)\temps{4.92}{6.19}{15.7\cdot10}{118}{3.25\cdot10^4}\\ 
\mafig{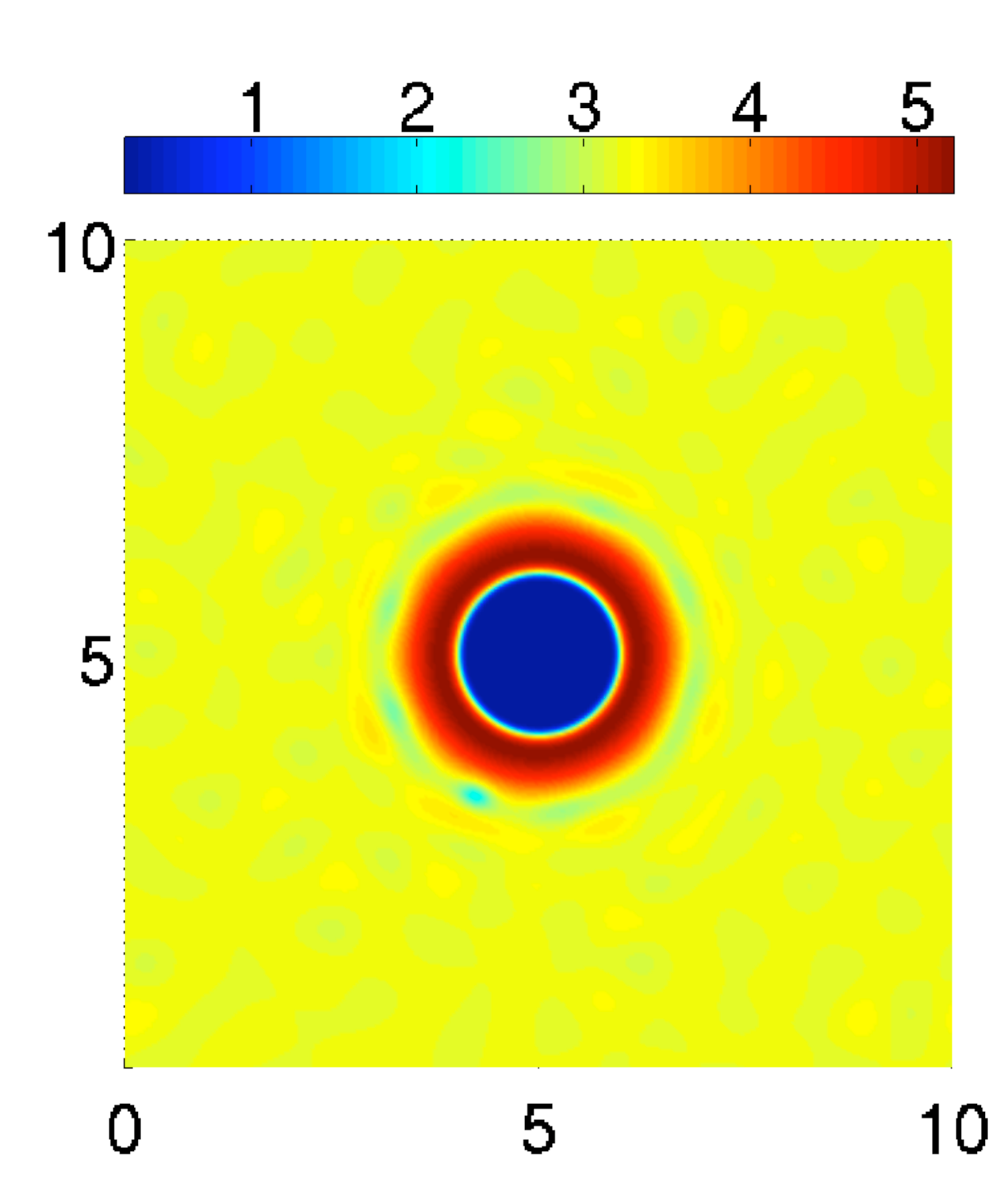} 
\mafig{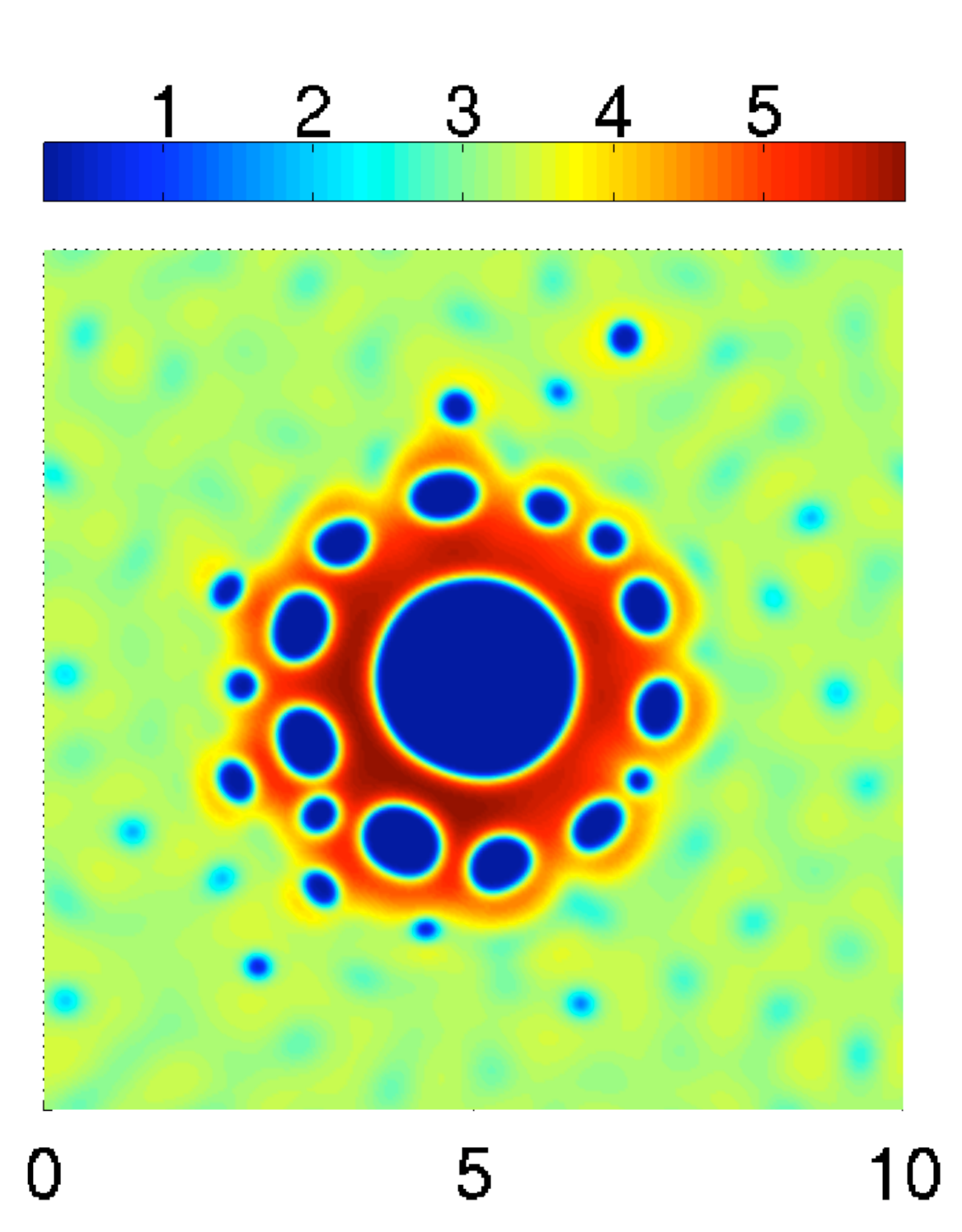} 
\mafig{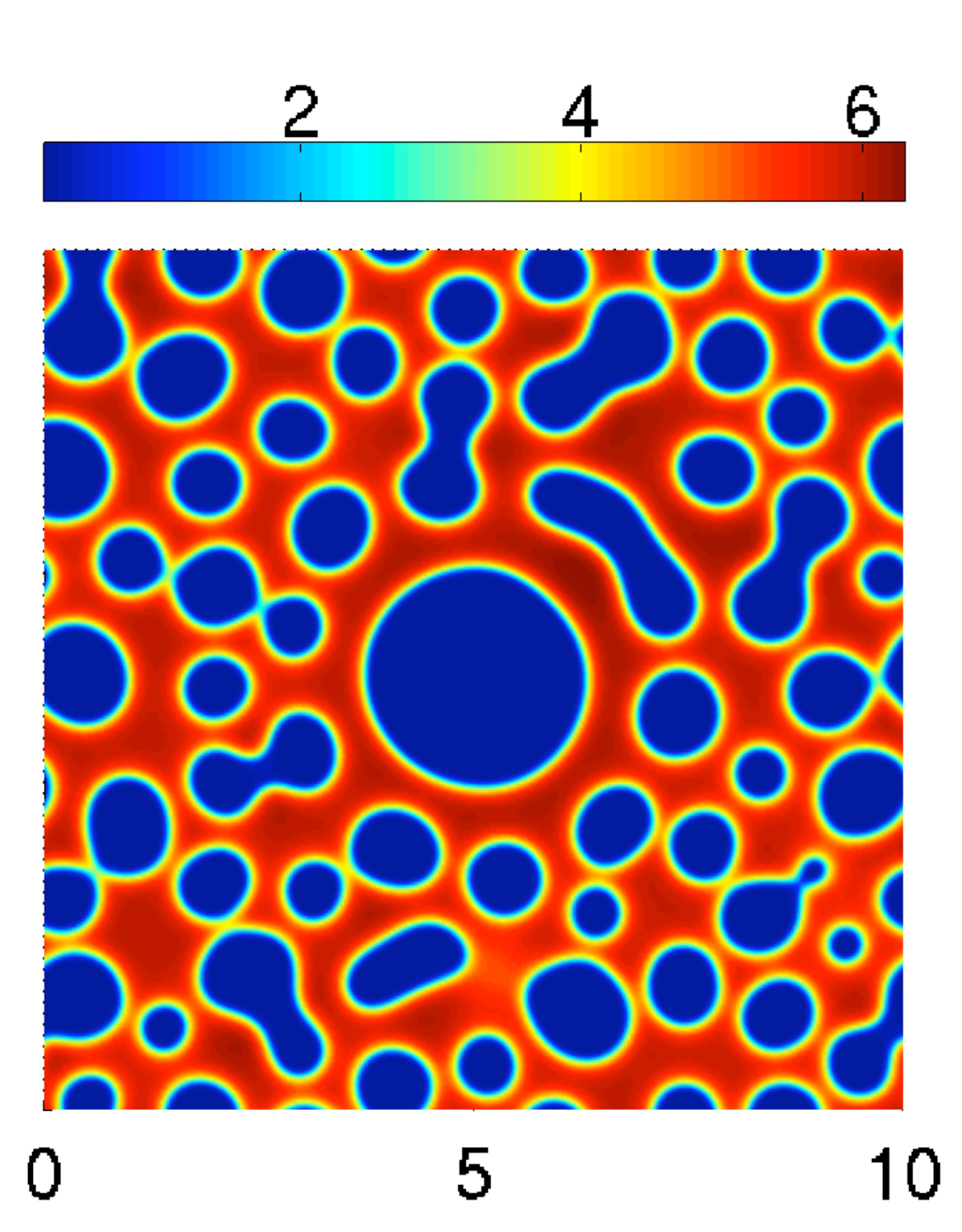} 
\mafig{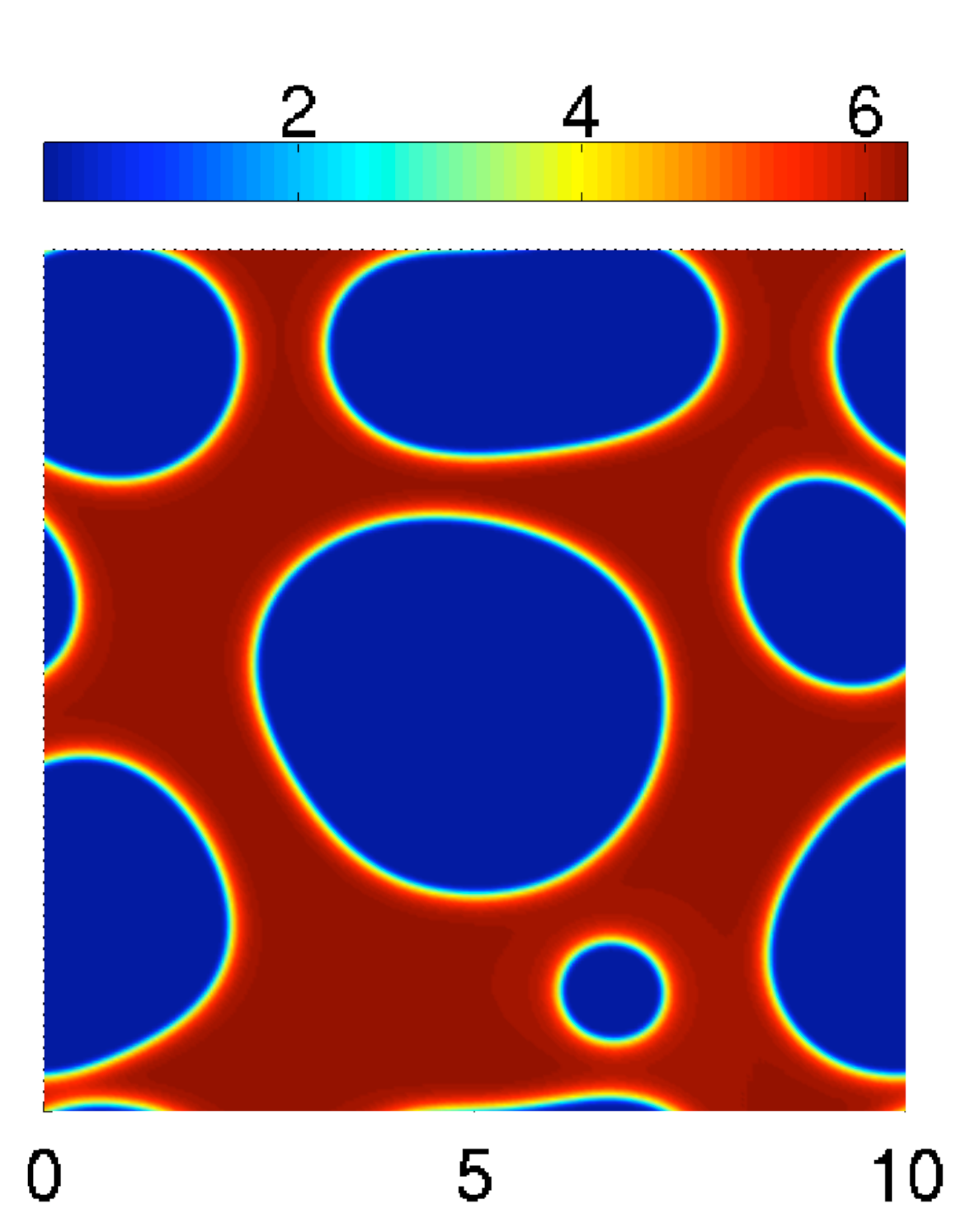} 
\mafig{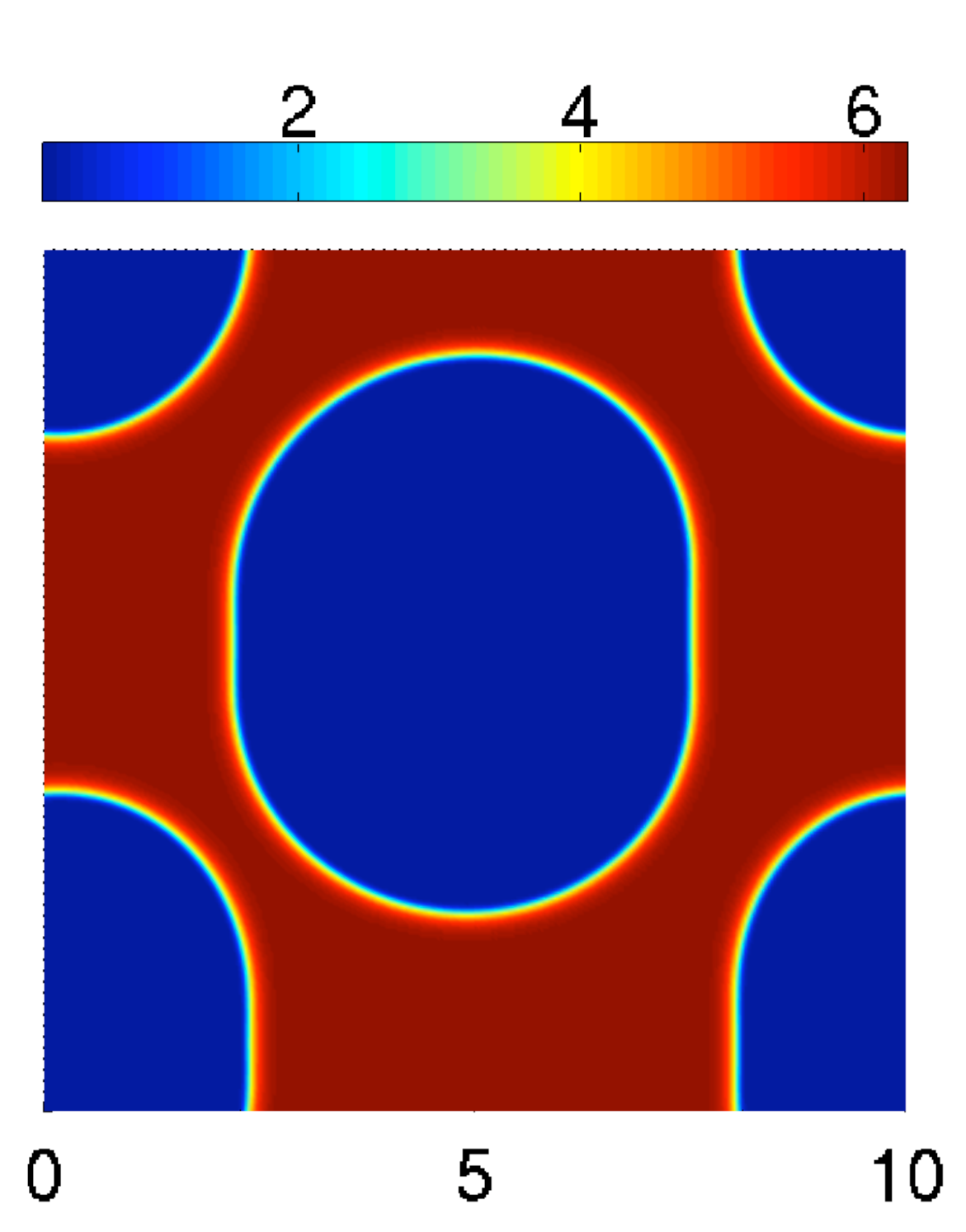}\\ 
(b)\temps{4.29}{5.69}{10.4}{120}{1.14\cdot10^3}\\ 
\caption{ Snapshots from the evolution of dewetting thin films in the 
  (a) surface-instability-dominated regime at $H=2.4$ and (b) in the 
  nucleation-dominated regime at $H=3.2$.  The initial condition 
  is as in Fig.~\ref{fig:dew3dd} with an added roughness
  of 0.1\% of $H$. The domain size is \(10 
  \lambda_m\times 10 \lambda_m\) and all other settings are as in 
  Fig.~\ref{fig:dew3dd}.} \label{fig:dew3dp2}
\end{figure} 
 
\begin{figure} 
\centering 
\mafig{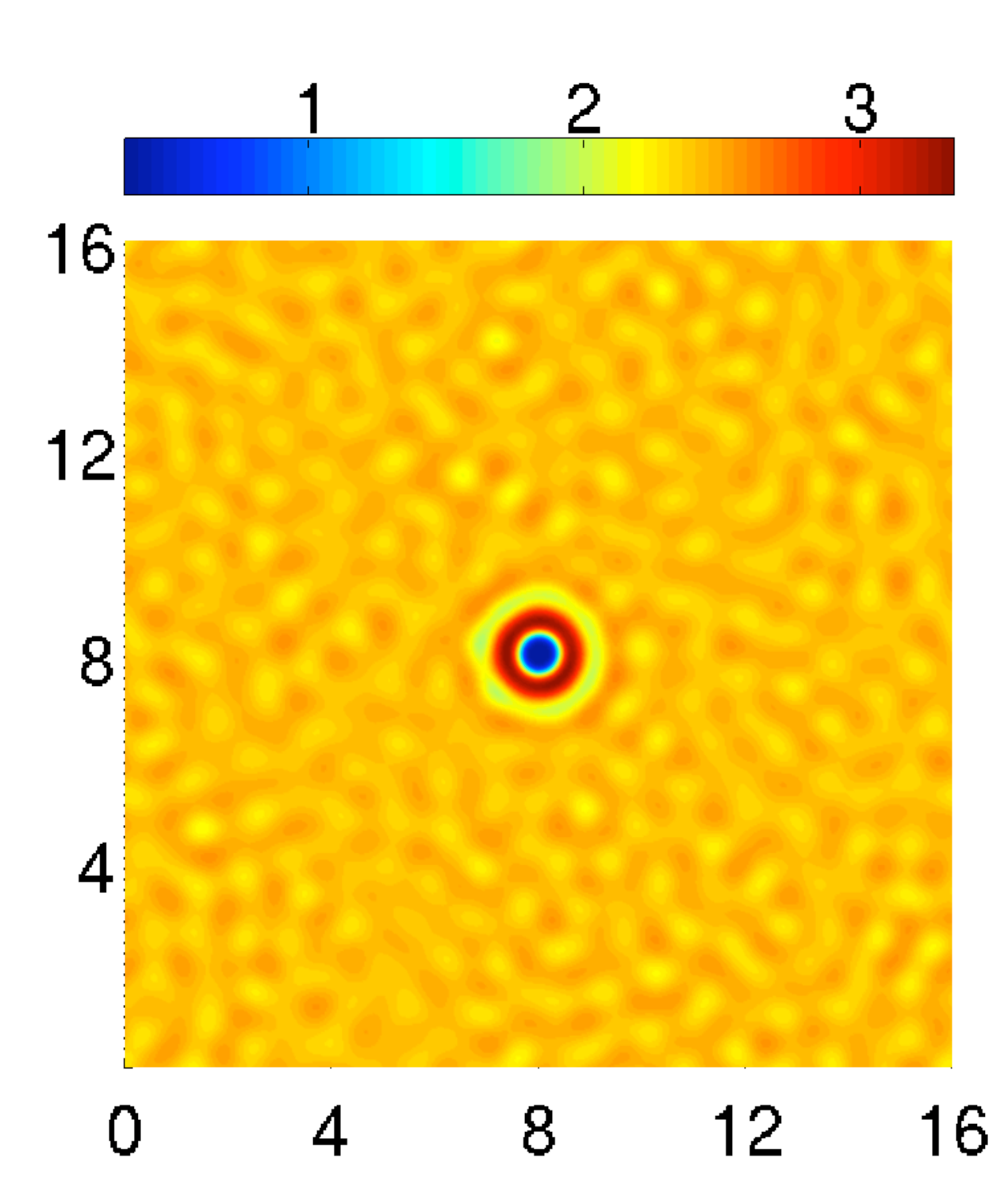} 
\mafig{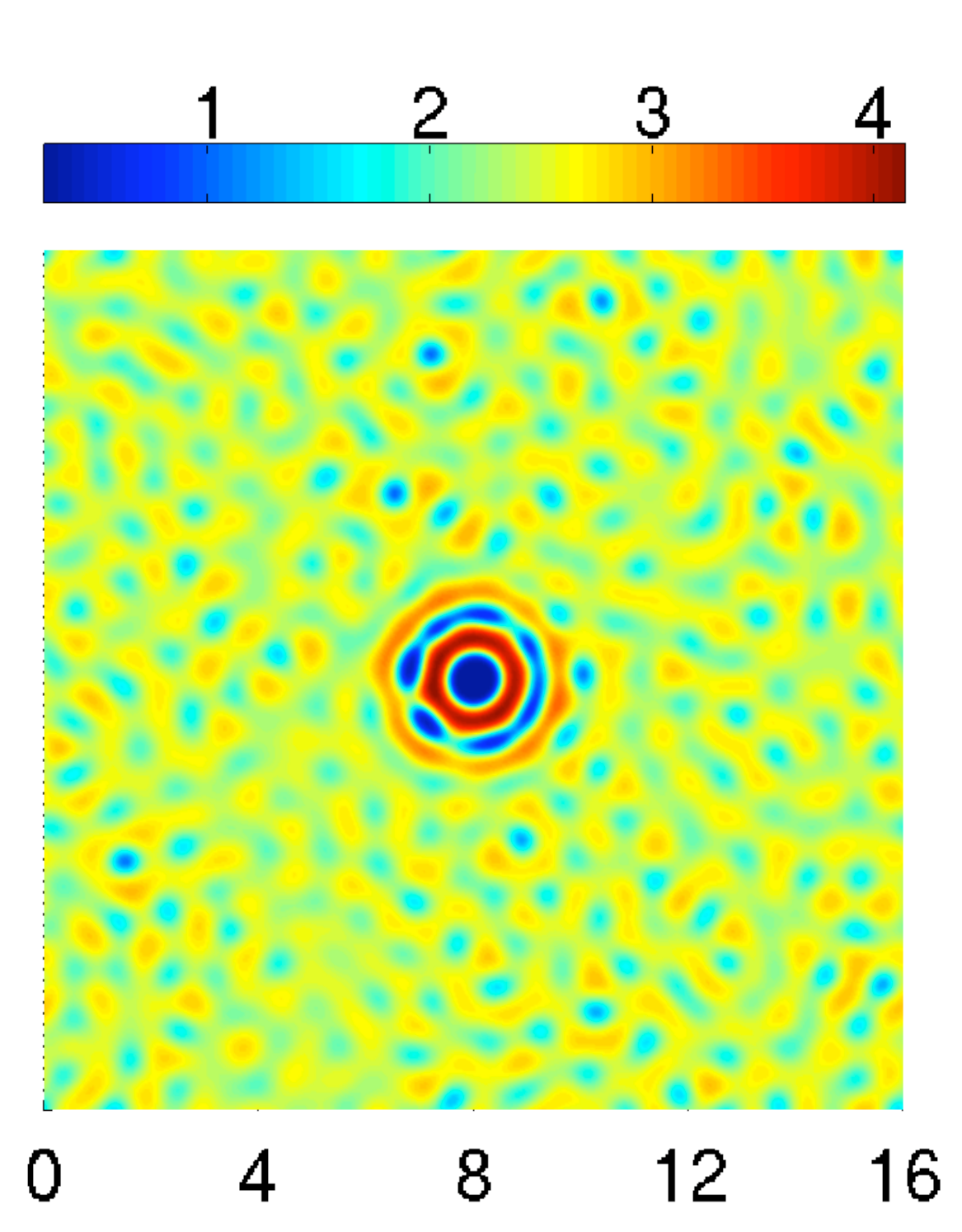} 
\mafig{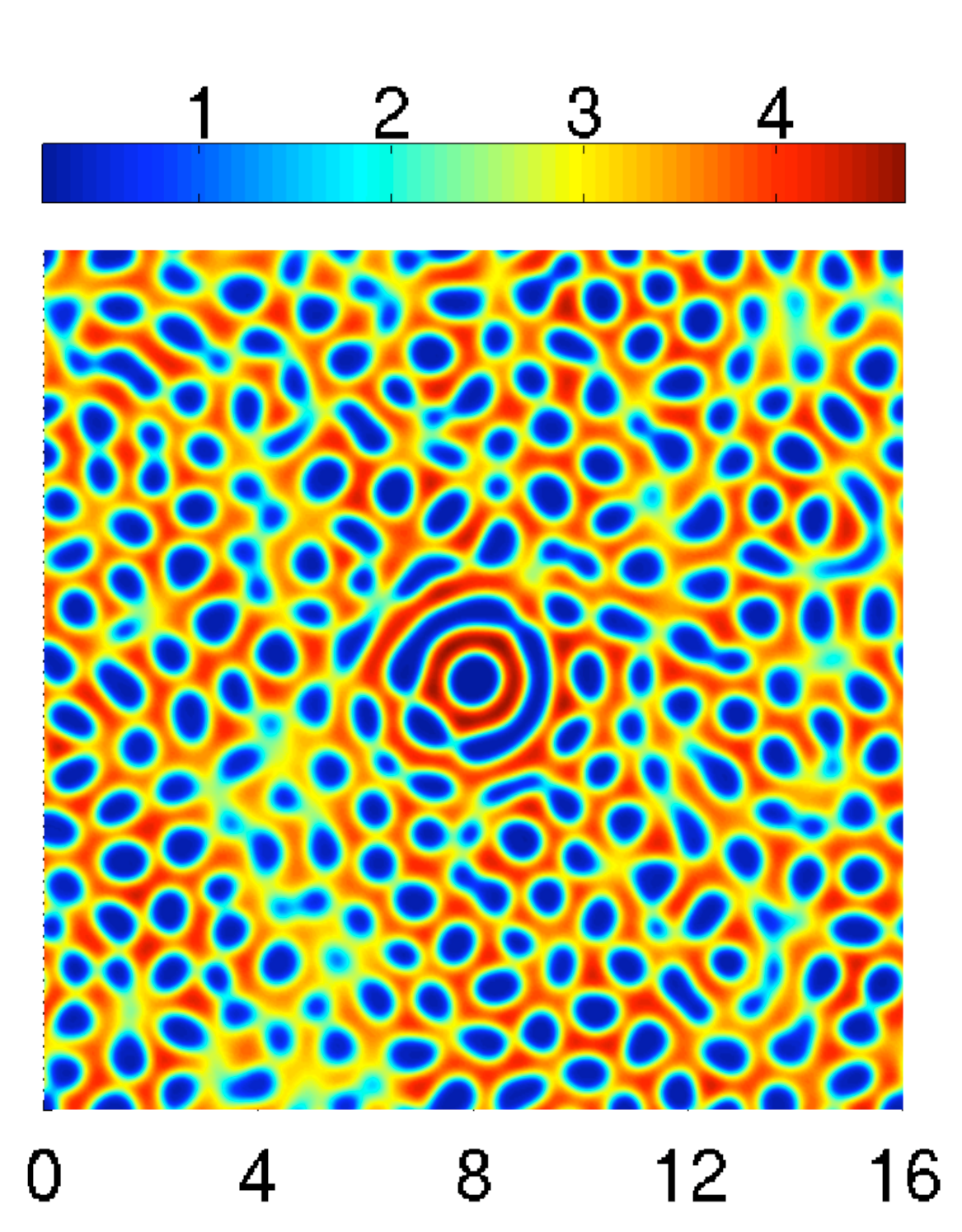} 
\mafig{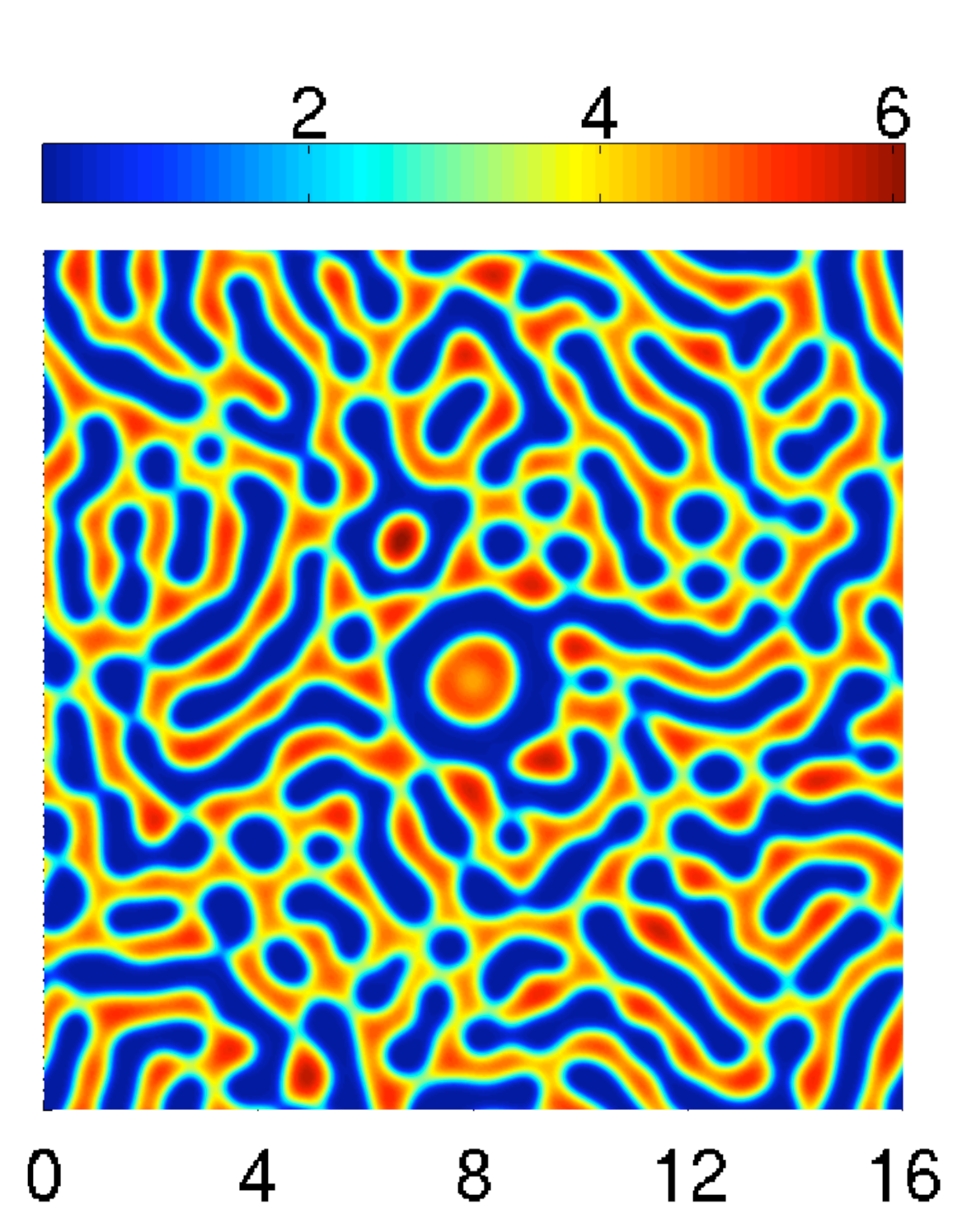} 
\mafig{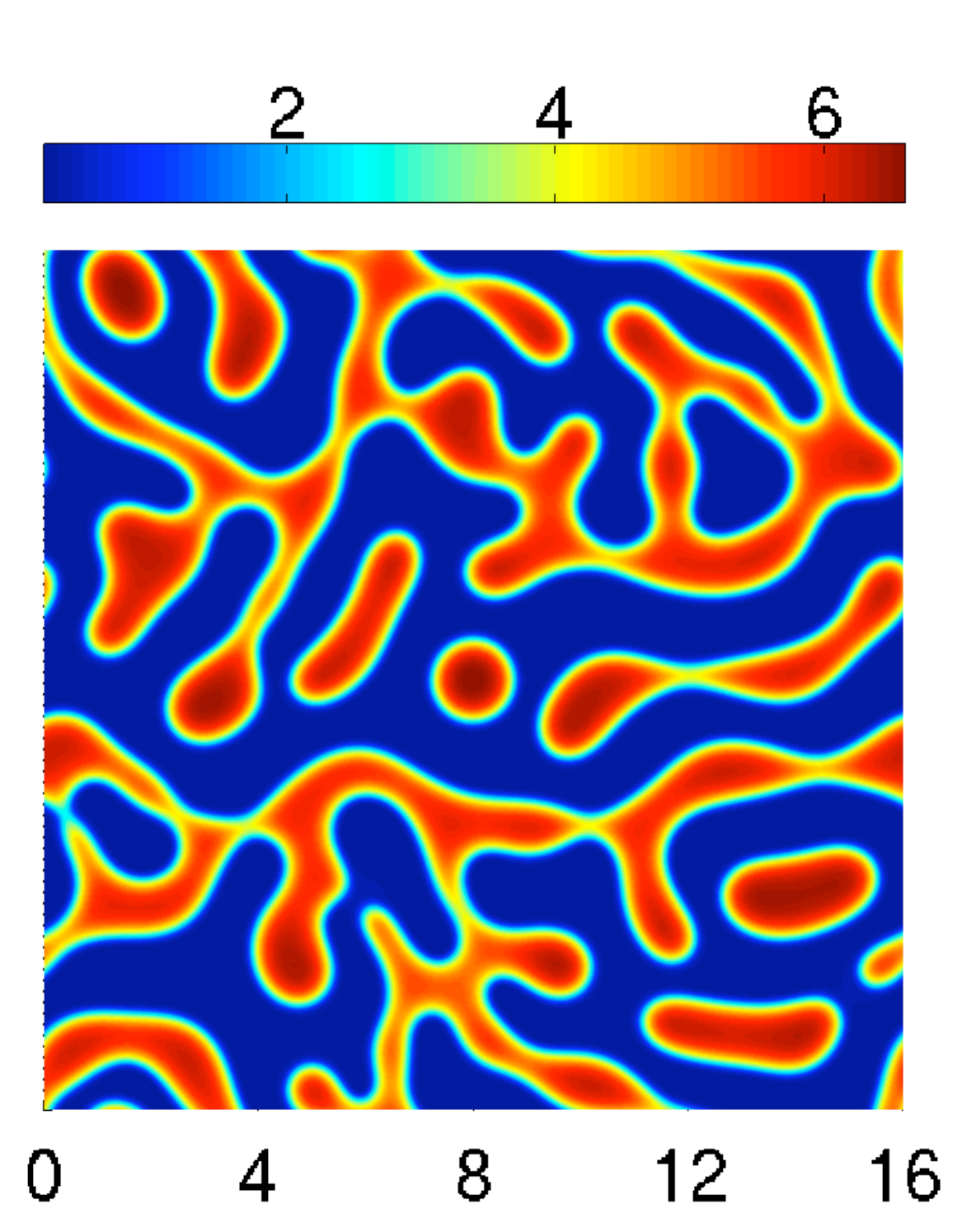}\\ 
(a)\temps{3.42}{4.84}{6.51}{12.5}{34.1}\\ 
\mafig{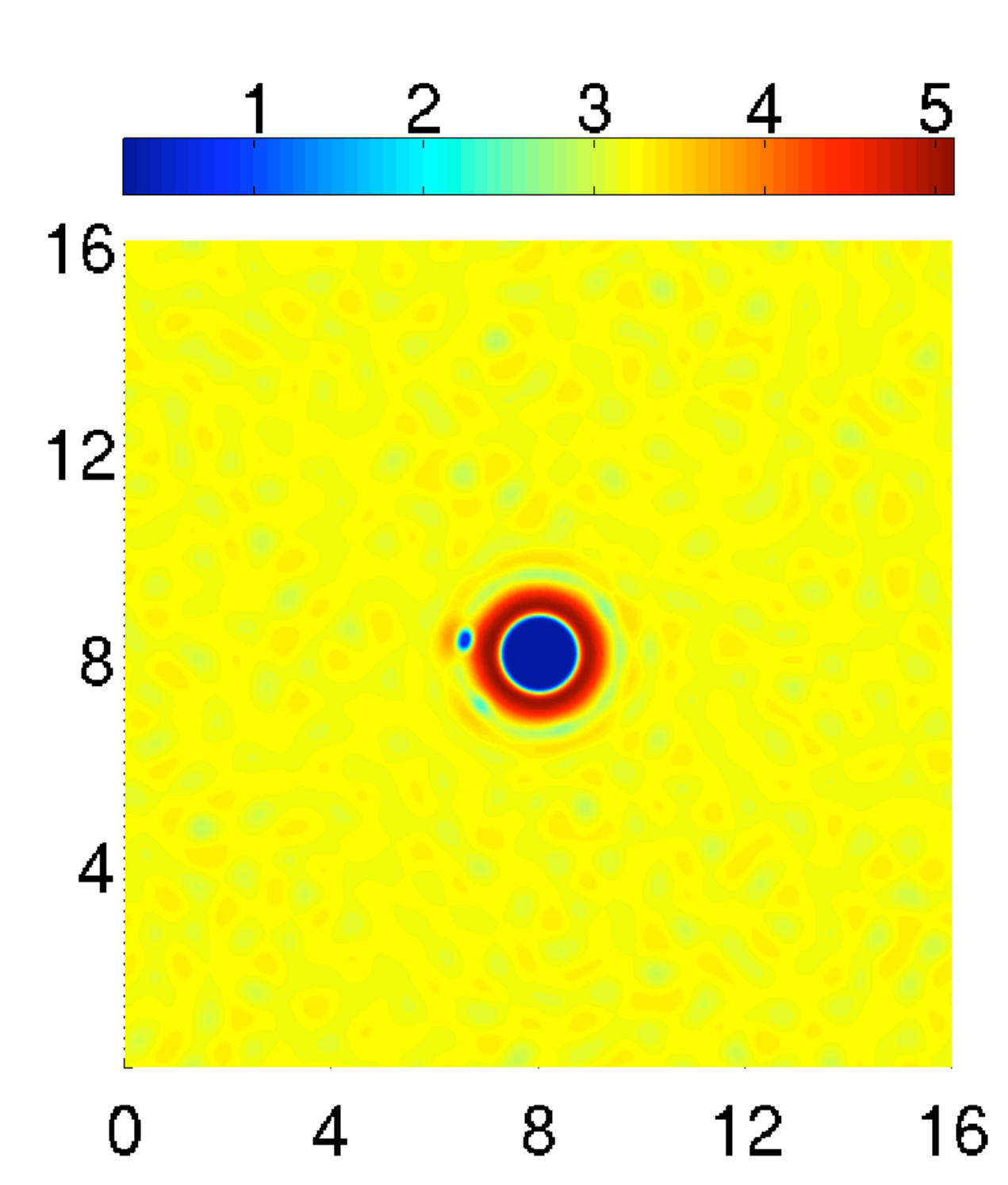} 
\mafig{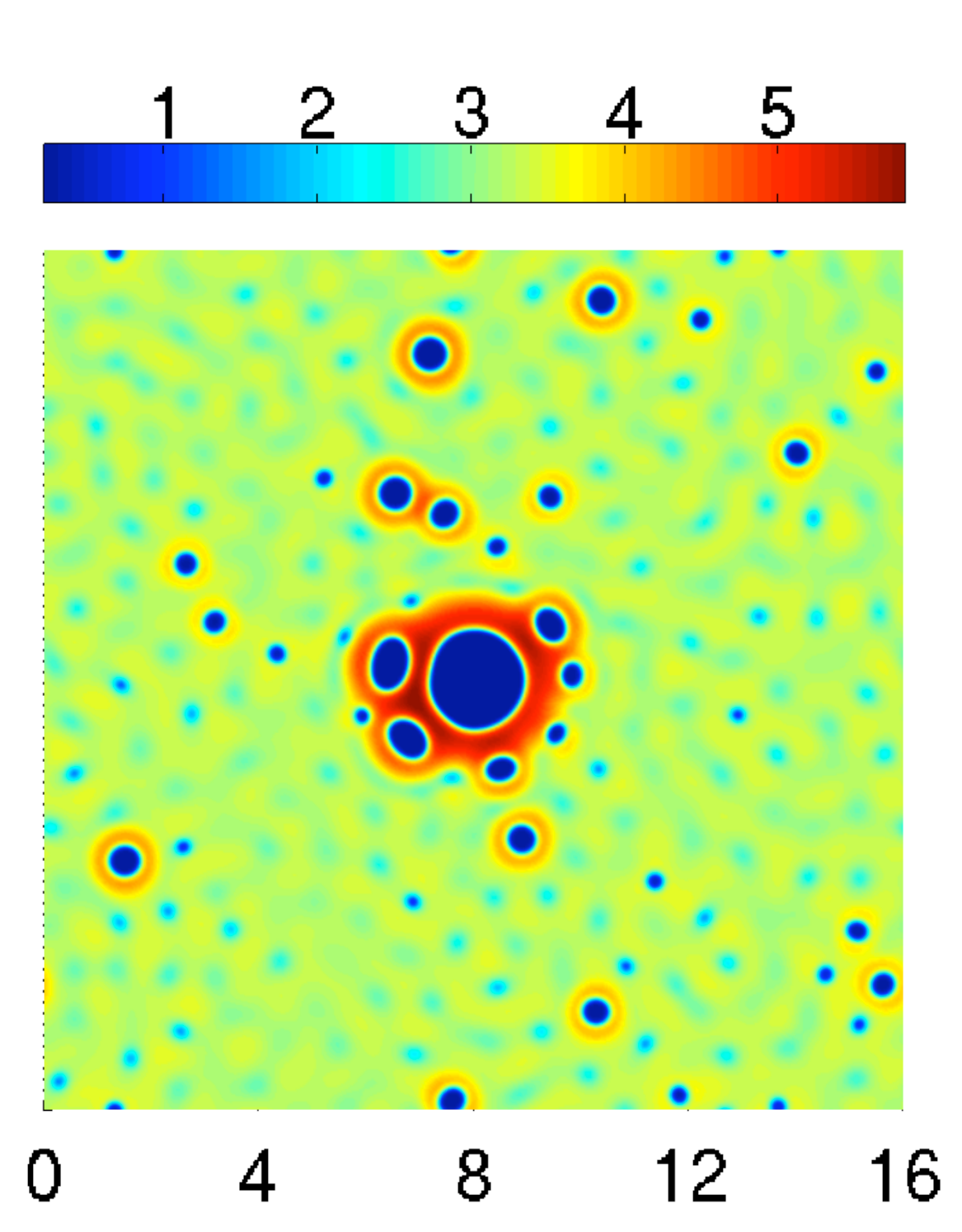} 
\mafig{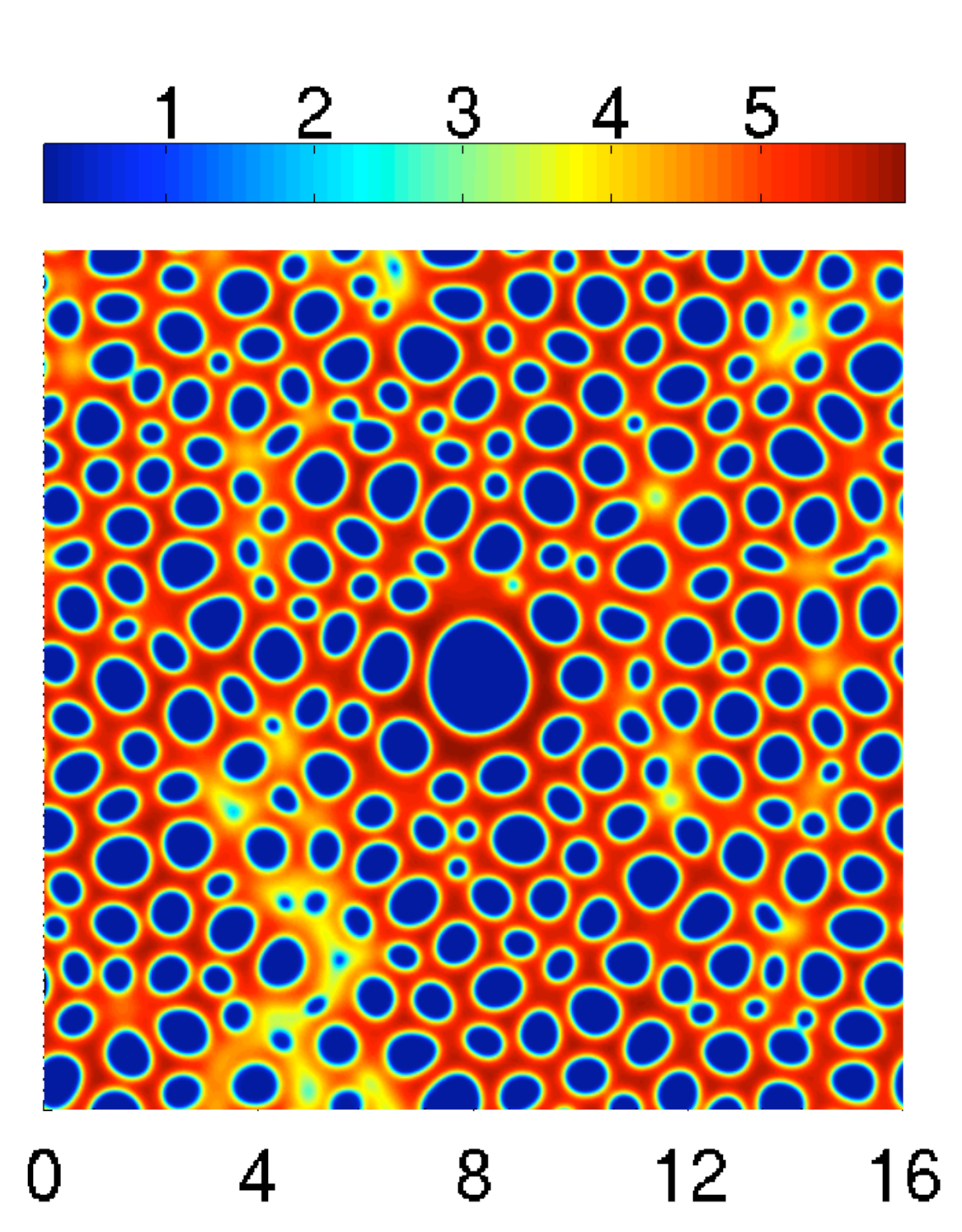} 
\mafig{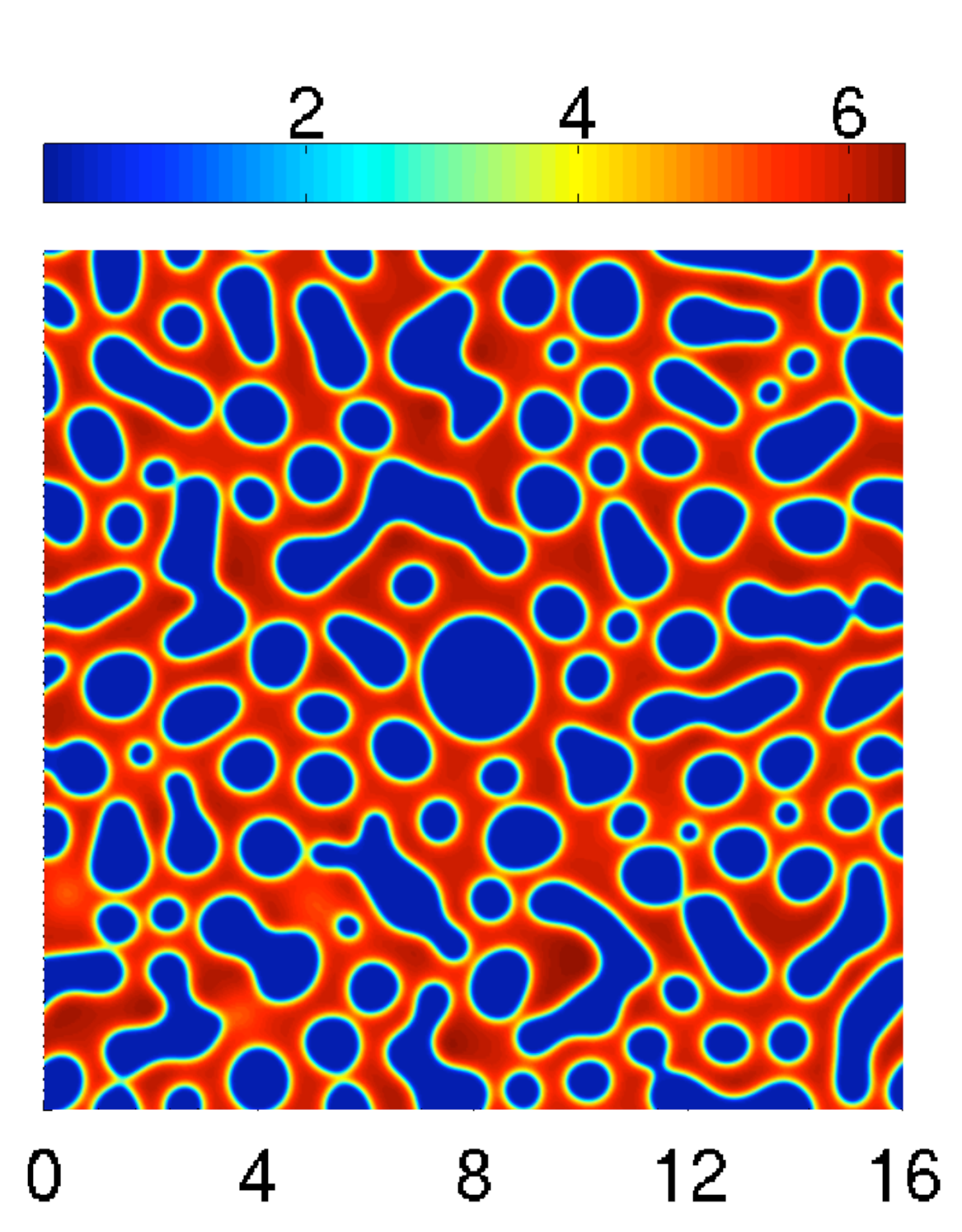} 
\mafig{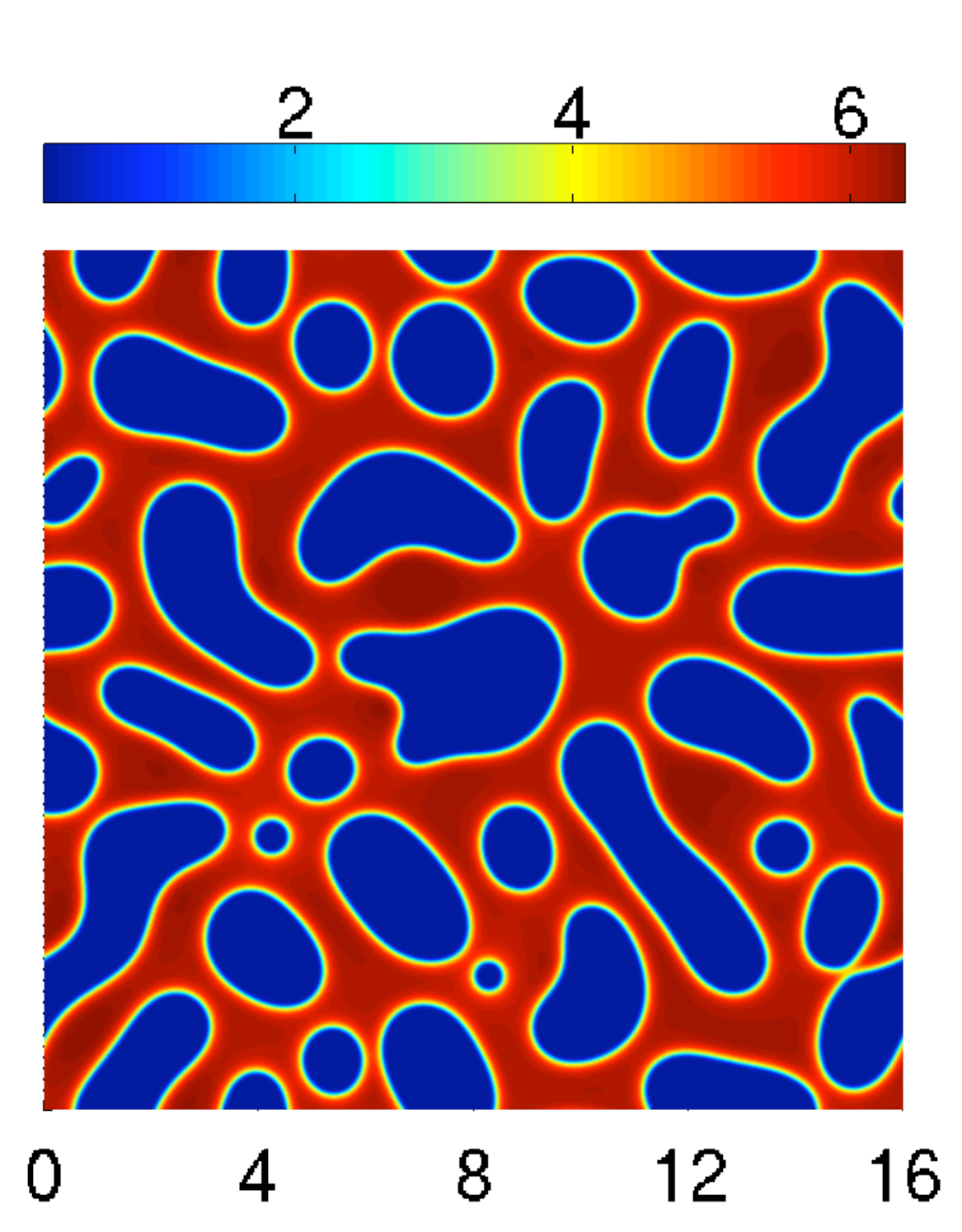}\\ 
(b)\temps{3.03}{4.09}{5.21}{10.7}{32.0}\\ 
\caption{Snapshots from the evolution of dewetting thin films for (a) 
  $H=2.4$ and (b) $H=3.2$. The initial condition corresponds to the 
  one in Fig.~\ref{fig:dew3dd} with an added roughness of of 
  1.0\% of $H$. The domain size is \(16 \lambda_m\times 16 \lambda_m\) 
  and all other settings are as in Fig.~\ref{fig:dew3dd}.}\label{fig:dew3dp} 
\end{figure} 
 
The result shows that the numerical noise is small enough to not
  break the radial symmetry during the time integration. Adding,
  however, an initial roughness  the symmetry is rapidly destroyed
  (see below).  This explains why normally in dewetting experiments
  with very thin films that are affected by thermal noise and other
  tiny perturbations no such regular structures are observed.
However, recent experiments with electrically destabilized thicker
films (less prone to noise) show regular ring structures when an
inhomogeneous electrical field is applied in such a way that it plays
the role of our initial radially symmetric defect (see Figs.~2 and ~4
of Ref.~\cite{CZDC05}). 
 
To determine the influence of noise of the relative importance
  of nucleation and surface instability we perform several simulations
  using different initial film roughness $\zeta$. In particular, we
  add a roughness of $\zeta=0.1$\% (Fig.~\ref{fig:dew3dp2}) and
  $\zeta=1$\% (Fig.~\ref{fig:dew3dp}) of the mean film thickness.
  The larger $\zeta$ becomes, the less time has the radial structure
  to evolve because the roughness 'accelerates' the isotropic surface
  instability.  For small $\zeta=0.1$\% the radial symmetry is
  appreciable quite some time even into coarsening
  (Fig.~\ref{fig:dew3dp2}). However, coarsening eventually 'washes
  out' any memory of the initial defects. In particular, for $H=2.4$
  the second ring is still complete (at $t=4.92$), but already the
  next depression is not radially symmetric but resembles a ring-like
  assembly of holes (at $t=6.19$). For $H=3.2$, already the depression
  outside the first ring emerges as a circular hole pattern, i.e., as
  a typical secondary nucleation pattern (cf.~\cite{Beck03}).  In both
  cases the initial defect has no further influence on the structure.
  The remaining area is covered by typical spinodal structures.
For the stronger initial roughness ($\zeta=1$\%, Fig.~\ref{fig:dew3dp}) the 
initial defect is of minor influence, i.e., 
the central radial structure only dominates during a very short time  
(till about $t=3\dots4$) and is later homogenized through coarsening. 
Everywhere else the surface instability dominates.
\section{Depinning of a drop on a heterogeneous substrate 
\label{sec:het}} 
The second problem we focus on is the depinning of drops. If a lateral
force is applied on films/drops on a homogeneous substrate [$P\neq0$
in Eq.~(\ref{eq:lub})], one only finds traveling surface waves or
sliding drops \cite{Thie01,ThKn03}. No steady-state solutions exist,
except for the flat film.  On a heterogeneous substrate, however, the
heterogeneity (e.g., chemical or topographical defect) can pin a
drop.   Here, we consider
wettability defects in the form of stripes, i.e., we use disjoining
pressure (III) that is modulated in the $x$-direction
(cf.~Eq.~(\ref{eq:dispin})).  
The resulting wettability profile  $\xi(x)$ is presented in the lower panels of Fig.~\ref{fig:stat2d}.
The parameter $\epsilon$ represents the amplitude of the
heterogeneity, i.e., the wettability contrast.  If $\epsilon>0$
[$\epsilon<0$] the defect is less [more] wettable than the surrounding
substrate, i.e., the defect is hydrophobic [hydrophilic].
 
When the lateral driving is increased the pinned drops are deformed
and their center of mass slightly shifts until at a critical driving
$P_c$ the drop depins. The analysis of the steady states and the
depinning bifurcation is tackled using the continuation approach
developed above in Section~\ref{sec:cont}.  In the 1d case, results
are already available: In Ref.~\cite{ThKn06,ThKn06b} the continuation
package AUTO \cite{AUTO97} and an explicit time-stepping algorithm
with adaptive time-step were used.  We employ this case in
Section~\ref{sec:pin2d} to validate our continuation code.  As the
results for identical parameters are identical to the ones of
Ref.~\cite{ThKn06,ThKn06b}, we here present results for larger drops.
In Section~\ref{sec:pin3d} we explore the 2d case which can not be
treated using AUTO because the governing equations are not equivalent
to an ODE system.
 
\begin{figure} 
\centering 
 \includegraphics[width=8cm]{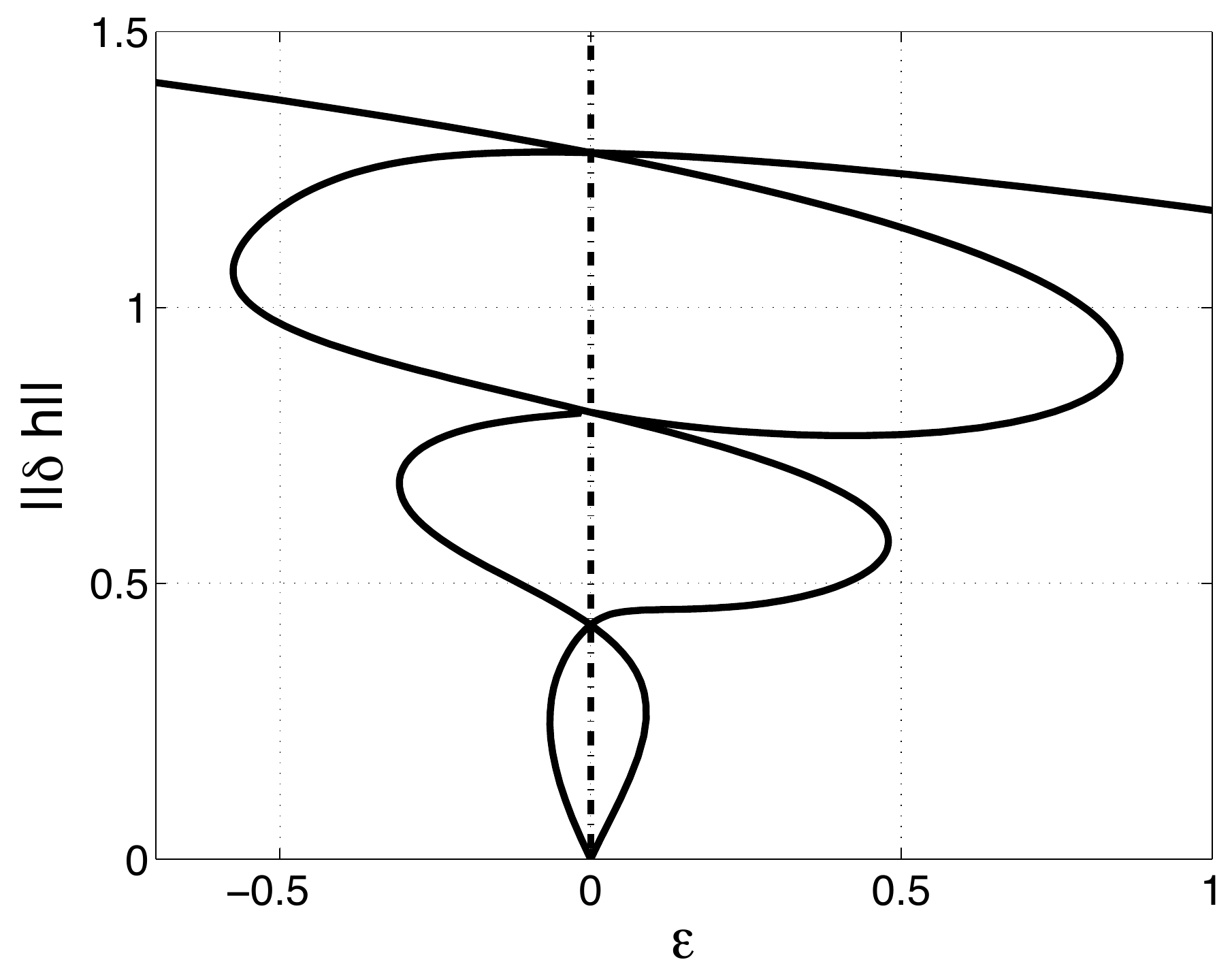} 
 \caption{Bifurcation diagram for 1d steady drop solutions
   on an horizontal heterogeneous substrate ($P = 0$). Shown is the
   $L^2$-norm $||\delta h||$ versus the defect strength
   $\epsilon$. The used disjoining pressure~(III) is given by
   Eq.~(\ref{eq:dispin}) with $b=0.1$; and the domain size is
   $L=50$. }\label{fig:coneps}
\end{figure}

\subsection{The one-dimensional case}\label{sec:pin2d} 
For a homogeneous substrate without lateral driving ($\epsilon=0$ and
$P=0$), the critical wavelength for a film of thickness $H=1.5$ is
$\lambda_{c}\simeq15$, i.e., for a domain size of $L=50$ there exist
at least steady states containing one, two or three drops. They
bifurcate from the flat film at $n \lambda_c$ with $n=1,2,3$,
respectively. If the primary bifurcation is subcritical there might be
more solutions (cf.~Ref.~\cite{ThKn04}).
 
\begin{figure} 
{\centering 
\includegraphics[width=6cm]{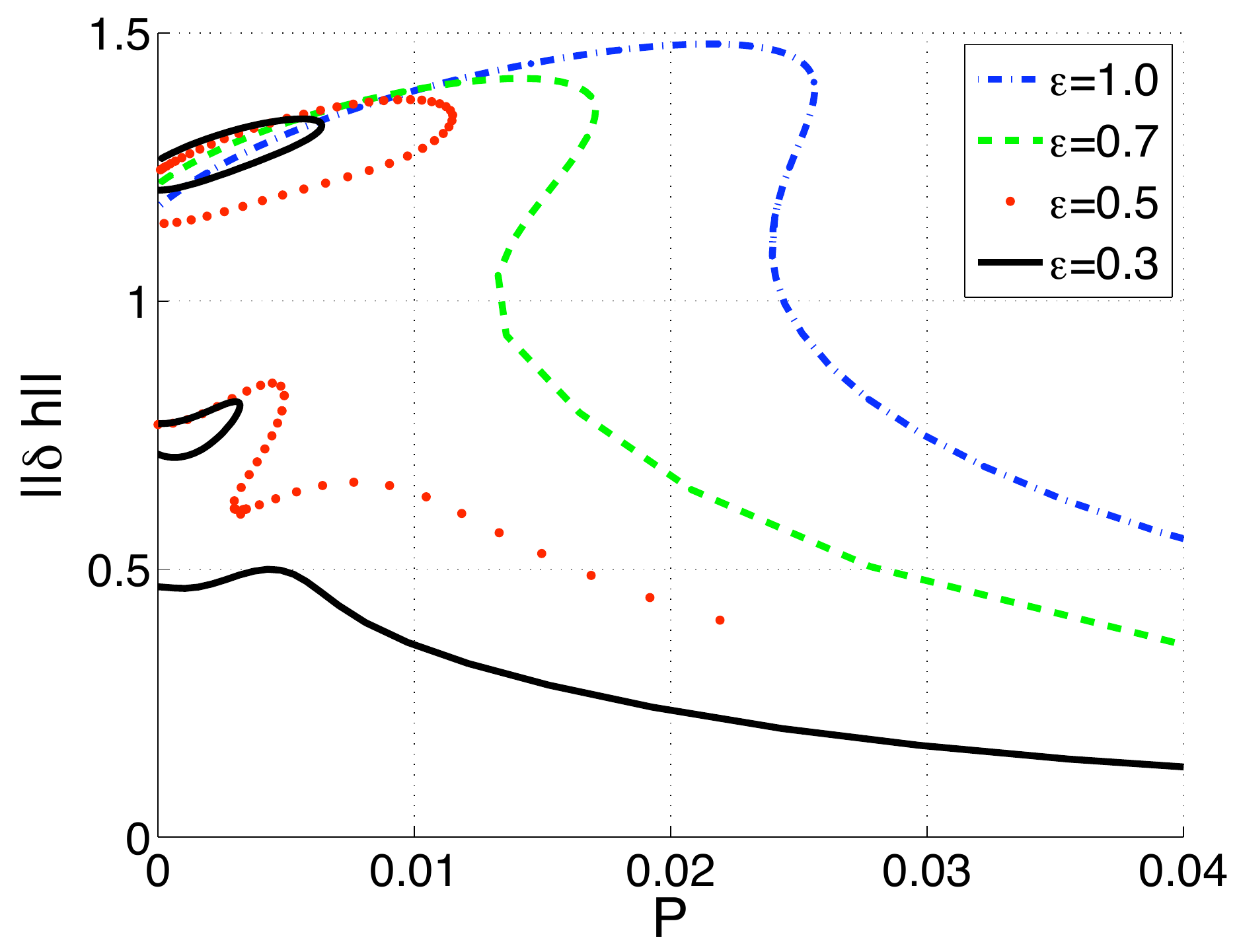} 
\includegraphics[width=6cm]{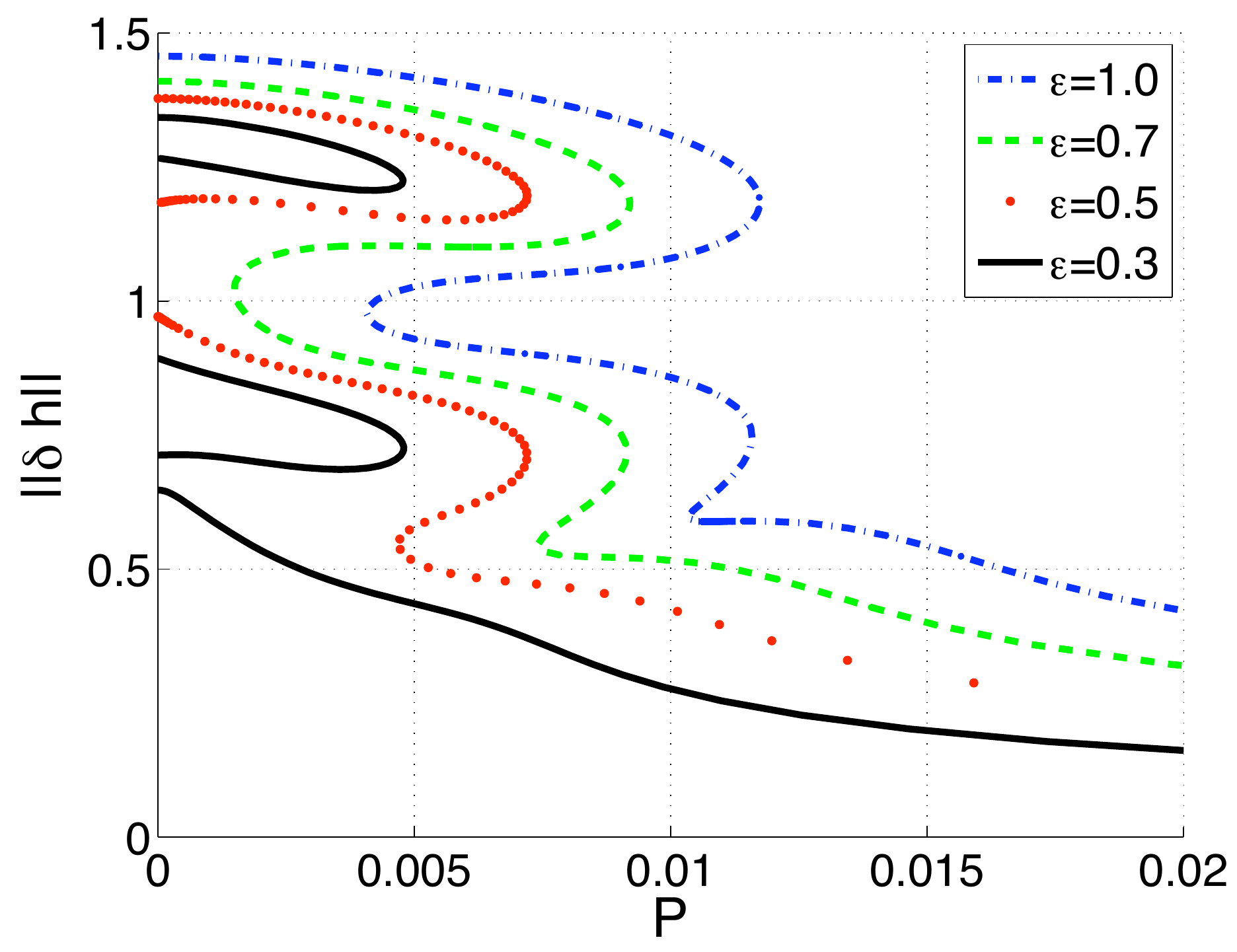} 
} 
\hspace*{3.0cm}(a) \hspace{6.3cm} (b) 
\caption{Bifurcation diagrams for families of pinned 1d steady
  solutions representing one-, two- and three-drop states for (a)
  hydrophobic and (b) hydrophilic defects of various strength
  $\epsilon$ as given in the legend. Shown is the $L^2$ norm $||\delta
  h||$ in dependence of the lateral driving force $P$ The remaining
  parameters are as in Fig.~\ref{fig:coneps}.}\label{fig:con1d}
\end{figure} 

To determine the various steady-state solutions for a heterogeneous
substrate without lateral driving we use continuation when varying the
heterogeneity strength $\epsilon$ for a heterogeneity period equal to
the domain size (Fig.~\ref{fig:coneps}). Branches of steady-state
solutions cross the axis $\epsilon=0$ three times, corresponding to
the one-, two- and three-drop solution for a homogeneous substrate.
For a strong heterogeneity ($|\epsilon|$ large) only the single drop
solution remains that is the most interesting solution for a study of
depinning.
However, for smaller $|\epsilon|$ up to seven steady states can exist
corresponding to various stable and unstable one-, two, and three-drop
equilibria.  As Ref.~\cite{ThKn06} studies a smaller domain ($L=25$)
their Fig.~6 only shows one crossing of the axis
$\epsilon=0$. Studying the stability of the equilibria one finds that
only the uppermost branch corresponds to stable solutions.  All other
branches terminate in saddle-node bifurcations.

\begin{figure} 
{\centering 
\includegraphics[width=6cm,height=3cm]{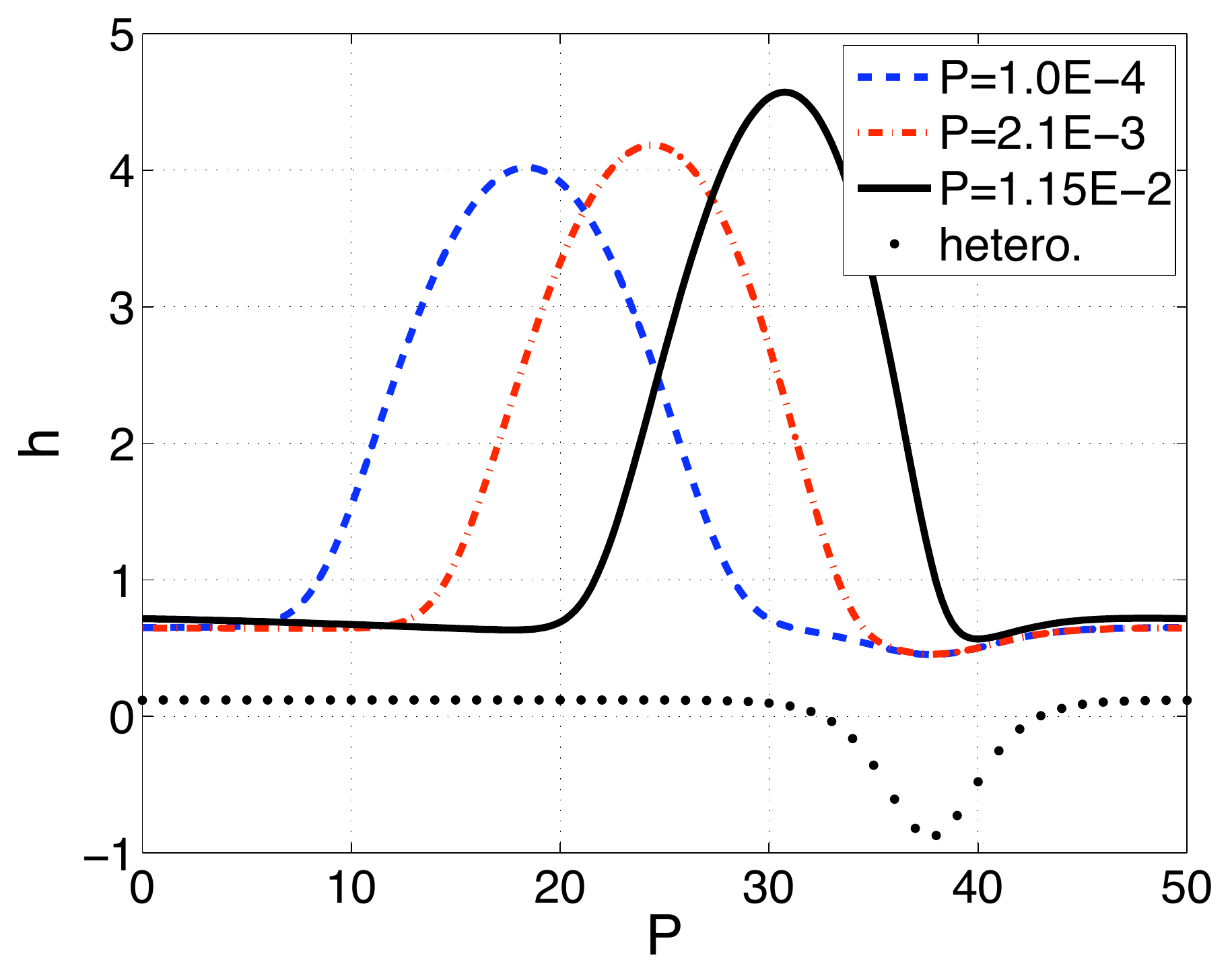} 
\includegraphics[width=6cm,height=3cm]{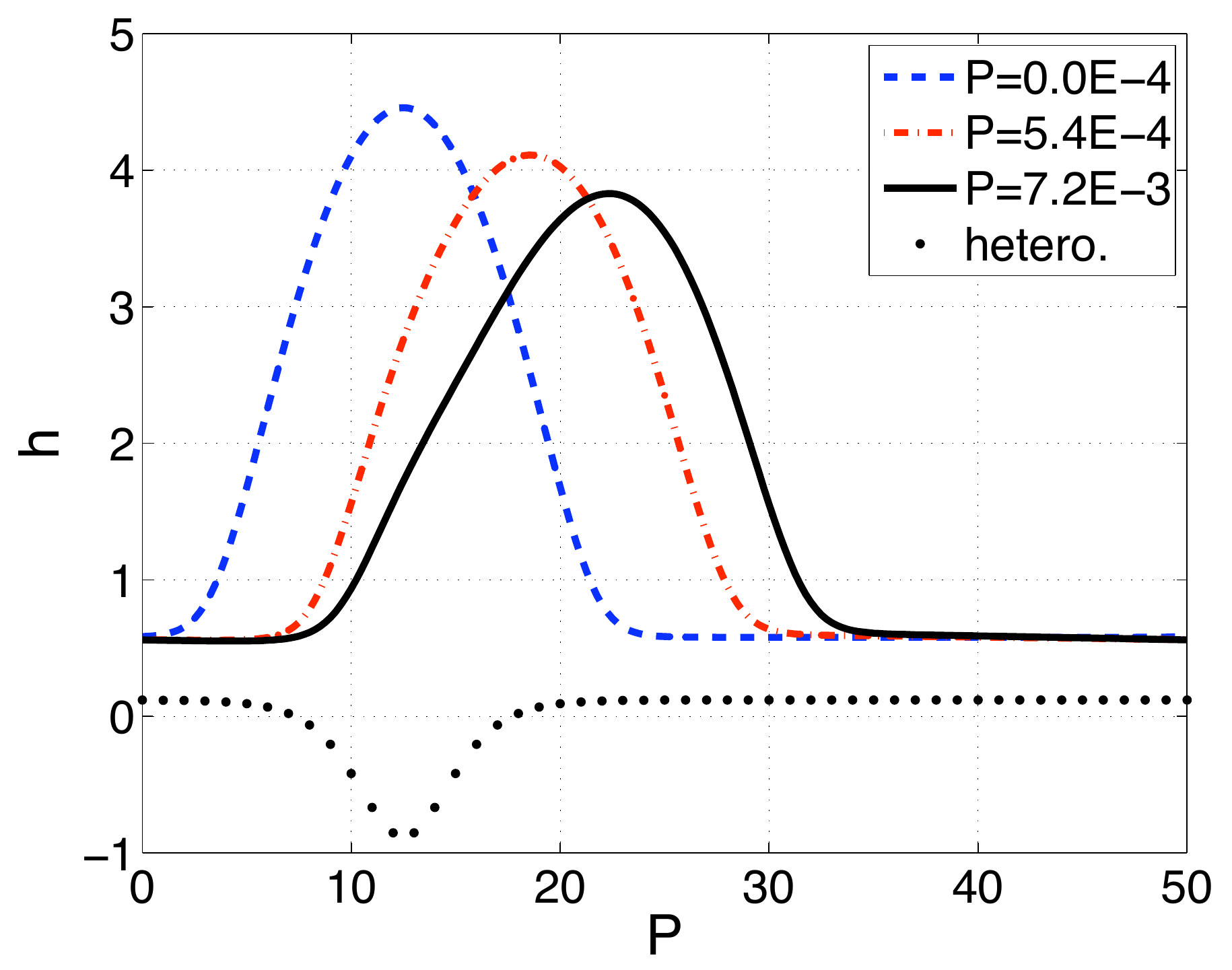} 
} 
\hspace*{3.0cm}(a) \hspace{6.3cm} (b) 
\caption{Selected steady drop profiles on a heterogeneous substrate 
  for different driving forces $P$ as given in the legend. Shown are 
  (a) the hydrophobic case with defect strength $\epsilon = 0.5$, and 
  (b) the hydrophilic case with $\epsilon = - 0.5$.  The respective 
  solid line represents the profile at depinning. The lower part of 
  the panels gives the profile $\xi(x)$ of the heterogeneity. 
} 
\label{fig:stat2d} 
\end{figure} 
 
Next, all steady-state branches are tracked when increasing the
lateral driving force $P$ from zero for various fixed $\epsilon$.
Bifurcation diagrams and selected profiles of pinned drops are given
in Figs.~\ref{fig:con1d} and \ref{fig:stat2d}, respectively.  The
various branches found for $P=0$ continue to exist for small driving
as 'pinned solutions'. However, at critical values of the driving most
'annihilate' each other. Physically speaking, the heterogeneity can
not retain the drops any more and they start to slide, i.e., they
depin.  Beyond a critical value $P_c$ no steady drop does exist as
even the stable single drop depins (Fig.~\ref{fig:con1d}). The
  bifurcation at $P_c$ is a sniper (Saddle Node Infinite PERiod)
  bifurcation. At $P_c$ a branch of space- and time-periodic solutions
  emerges (not shown) that corresponds to drops sliding on the
  heterogeneous substrate.  The temporal period diverges as one
approaches $P_c$ from above and the drop motion resembles stick-slip
behavior (cf.~Refs.~\cite{ThKn06,ThKn06b}.
The obtained results indicate that our continuation
algorithm is well capable to follow stable and unstable steady states
in the 1d case.  The saddle-node bifurcation has been well detected
and as expected no numerical stability problems have been encountered
near the bifurcation. Next, the continuation algorithm has to prove
its capabilities in the 2d case.
 
\begin{figure} 
\centering 
\includegraphics[width=7cm]{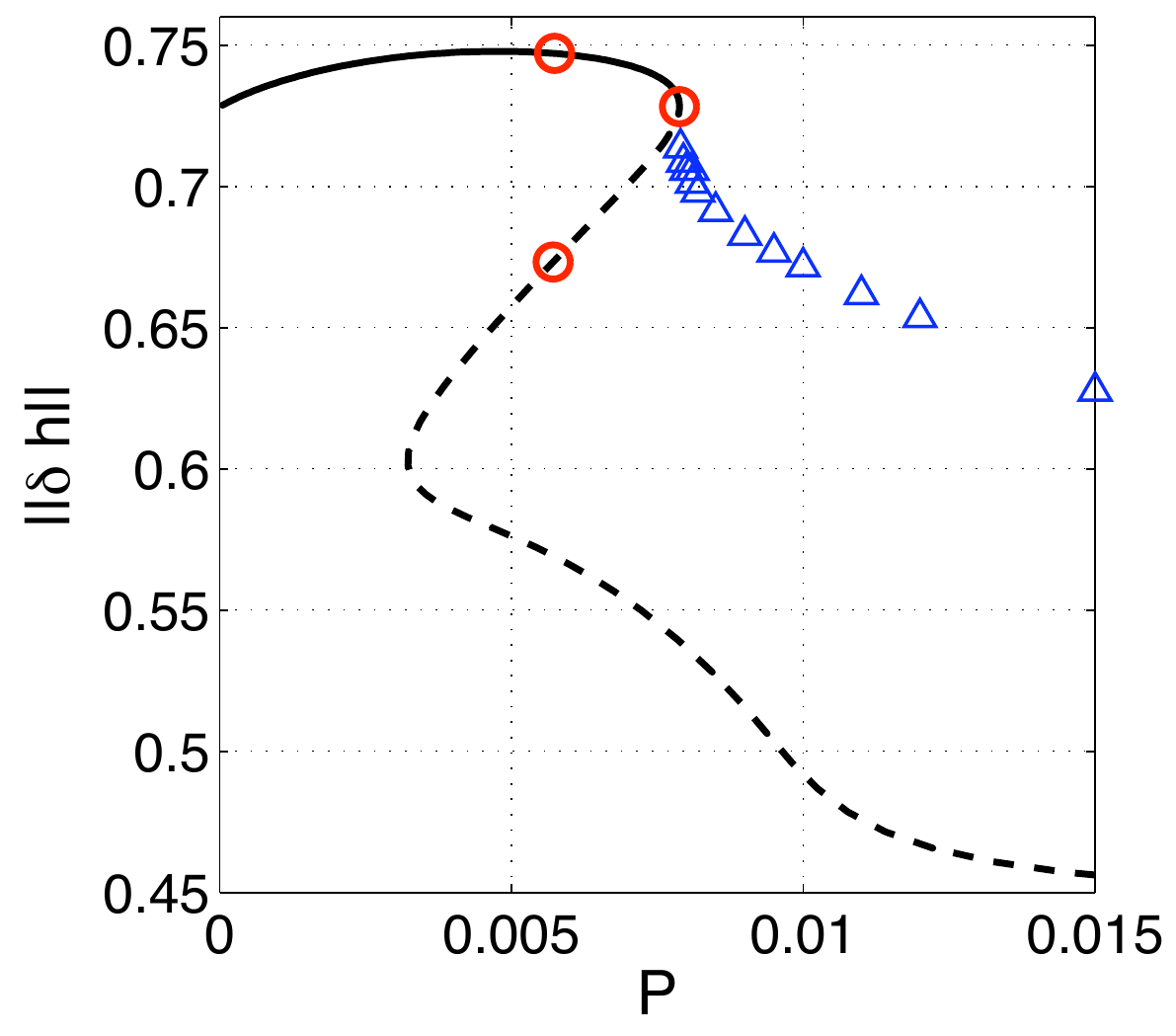} 
\caption{Bifurcation diagram for the depinning of 2d drops
    from a hydrophobic line defect given by Eq.~(\ref{eq:dispin}) with
    $\epsilon=0.3$. Shown is the $L^{2}$ norm $||\delta h||$ in
    dependence of the driving force $P$ for steady state solutions
    (solid line). For depinned sliding drops the time-averaged norm
    is given (symbols '$\Delta$').  The domain size is $40\times40$. The
    circles indicate the steady states represented in  Fig.~\ref{fig:ss3d}. All other parameters are as in
    Section~\ref{sec:pin2d}}\label{fig:cond3d}
\end{figure} 
\begin{figure} 
{\centering 
\includegraphics[width=6.4cm]{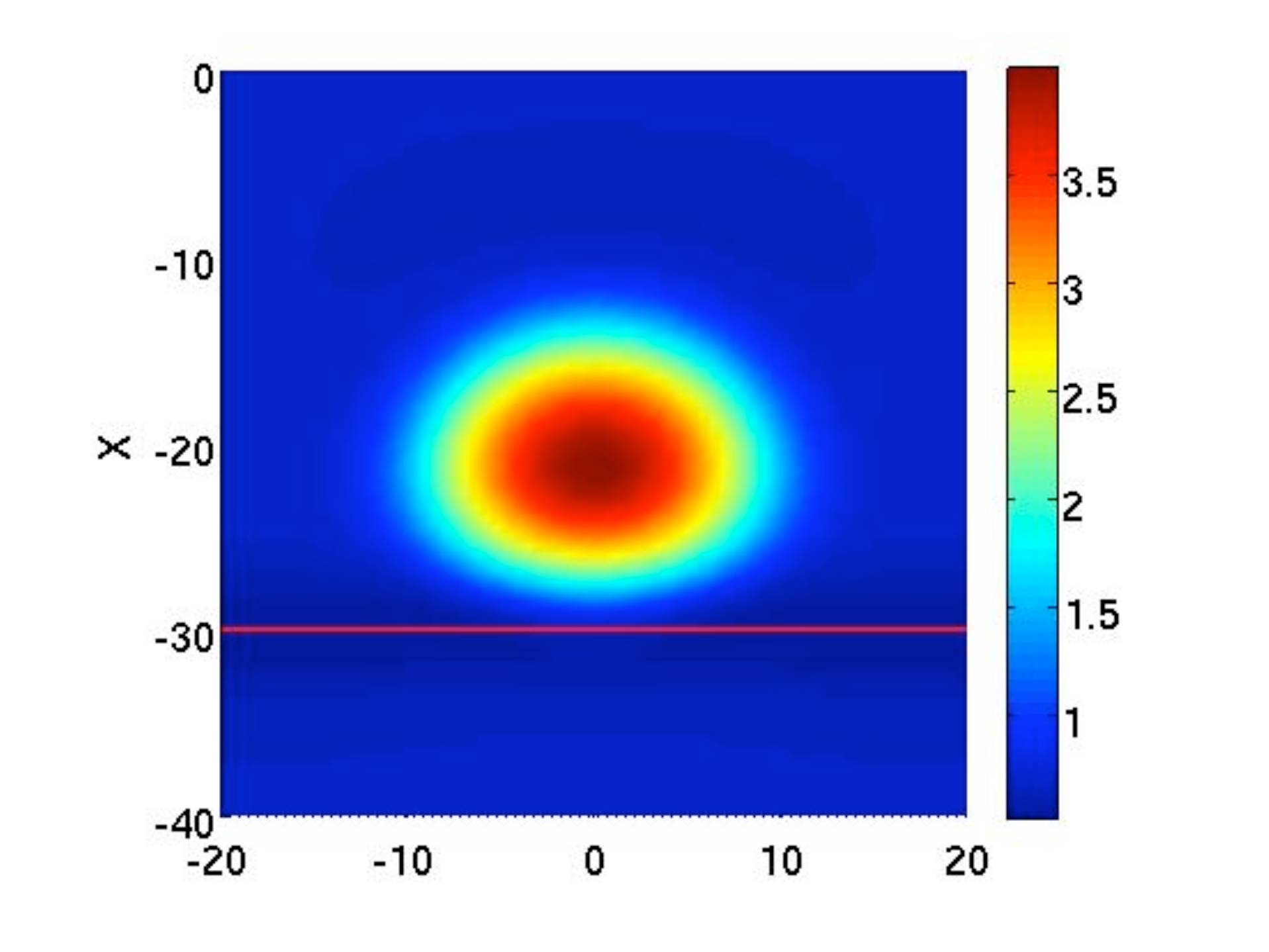} 
\includegraphics[width=6.4cm]{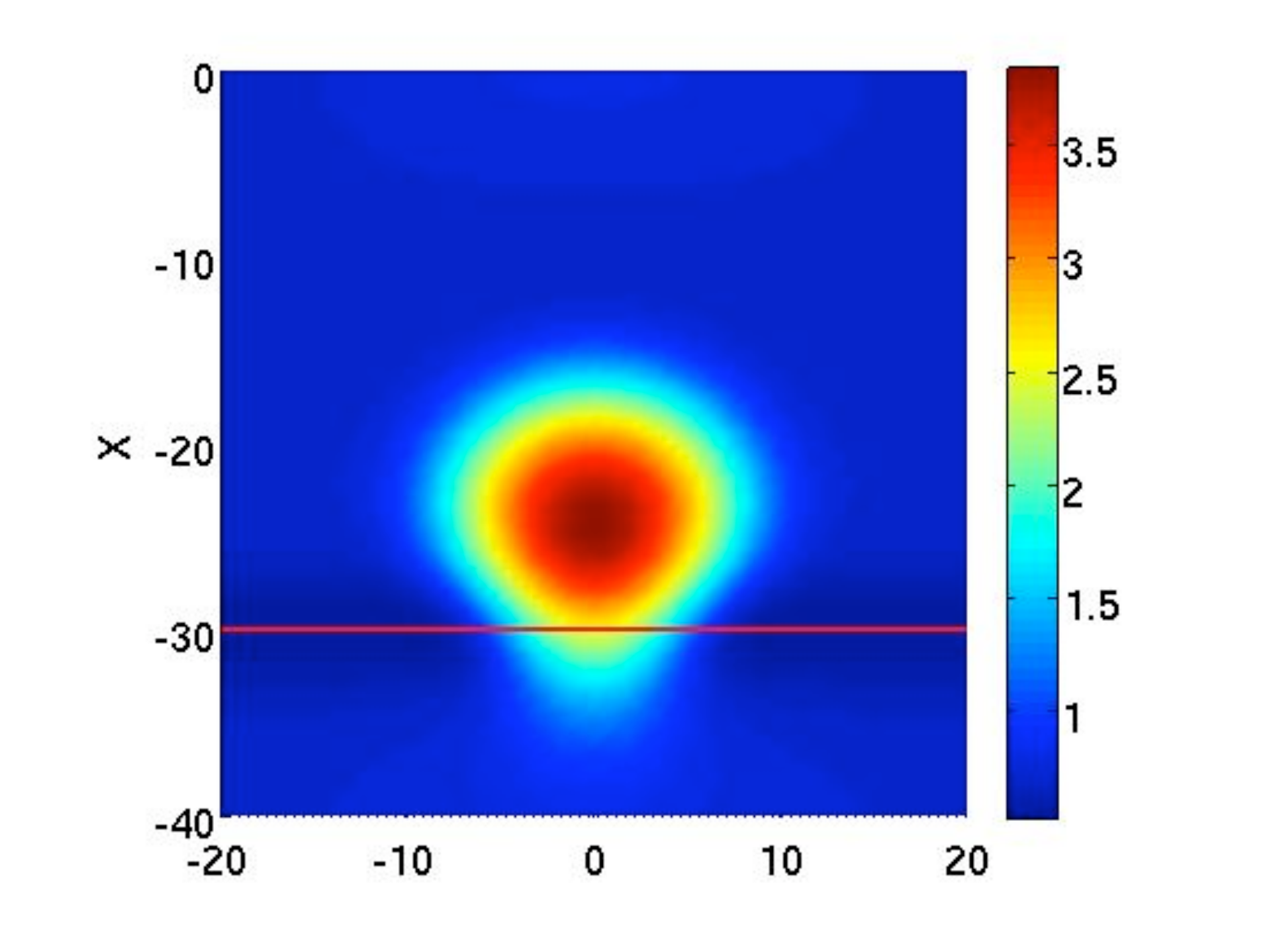} 
} 
\hspace*{2.9cm}y \hspace{6.2cm} y\\ 
\hspace*{3.0cm}(a) \hspace{6cm} (b) 
\caption{Shown are contours of the pinned (a) stable and (b) unstable 
  steady drop solutions at $P = 5.68 \cdot 10^{- 3}$ (leftmost circles in 
  Fig.~\ref{fig:cond3d}).  The thin horizontal line marks the maximum 
  of the heterogeneity. The remaining parameters are as in 
  Fig.~\ref{fig:cond3d}.} 
\label{fig:ss3d} 
\end{figure} 

\subsection{The two-dimensional case}\label{sec:pin3d} 
In the 2d case we consider stripe-like wettability defects.  In
  particular, we employ an $x$-dependent heterogeneity profile and
  choose the lateral driving force $P$ as well in $x$-direction. In
  this way a hydrophobic stripe ($\epsilon>0$) blocks a drop at its
  front end whereas a hydrophilic one ($\epsilon<0$) holds it at the
  back end.
  Fig.~\ref{fig:cond3d} presents the bifurcation diagram for
  $\epsilon=0.3$ obtained when continuing the steady single stable
  drop solution for increasing $P$. Stable (unstable) states are given
  as solid (dashed) lines. The stable blocked drop increases its norm
with increasing driving as it is 'pressed' against the defect. The
drop finally depins at a saddle-node bifurcation ($P_c\approx0.003$)
where the stable branch turns and loses stability.  Time simulations
beyond depinning show a typical stick-slip motion of drops with a
period that diverges when the bifurcation is approached from above.
The time averaged norm for selected values of $P$ is indicted by
triangles in Fig.~\ref{fig:cond3d}. The results indicate that
depinning occurs via a sniper bifurcation.
 
Examples of steady stable and unstable drop profiles are given in
Figs.~\ref{fig:ss3d}(a) and (b), respectively. The stable drop sits
behind the line of minimal wettability.  As it is squeezed against the
defect by the driving force it has an oval shape.  In contrast, the
unstable drop, has a 'forward protrusion' that crosses over the
minimum of wettability. This solution represents a depinning threshold
for $P<P_c$, i.e., adding a small perturbation will either let the
drop retract its advancing protrusion to again settle behind the
defect {\it or} trigger a depinning event that sends the drop sliding
down to the next defect where it is pinned again. For a more detailed
analysis of the depinning process in 2d see Ref.~\cite{BHT09}.

Finally, let us come back to the computational scheme. Apart from the
increase in the system size $N$, in the 2d case one encounters a new
difficulty related to the translation symmetry in $y$-direction.  It
implies that for each solution there exist a continuum of solutions
obtained by translation in $y$-direction that has to be avoided by the
continuation algorithm.  Neglecting numerical noise, the solutions
possess a left-right reflection symmetry $y\rightarrow -y$
(Fig.~\ref{fig:ss3d}). Therefore, the Jacobian is an equivariant
operator for this left-right reflection.  Thus, the action of the
Jacobian on left-right symmetric vectors results in vectors with the
same symmetry. This effectively excludes any translation in the
$y$-direction.  However, when the solution is close to the trivial
flat film state even small numerical noise becomes relevant and the
leading eigenvalues are very small as well.  In the above numerical
example we observe related problems for large driving at $P>0.1$ when
the steady solution corresponds to very shallow rivulet. The
continuation algorithm might then stay on the orbit of solutions
related by translation. The problem can be easily overcome by fixing
the maximum at a particular point, although, this might cause problems
in situations where various maxima may coexist. To avoid any
ambiguity, we use a similar technique as for the problem of mass
conservation: in the Krylov step we project the basis vectors
orthogonally to the translation mode $\partial_y h_0$. This does not
change the structure of the algorithm and is performed at negligible
cost (see app.~\ref{sec:krylov}).

\section{Conclusion} 
\label{sec:conc} 
We have presented (i) a time integration scheme based on exponential
propagation and (ii) a continuation algorithm employing the Cayley
transform for the highly non-linear thin film equations.  These
equations contain differential operators till fourth order.  To avoid
severe stability restrictions on the time-step, a linear term may be
treated implicitly.  However, for the thin film equations the
difficulty is to find a relevant linear operator. In consequence we
use the Jacobian matrix at each time-step.  In this framework, an
exponential propagation scheme is more efficient than a semi-implicit
scheme \cite{HLS98,Tokman06}. The method is based on an exact solution
of the linear problem for each time-step and involves the
determination of the exponential of a matrix. To do this in an
economic way, the linear operators may be reduced to small Krylov
subspaces.  However, for the thin film equation we need a dimension of
the Krylov subspaces of about 100 -- much larger than necessary for
problems involving only a second order operator \cite{FTDR89}.  For
better convergence of the Krylov reduction, we have coupled the
Arnoldi algorithm with the Cayley transform that is performed using an
ILU-factorization.  In consequence, this Cayley-Arnoldi method allows
us to take larger time-steps and, furthermore, well estimates the
leading eigenvalue. This facilitates an effective adaption of the
time-step to the changing characteristic time scale of the
dynamics. This has led to a major improvement in simulations of one-
and two-dimensional thin film dynamics that involve multiple time
scales as, e.g., coarsening dynamics for dewetting films or the
stick-slip motion close to depinning transitions.

We have also developed an algorithm for the continuation (or
path-following) of steady states that is based on the tangent
predictor - secant corrector scheme. Both tasks -- time stepping and
continuation -- can be performed using the same Cayley-Arnoldi
algorithm.  The advantage of this approach is the possibility to
perform all tasks arising in a bifurcation analysis
simultaneously. This includes the computation of the kernel of the
Jacobian to detect bifurcations and the stability analysis of the
steady states.
The developed algorithms have been used to study the bifurcation
structure and time evolution of (i) dewetting thin films and (ii)
depinning drops for physically two- and three-dimensional settings,
that correspond to 1d and 2d thin film equations, respectively.

For the dewetting film we have investigated different pathways for the
initial 'rupture', i.e., the short-time evolution.  The focus has been
on the competition of the surface instability and the nucleation at
defects. The long-time coarsening dynamics has as well been studied.
For the 1d case, we could follow the coarsening process till $10^6
\tau_m$ where $\tau_m$ is the characteristic time of the surface
instability.
In the two-dimensional case, we have found that the short-time
dewetting process induced by a radially symmetric finite defect
conserves the radial symmetry until disturbed by the boundaries of the
square domain. This indicates the very small influence of round-off
errors in our algorithm. The coarsening then proceeds via a cascade of
ring contractions. However due to the importance of thermal
fluctuations such regular structures are normally not observed in
dewetting experiments with very thin films.  However, they are
observed for thicker dielectric films in a capacitor when an
inhomogeneous electrical field plays the role of the finite defect
\cite{CZDC05}.  Adding noise to the initial conditions we recover the
'classical' dewetting structures.  Using the
Cayley-Arnoldi method for a system size of about $10 \lambda_m\times
10 \lambda_m$, one is able to simulate the coarsening dynamics till
reaching a single drop, i.e., the globally stable solution
(Fig.~\ref{fig:dew3dp2}).  Beside the time evolution we have employed
continuation to study 2d steady state solutions corresponding to
square and hexagonal arrangements of drops or holes.

Second, we have studied the depinning of ridges (1d case) and drops
(2d case) from substrate heterogeneities. Pinned steady solutions have
been followed using our path-following algorithm. In particular, we
have used as continuation parameters the wettability contrast and the
lateral driving force.  In the 1d case we have successfully reproduced
results obtained in Refs.~\cite{ThKn06,ThKn06b} using the package AUTO
\cite{AUTO97}.  We have as well tracked 2d stable and
unstable steady drop solutions. This has been supplemented by a study
of the evolution of time-dependent solutions beyond depinning.  This
has led us to the conclusion that in the 2d case depinning occurs as
in the 1d case via a sniper bifurcation and that beyond (but close to)
the bifurcation the sliding drops show stick-slip behavior.

Although our approach improves the time integration and path-following
for thin film equations, the employed constant equidistant finite
difference spatial discretization remains very basic.  The weakness of
such a regular discretization appears, for instance, when tracking
large drops pinned at hydrophobic defects.  For increasing driving
force the stable drop becomes strongly localized at the defect. The
unstable solution for the same driving force is very close to the
stable one as already observed in the 1d case for large drops (Fig.~29
of Ref.~\cite{ThKn06}). To clearly distinguish the two solutions
numerically a higher accuracy is required.  This can be achieved using
an adaptive mesh. In our particular system, for instance, the
continuation procedure applied to a drop on a $50\times50$ square
domain breaks down near the depinning bifurcation.  This accuracy
problem is similar to the one encountered in the last stages of
coarsening in Section~\ref{sec:dew2d} where a mesh refinement near the
drop edges would be beneficial.   

We have restricted our study to thin film equations that describe
films and drops on partially wettable homogeneous and heterogeneous
substrates. Beside the curvature pressure, the disjoining pressure has
been the only term resulting from the underlying free energy.
However, the method can be applied to various thin film systems
involving other contributions to the free energy. This includes, for
instance, thin films of dielectric liquid in a capacitor
\cite{MPBT05,VSKB05,BSTR09}, heated thin films
\cite{OrRo94,BPT03,ThKn04}, films with an effective thickness
dependent surface tension caused by high-frequency vibration
\cite{LMV01,TVK06} and film flows in a time-dependent ratchet
potential \cite{JHT08}.
The algorithms can also be applied to multilayer problems in the 
long-wave framework like, for instance, in bilayer dewetting 
\cite{PBMT05,BGS05,PBMT06} or to systems involving a single 
thin film equation coupled to a (reaction-)diffusion equation for a 
reactive \cite{DaPi84,PTTK07b} or non-reactive \cite{JeGr92,MaCr01} 
surfactant or adsorbate at the substrate \cite{TJB04,JBT05}.  For 
those problems the overall structure of the equations does not 
change. Only the system size $N$ is multiplied by the number of 
equations.  The time-stepping and continuation code can be applied 
without major change. 
 
In general, we expect the method to be as well relevant for closely
related evolution equations like, for instance, the (driven)
Cahn-Hilliard equation \cite{CaHi58,GNDZ01} or the
Kuramoto-Sivashinsky \cite{KNS90,BKOR92} equation which both contain
the Bilaplacian operator.  However, in contrast to the lubrication
equation for those equations the mobility function $m(h)$ multiplying
the Bilaplacian is often a constant.  As the presence of the
non-constant mobility function has been one reason for our choice of
the Jacobian matrix as linear operator at each time-step, for the
constant-mobility Cahn-Hilliard and the Kuramoto-Sivashinsky equation,
this choice is not crucial and a classical semi-implicit scheme with
$\mathbf{L}=\Delta^2$ as linear operator might result in a viable
scheme. The efficiency of both time integration schemes has to be
compared to decide which is the most powerful method.
 
The presented continuation algorithm for steady-state solutions, can
be improved in a straight forward manner by adapting it for stationary
states, i.e., for traveling waves or sliding drops. These can be seen
as steady-state solution in a co-moving frame. For the case of a
driving force in $x$-direction, solutions are also invariant w.r.t.\
translation in $x$-direction. The resulting problems can be overcome
using the same technique as above for the translational invariance in
the $y$-direction. Thus, the Cayley-Arnoldi method can be applied
without major change. The extended algorithm would be applicable,
e.g., to the study of the morphological transitions observed for
sliding drops on inclined homogeneous substrates \cite{PFL01,SRLL05}.
 
\appendix

\section{Krylov reductions}\label{sec:krylov} 
\subsection{Arnoldi-Krylov}
The approximation of $v_g=G(\jacobe\tau)b$ and
$v_e=\exp(\jacobe\tau)c$ is a crucial step of the time integration
algorithm (\mysec{time}).  We propose to use a Krylov reduction as
usually employed for sparse operators.  The aim is to obtain an
accurate approximation such that the time-step is only limited by the
order of the scheme.  As the technique works similarly for $v_g$ and
$v_e$ we only focus on $v_g$ (corresponding to the second order linear
scheme \myeq{ulin}).
The Krylov reduction employs that the series of subspaces
\begin{equation}
K_m=\mbox{span}\left\{b,\jacobe b,\jacobe^2 b,...,\jacobe^{m-1}b\right\}
\label{eq:ks}
\end{equation}
converges to a finite dimension Krylov subspace $K_M$ which contains
$v_g$. The method is only efficient if $K_m$
is a good approximation of $v_g$ for $m\ll N$. 
The Arnoldi method is used to construct an orthonormal basis $V_m$ of the
subspace $K_m$. The resulting approximated Jacobian matrix $\jacobe_m$,
is a $m\times m$ upper Hessenberg matrix
\begin{equation}
\jacobe_m=V_m^t\jacobe V_m\label{eq:jm}
\end{equation}
that in this form can be used to approximate
\begin{eqnarray}
v_g=G({\jacobe\tau})b&\simeq&V_mG(\jacobe_m\tau) V_m^t b. \label{eq:approx_krylov}
\end{eqnarray}
Since the dimension $m$ is small, the classical QR algorithm is
a reliable and efficient method to diagonalize $\jacobe_m$, to obtain the
matrix $D_m$ and to compute $G(\jacobe_m\tau)$. The resulting
approximation of the vector $v_g$ is
\begin{eqnarray}
v_g&\simeq&V_m\mathbf{P_m}G(D_m\tau)\mathbf{P_m^{-1}}V_m^t b
\end{eqnarray}
where $\mathbf{P_m}$ is the matrix of the eigenvectors of
$\jacobe_m$. The columns of the rectangular matrix $(V_m\mathbf{P_m})$
are the Ritz vectors, i.e., the approximated eigenvectors of
$\jacobe$.
Since the dynamics is dominated by the rightmost eigenvalues of
$\jacobe$ a good Krylov approximation results in Ritz vectors that are
close to the rightmost eigen-directions. We call them the 'wanted'
eigen-directions.  However, the Krylov-Arnoldi algorithm first
converges to the 'unwanted' eigenvalues of largest modulus that are
situated in the leftmost spectrum and are the main reason for
stability problems. To improve the Krylov-Arnoldi method we
  adapt algorithms normally used to estimate the rightmost spectrum:
  We transform the spectrum of $\jacobe$ in such a way that the
wanted eigen-directions are associated to the eigenvalues of largest
modulus. By applying Krylov-Arnoldi to the transformed operator the
wanted eigen-directions are selected after a few steps.  For
  asymmetric sparse systems two transformations can be used:
  Chebyshev acceleration and shift invert Cayley transform. We
  discuss their efficiency in the next section.

Note, that the Krylov reduction allows us to introduce supplementary
requirements on the discretized space in a simple way. For instance,
to preserve the volume one suppresses the direction corresponding to a
variation of volume in the Arnoldi step. Thereby, one ensures that the
height $h$ remains in the Euclidian space $\{H+E_0\}$ (\mysec{space})
during the time-stepping. In a similar manner, the directions related
to translation invariance may be suppressed during the continuation
algorithm to problem mentioned in \mysec{pin3d}.
\subsection{Chebyshev acceleration} 
\myeq{approx_krylov} can be interpreted as a
polynomial approximation since the basis $V_m$ is a sum of elements of
$K_m$ (\myeq{ks}) \cite{Saad92}:
\begin{equation}
v_G\simeq p_{m}(\jacobe\tau)b
\label{eq:poly}
\end{equation}
where $p_m$ is a polynomial of degree $m-1$.  Chebyshev acceleration
determines an optimal polynomial to accelerate the convergence.  In
\cite{Manteuffel1977} it is shown that scaled and translated
Chebyshev polynomials have certain optimal convergence properties.
The main reason of the success of the Chebyshev polynomial is the
possibility to decrease some 'unwanted' eigen-directions contained in
an ellipse.  This property is employed to compute the rightmost
eigenvalues of large sparse non-symmetric matrices
\cite{Ho1990,Sadkane1993}.  However, in our case the application of
the algorithms presented in \cite{Meerbergen1996} does not
significantly improve the convergence of the Krylov approximation.
Indeed, according to \cite{Sadkane1993} the rapidity of the
convergence is directly affected by the accumulation of the rightmost
eigenvalues.  In consequence, a polynomial approximation does not
result in the improvement of the Krylov reduction.
\subsection{Cayley transform} 
Unlike the Chebyshev method, the Cayley transform is not polynomial
but rational. Let us introduce the transformed operator $\mathbf{C}$:
\begin{equation} 
\mathbf{C}=\mathbf{J}_c^{-1}=(\mathbf{J}-c\mathbf{I})^{-1} \label{eq:cayley} 
\end{equation} 
where $c$ is an arbitrary real constant. The matrix
$\mathbf{C}$ contains the eigenvectors of $\jacobe$ but has a
different spectrum. If the chosen $c$ is larger than the leading
eigenvalue $\lambda_{\rm max}$ of $\mathbf{J}$, the spectrum of
$\mathbf{C}$ falls into the band $\left[({\rm Re}(\lambda_{\rm
    max})-c)^{-1};0\right[$. In consequence, the wanted
eigen-directions of $\jacobe$ correspond to the eigenvalues of
$\mathbf{C}$ with the largest modulus. Therefore, the orthonormal
basis $V_m$ constructed using the Arnoldi procedure within
$\mathbf{C}$ should converge after a few steps to the wanted
eigen-directions. One introduces the approximate operator $\mathbf{C_m}$
\begin{equation} 
\mathbf{C_m}={V_m^t}\mathbf{C}{V_m}\label{eq:redc} 
\end{equation}  
and diagonalizes $\mathbf{C_m}$ using the QR method (similar to the Arnoldi-Krylov reduction):
\begin{equation} 
\mathbf{C_m}=\mathbf{P_m}\mathbf{D_{cayley}}\mathbf{P_m^{-1}}.
\end{equation}
Finally according to the definition (\ref{eq:cayley}) and the Krylov
reduction (\ref{eq:redc}) we obtain the approximation of $v_g$
\begin{equation}
v_g=G(\jacobe\tau)b \simeq V_m\mathbf{P_m}G(\mathbf{D_{cayley}^{-1}+c\mathbf{I}})\mathbf{P_m^{-1}} V_m^tb.
\end{equation}
The efficiency of the algorithm depends strongly on the choice of
$c$. Indeed if $c\gg {\rm Re}(\lambda_{\rm max})$, then the Cayley transform is
not relevant any more since $\mathbf{C}\simeq-c^{-1}\mathbf{I}$. In
addition, if $c$ is close to an eigenvalue of $\jacobe$, the operator
$\jacobe-c\mathbf{I}$ becomes singular and the method diverges.
Numerical calculations show, that for moderately large
values of $c$ (as compared to Re$(\lambda_{\rm max})$) the
accuracy decreases notably. However, a choice of $c$ close to
Re$(\lambda_{\rm max})$ leads to very accurate results even if
$c<{\rm Re}(\lambda_{\rm max})$.  
In consequence, a good choice is to take a constant $c^k$ at time-step
$k$ that is slightly larger than Re$(\lambda_{\rm max}^{k-1})$ as
estimated in the previous time-step $k-1$. The constant $c^k$ could
then be smaller or larger than Re$(\lambda_{\rm max}^{k})$.  In rare
cases it might be very close to an eigenvalue of the Jacobian matrix
$\jacobe$. However, such a degeneracy would be detected automatically
by the presence of huge rightmost eigenvalues of the matrix
$\mathbf{C}$. So, in that singular case the time-step is performed
again with a larger value of $c$.

The difficulty of the method remains the evaluation of the action of
$\mathbf{C}$, defined as the inverse of $\jacobe_c$, on the vector
$b$.  It is obtained using an Incomplete LU-factorization (ILU) on
$\jacobe_c$ which is a powerful method for sparse band-matrices.  The
cost of this factorization is about $O(N^{3/2})$. As all other
operations are $O(N)$ we expect that the ILU slows down the algorithm
for large systems.
\subsection{Comparison} 
We next compare the Krylov-Arnoldi and Cayley-Arnoldi methods for the
estimation the vector $v_g$. Even though the Krylov reduction does not
depend on the chosen time-step $\tau$, we expect that the
approximation of $v_g$ does.  We aim at a Krylov reduction that is
accurate enough to not add a restriction on the time-step $\tau$. For
a second order linear scheme \myeq{ulin} the relevant time scale $\tau_\lambda$ is
the inverse of the leading eigenvalue. This value is employed in the different
  numerical convergence tests.

In the one-dimensional case, $v_g$ may be computed by a direct method
as, for instance, a QR-diagonalization. The latter is taken as
reference value $v_{\rm ref}$ for the relative error. Unfortunately,
for the two-dimensional case, the system may be large,
i.e.~$N=O(10^5)$, and a QR-diagonalization dramatically increases the
CPU cost and memory requirements (being proportional to $N^3$).  Thus,
in this case, the reference solution $v_{\rm ref}$ is computed using
the Cayley-Arnoldi method for a Krylov subspace dimension $m$ large
enough to obtain convergence.

\begin{figure} 
{\centering 
 (a)\includegraphics[height=6cm,width=6cm]{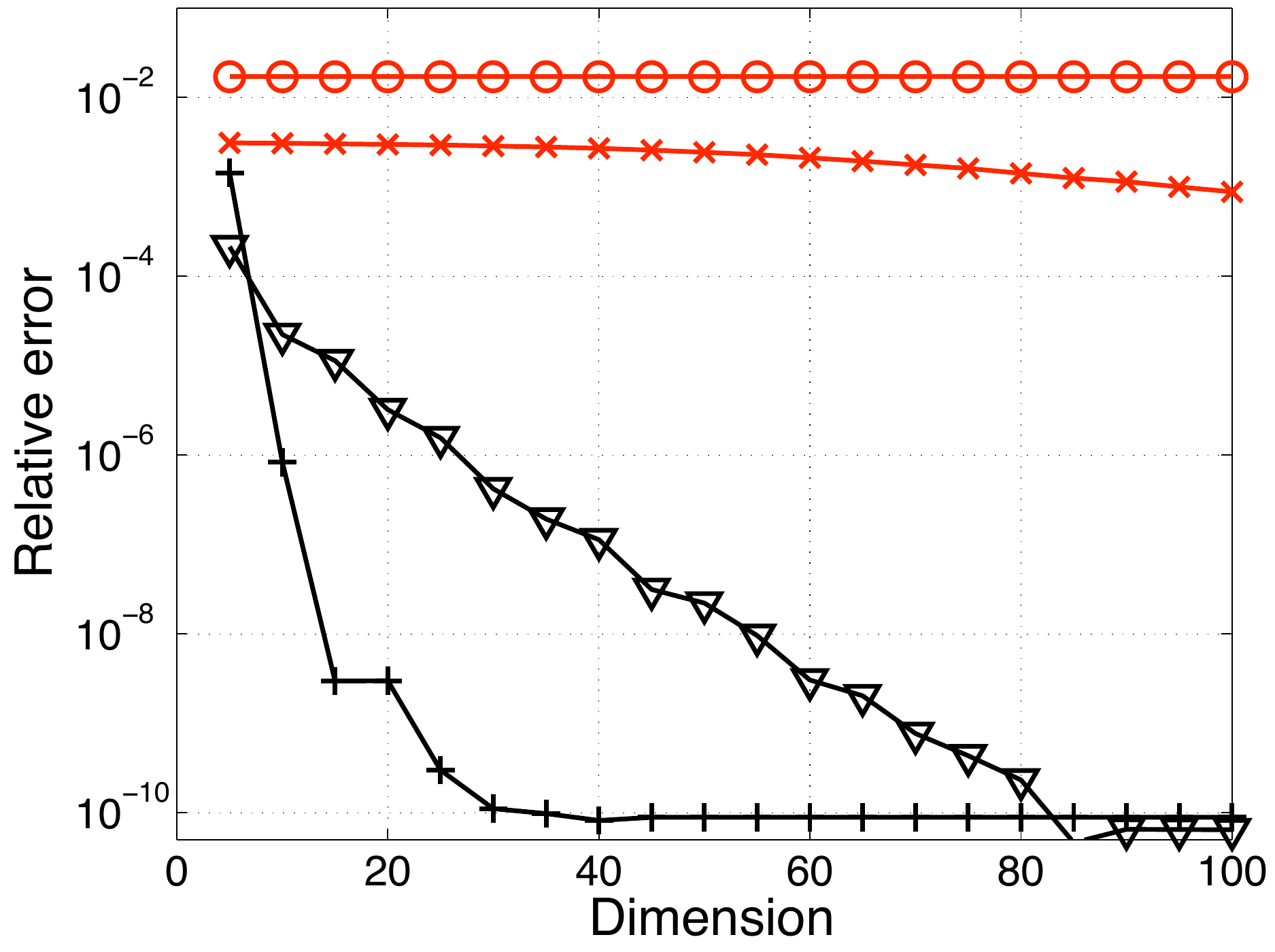} 
 \includegraphics[height=6cm,width=6cm]{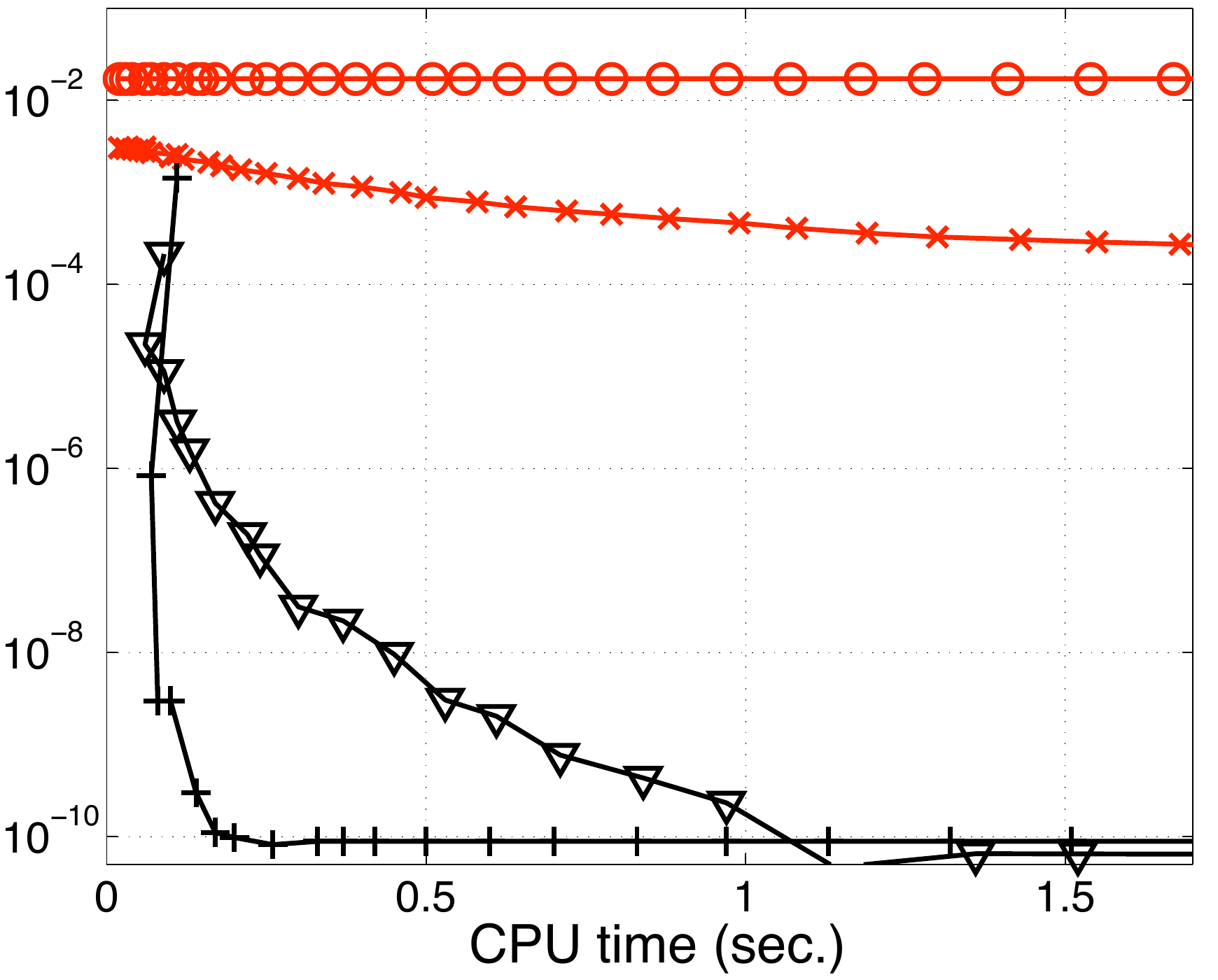}(b)} 
\caption{Convergence of the time-stepping algorithm in dependence of
  (a) dimension $K$ of Krylov subspaces and (b) CPU time needed to
  determine the profile variation $u$ for one time-step. The example
  used is the 1d dewetting problem (section~\ref{sec:dew2d}) with parameters as in
  Fig.~\ref{fig:dew2dp}(b).  Results are given for two sets of
  computational parameters -- set~1 has time-step
  $\tau_1/\tau_m=0.08$, initial profile $h_1$ at time $t_1=10\tau_m$;
  and set~2 has time-step $\tau_2=100\tau_m$, initial profile $h_2$
  at time $t_2=2\cdot 10^3\tau_m\simeq\tau_\lambda$.  Convergence is
  shown for the simple Arnoldi method (symbols
  ``$\color{red}{\times}$'' for set~1, ``{\color{red}{o}}'' for
  set~2), and the Cayley-Arnoldi method (symbols ``$\nabla$'' for
  set~1, ``$+$'' for set~2).  }
\label{fig:val1d}
\end{figure}

\begin{figure} 
{\centering 
(a)\includegraphics[width=6cm,height=6cm]{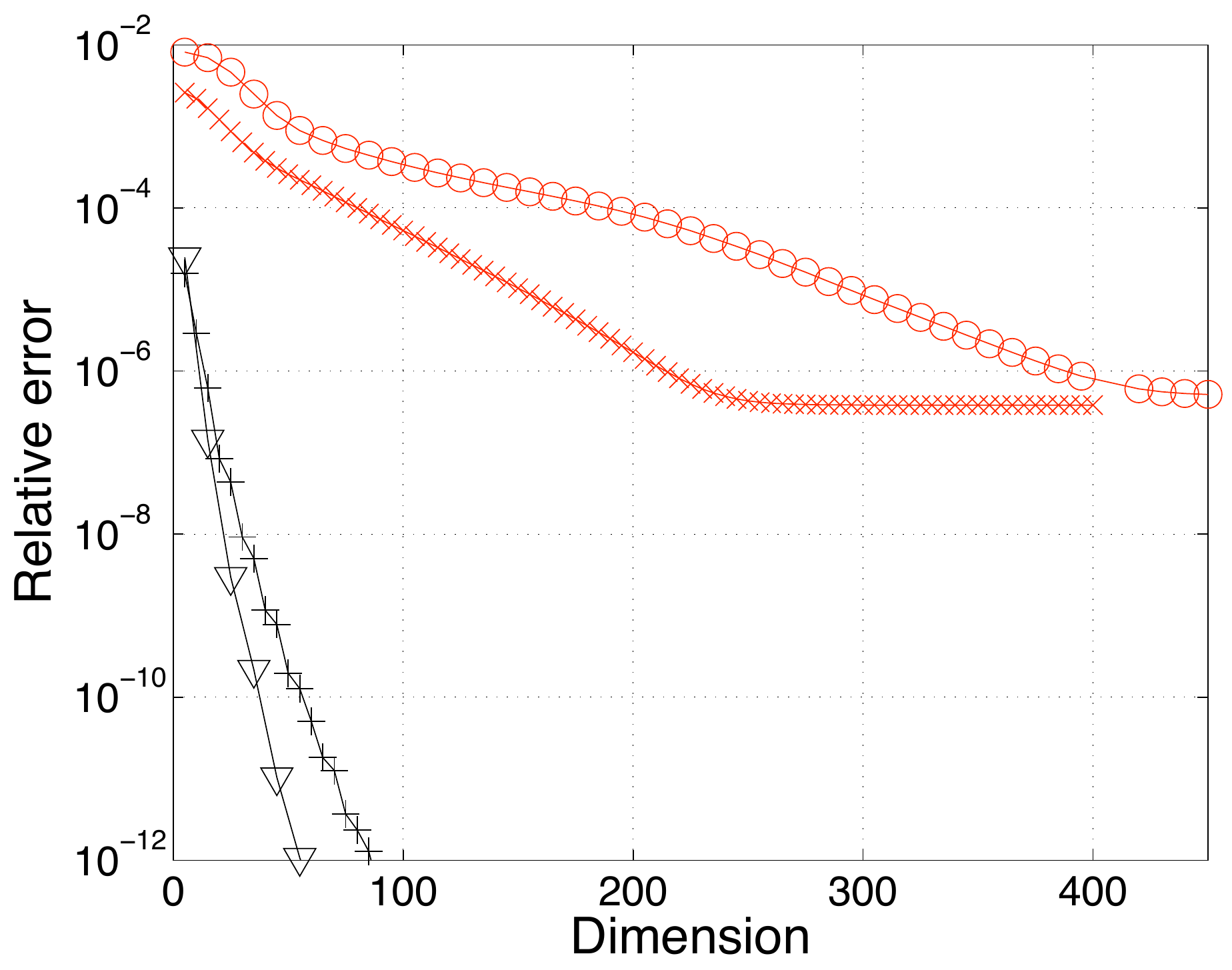} 
\includegraphics[width=6cm,height=6cm]{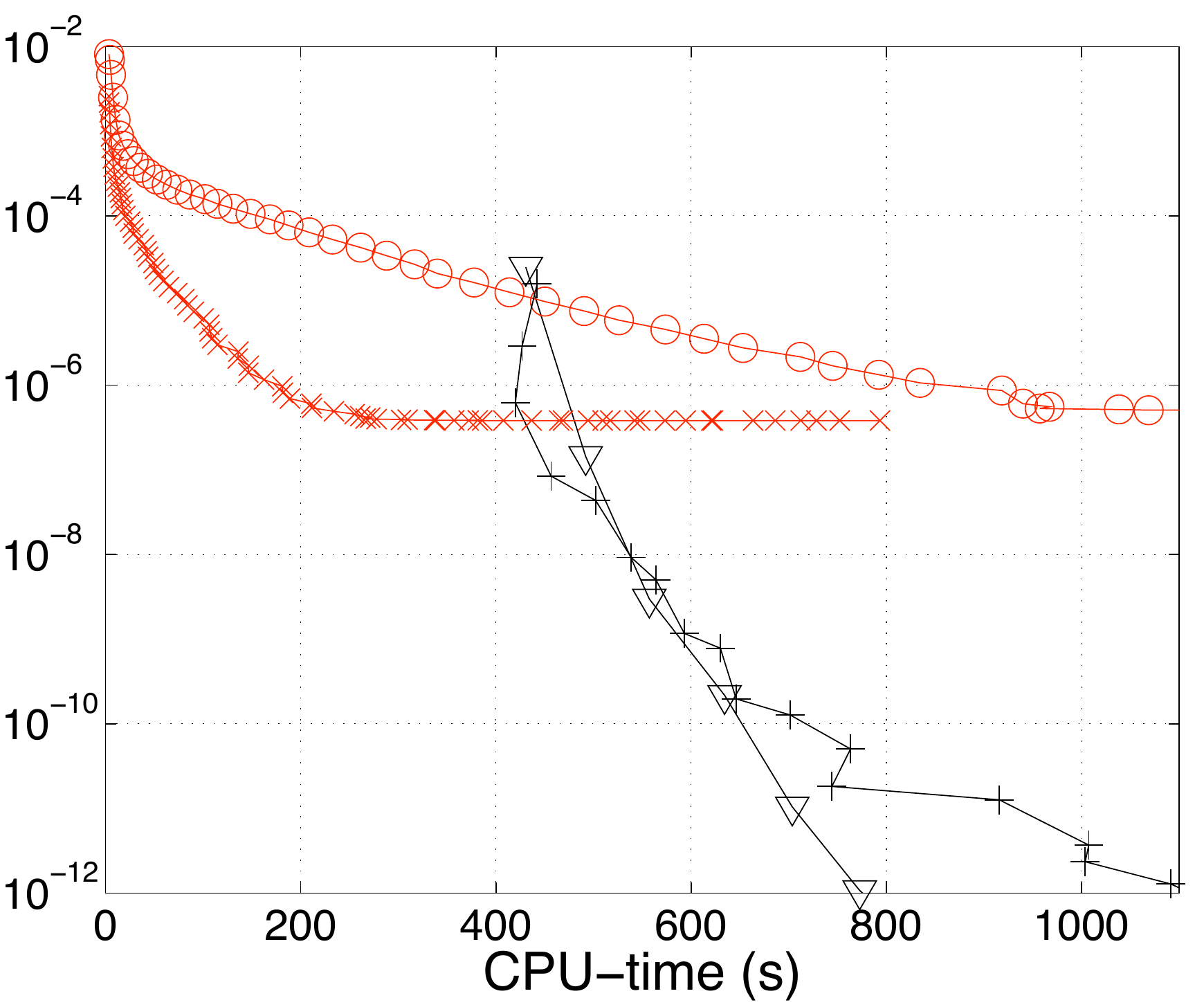}(b)} 
\caption{Convergence of the time-stepping algorithm in dependence of
  (a) dimension $K$ of Krylov subspaces and (b) CPU time needed to
  determine the profile variation $u$ for one time-step. The example
  used is the 2d dewetting problem (section~\ref{sec:dew3d}) with
  parameters as in Fig.~\ref{fig:dew3dp2}(b).  Results are given for
  two sets of computational parameters -- set~1 has time-step
  $\tau_1=0.05\tau_m\simeq\tau_\lambda$, initial profile $h_0$ at time
  $t=22\tau_m$; and set~2 has time-step
  $\tau_2=0.15\tau_m\simeq\tau_\lambda$, and $h_0$ as set~1.
  Convergence is shown for the simple Arnoldi method (symbols
  ``$\color{red}{\times}$'' for set~1, ``{\color{red}{o}}'' for
  set~2), and the Cayley-Arnoldi method (symbols ``$\nabla$'' for
  set~1, ``$+$'' for set~2).  }
\label{fig:val2d}
\end{figure}

Figs.~\ref{fig:val1d} and \ref{fig:val2d} present selected results for
the convergence of the time-stepping scheme for the dewetting problem
in the 1d and 2d case, respectively.  Both panels (a) shown an
extremely slow convergence of the classical Krylov method. Krylov
subspaces with dimensions $m\approx100$ have still a relative error of
$10^{-4}$.  An accuracy of about $10^{-6}$ is obtained by a
$800$-dimensional Krylov subspace (\myfig{val2d}(a)).  In contrast,
with the Cayley-method, the same accuracy is obtained with Krylov
subspaces of much small dimensions $m\approx10\dots30$.  For a
time-step $\tau$ smaller than the characteristic time $\tau_\lambda$,
the convergence is notably improved for the classical Krylov algorithm
only. This indicates that for this Krylov reduction the eigenvalues
are badly approximated contrary to a Krylov algorithm using the Cayley
transform.
We find that the efficiency of the two methods is equivalent for
$N\simeq10^5$.  By 'efficiency' we mean the ratio between the
time-step $\tau$ and the CPU time cost for a relative error of
$10^{-6}$.
 
In conclusion, the Cayley transform emerges as a powerful method to
perform an accurate Krylov reduction.  However, its efficiency is
limited by the ILU-factorization which considerably slows down the
time-step for large system $N>10^5$. In this case the simple Arnoldi
procedure is preferable and the accuracy of the time integration
scheme is determined by the Krylov approximation.  Finally, the
Cayley-Arnoldi reduction may be applied to higher order schemes such
as those presented in \cite{HLS98,Tokman06}. Since it is not required
to use another ILU-factorization this can be done at negligible cost.

Note that the estimate of the leading eigenvalue $\lambda_m$ 
  in the Cayley-Arnoldi reduction allows one to evaluate the time-step $\tau$.
  One assumes that the relative error $\epsilon_r$ depends on the relative
  profile variation defined by
$\epsilon_{\rm var}=||u^{(1)}||/||h||$, and approximated using Eq.~(\ref{eq:ulin}) with the
Jacobian matrix $\jacobe$ replaced by the scalar $\lambda_m$. This
gives the time-step
\begin{equation} \label{eq:tau} 
\tau=\frac{1}{\lambda_m}\ln{\left( 1+ \lambda_m\epsilon_{\rm var}\frac{||h||}{||F(h)||}\right)}.
\end{equation} 
Here we fix the typical value of $\epsilon_{\rm var}$ at a few
percent. The relative error $\epsilon_r$ resulting from the estimate
(\ref{eq:tau}) is rarely larger than a $10^{-6}$.  
%

%
\section{Comparison to classical algorithm}
\label{sec:concalgo} 
$\jacobe$ using the ILU factorization.  As the complexity of
  this task  is $O(N^{3/2})$ the CPU cost is a
more important issue than in other schemes.

If one implemented a classical implicit time integrator one would need
to solve a linear system involving the matrix $\jacobe$. For instance,
a backward Euler scheme leads to the system
$(\mathbf{I}+\jacobe\tau)u=b\tau$, where $b$ is a known vector.  A
typical choice for the latter is the use of an iterative Krylov method
as, e.g., the GEMRES method which is a good candidate for asymmetric
matrices.  The efficiency of this method does strongly depend on the
knowledge of an effective preconditioner.  Without preconditioner, the
inversion for a simple semi-implicit scheme (backward Euler) converges
very slowly and almost fails to obtain the wanted tolerance.  
  Since no general preconditioner exists for the asymmetric matrix
  $\jacobe$, its ILU factorization appears to be the only systematic
  way to construct an effective preconditioner.  Therefore, a clear
analogy exists between the exponential scheme and a semi-implicit
scheme.  On the one hand the Arnoldi algorithm to compute
$G(\jacobe\tau)$ is equivalent to the use of an iterative method
without preconditioner to solve the linear system in an implicit
scheme. On the other hand, the Cayley-Arnoldi algorithm to compute
$G(\jacobe\tau)$ is equivalent to an iterative method with $LU$
preconditioner to solve the linear system in an implicit scheme.
Judging the complexity of the algorithm, a classical implicit and an
exponential propagation algorithm are roughly equivalent schemes at
the same order.  However, exponential schemes seem to have two
advantages:
\begin{itemize}
\item The $G$ and $\exp$ functions have a better leftmost spectrum
  filtering property than rational functions.
\item Krylov techniques converge faster
  when employed for the evaluation of $G(\jacobe\tau)b$ and $\exp(\jacobe\tau)$ than
   when employed for the solution of a linear system \cite{HLS98}.
\end{itemize}
Furthermore, the Cayley-Arnoldi exponential propagation method used
has the ability to give an estimate of the leading eigenvalues what
the implicit method is not able to do.  This information allows for an
adaptive time-step that constitutes a major advantage when studying
dynamics characterized by several time scales.  In addition, it is an
important information that allows one to perform non-standard stability
analysis as, e.g., employed by M\"unch \cite{Muench05} to detect a
fingering instability in dewetting.

Although the shift Cayley-transform is a standard method to find the
rightmost eigenvalues of a operator \cite{WhBa06} its application in a
continuation algorithm is less common.  However, when solving a linear
system within the Newton algorithm one faces similar problems as
discussed for time-stepping. Thus, an effective solution method
requires the use of the $LU$ preconditioner.  However, our algorithm
does not use the ILU factorization for a direct inversion, but rather
to find the rightmost eigenvalues. The advantage of this approach is
the ability to determine the tangent direction even close to a
saddle-node bifurcation.  Furthermore, at a bifurcation point, the
directions of (different) bifurcating branches can be found as set of
tangent directions.

We emphasize that the numerical difficulties encountered in the
time-stepping and the continuation task are both resolved using the
Cayley transform.  This reminds Ref.~\cite{TB00} where it is
pointed out that algorithms overcoming the numerical difficulties
  encountered during a time-stepping scheme can be adapted to perform
  the bifurcation tasks. The developed algorithms overcome two main
problems: (i) the operator has no 'simple' relevant linear part
(different spatial scaling), and (ii) the ``very bad'' conditioning of
the Jacobian. With the current computers, the Cayley-Arnoldi method
is the most efficient method for a moderate system size of
$N=O(10^5)$.  In consequence, the scheme is difficult to adapt for
three-dimensional PDEs.

\paragraph{Acknowledgments}  
We acknowledge support by the European Union via the FP7 Marie Curie
scheme [Grant PITN-GA-2008-214919 (MULTIFLOW)] and by the Deutsche
Forschungsgemeinschaft under grant SFB 486, project B13. We also thank
the Max-Planck-Institut f\"ur Physik komplexer Systeme in Dresden
(Germany) that hosted us during the early stage of the project. Finally, P.B. is grateful to L.~S.~ {\sc Tuckerman} for fruitful discussions.

\bibliographystyle{siam} 



\begin{thebibliography}{100}

\bibitem{BGS05}
{\sc D.~Bandyopadhyay, R.~Gulabani, and A.~Sharma}, {\em Stability and dynamics
  of bilayers}, Ind. Eng. Chem. Res., 44 (2005), pp.~1259--1272.

\bibitem{BSTR09}
{\sc D.~Bandyopadhyay, A.~Sharma, U.~Thiele, and P.~D.~S. Reddy}, {\em Electric
  field induced interfacial instabilities and morphologies of thin viscous and
  elastic bilayers}, Langmuir, 25 (2009), pp.~9108--9118.

\bibitem{Beck03}
{\sc J.~Becker, G.~Gr{\"u}n, R.~Seemann, H.~Mantz, K.~Jacobs, K.~R. Mecke, and
  R.~Blossey}, {\em Complex dewetting scenarios captured by thin-film models},
  Nat. Mater., 2 (2003), pp.~59--63.

\bibitem{BHT09}
{\sc P. Beltrame, P. H{\"a}nggi and U. Thiele},
{\em Depinning of three-dimensional drops from wettability defects},
Europhys. Lett., 86 (2009), p.~24006.

\bibitem{BCP01}
{\sc M.~Ben~Amar, L.~Cummings, and Y.~Pomeau}, {\em Singular points of a moving
  contact line}, C R Acad. Sci. Ser. IIB, 329 (2001), pp.~277--282.

\bibitem{BeBr97}
{\sc A.~L. Bertozzi and M.~P. Brenner}, {\em Linear stability and transient
  growth in driven contact lines}, Phys. Fluids, 9 (1997), pp.~530--539.

\bibitem{BGW01}
{\sc A.~L. Bertozzi, G.~Gr{\"u}n, and T.~P. Witelski}, {\em Dewetting films:
  {B}ifurcations and concentrations}, Nonlinearity, 14 (2001), pp.~1569--1592.

\bibitem{BMFC98}
{\sc A.~L. Bertozzi, A.~M{\"u}nch, X.~Fanton, and A.~M. Cazabat}, {\em Contact
  line stability and ''undercompressive shocks'' in driven thin film flow},
  Phys. Rev. Lett., 81 (1998), pp.~5169--5173.

\bibitem{BMS99}
{\sc A.~L. Bertozzi, A.~M{\"u}nch, and M.~Shearer}, {\em Undercompressive
  shocks in thin film flows}, Physica D, 134 (1999), pp.~431--464.

\bibitem{BeNe01}
{\sc M.~Bestehorn and K.~Neuffer}, {\em Surface patterns of laterally extended
  thin liquid films in three dimensions}, Phys. Rev. Lett., 87 (2001),
  p.~046101.

\bibitem{BPT03}
{\sc M.~Bestehorn, A.~Pototsky, and U.~Thiele}, {\em {3D} large scale
  {M}arangoni convection in liquid films}, Eur. Phys. J. B, 33 (2003),
  pp.~457--467.

\bibitem{Bonn09}
{\sc D.~Bonn, J.~Eggers, J.~Indekeu, J.~Meunier, and E.~Rolley}, {\em Wetting
  and spreading}, Rev. Mod. Phys., 81 (2009), pp.~739--805.

\bibitem{BKOR92}
{\sc H.~S. Brown, I.~G. Kevrekidis, A.~Oron, and P.~Rosenau}, {\em Bifurcations
  and pattern-formation in the regularized {K}uramoto-{S}ivashinsky equation},
  Phys. Lett. A, 163 (1992), pp.~299--308.

\bibitem{Burg01}
{\sc J.~M. Burgess, A.~Juel, W.~D. McCormick, J.~B. Swift, and H.~L. Swinney},
  {\em Suppression of dripping from a ceiling}, Phys. Rev. Lett., 86 (2001),
  pp.~1203--1206.

\bibitem{BuGo83}
{\sc E.~Buzano and M.~Golubitsky}, {\em Bifurcation on the Hexagonal Lattice and the Planar Benard Problem}, Phil. Trans. R. Soc. Lond. A, 308 (1983), pp.~617--667.

\bibitem{CaHi58}
{\sc J.~W. Cahn and J.~E. Hilliard}, {\em Free energy of a nonuniform system.
  1. {I}nterfacual free energy}, J. Chem. Phys., 28 (1958), pp.~258--267.

\bibitem{CHTC90}
{\sc A.~M. Cazabat, F.~Heslot, S.~M. Troian, and P.~Carles}, {\em Fingering
  instability of thin spreading films driven by temperature gradients}, Nature,
  346 (1990), pp.~824--826.

\bibitem{CZDC05}
{\sc L.~Chen, L.~Zhuang, P.~Deshpande, and S.~Chou}, {\em Novel polymer
  patterns formed by lithographically induced self-assembly ({LISA})},
  Langmuir, 21 (2005), pp.~818--821.

\bibitem{Conway92}
{\sc J.~Conway}, {\em The orbifold notation for surface groups}, in Groups,
  Combinatorics and Geometry, M.~W. Liebeck and J.~S. (eds.), eds., Proceedings
  of the L.M.S. Durham Symposium,, Durham, UK, July 5Ð15 1992.

\bibitem{CBH08}
{\sc B.~P. Cook, A.~L. Bertozzi, and A.~E. Hosoi}, {\em Shock solutions for
  particle-laden thin films}, SIAM J. Appl. Math., 68 (2008), pp.~760--783.

\bibitem{CrKn91}
{\sc J.~D. Crawford and E.~Knobloch}, {\em Symmetry and symmetry-breaking
  bifurcations in fluid-dynamics}, Ann. Rev. Fluid Mech., 23 (1991),
  pp.~341--387.

\bibitem{CrHo93}
{\sc M.~C. Cross and P.~C. Hohenberg}, {\em Pattern formation out of
  equilibrium}, Rev. Mod. Phys., 65 (1993), pp.~851--1112.

\bibitem{DaPi84}
{\sc Z.~Dagan and L.~M. Pismen}, {\em Marangoni waves induced by a multistable
  chemical reaction on thin liquid films.}, J. Colloid Interface Sci., 99
  (1984), pp.~215--225.

\bibitem{deGe85}
{\sc P.-G. de~Gennes}, {\em Wetting: {S}tatics and dynamics}, Rev. Mod. Phys.,
  57 (1985), pp.~827--863.

\bibitem{DeOr92}
{\sc R.~J. Deissler and A.~Oron}, {\em Stable localized patterns in thin liquid
  films}, Phys. Rev. Lett., 68 (1992), pp.~2948--2951.

\bibitem{DCM87}
{\sc B.~V. Derjaguin, N.~V. Churaev, and V.~M. Muller}, {\em Surface Forces},
  Consultants Bureau, New York, 1987.

\bibitem{DiKo02}
{\sc J.~A. Diez and L.~Kondic}, {\em Computing three-dimensional thin film
  flows including contact lines}, J. Comput. Phys., 183 (2002), pp.~274--306.

\bibitem{AUTO97}
{\sc E.~J. Doedel, A.~R. Champneys, T.~F. Fairgrieve, Y.~A. Kuznetsov,
  B.~Sandstede, and X.~J. Wang}, {\em AUTO97: Continuation and bifurcation
  software for ordinary differential equations}, Concordia University,
  Montreal, 2000.

\bibitem{ESR00}
{\sc M.~H. Eres, L.~W. Schwartz, and R.~V. Roy}, {\em Fingering phenomena for
  driven coating films}, Phys. Fluids, 12 (2000), pp.~1278--1295.

\bibitem{Ferm92}
{\sc M.~Fermigier, L.~Limat, J.~E. Wesfreid, P.~Boudinet, and C.~Quilliet},
  {\em 2-dimensional patterns in {R}ayleigh-{T}aylor instability of a
  thin-layer}, J. Fluid Mech., 236 (1992), pp.~349--383.

\bibitem{FTDR89}
{\sc R.~A. Friesner, L.~S. Tuckerman, B.~C. Dornblaser, and T.~V. Russo}, {\em
  A method for exponential propagation of large systems of stiff nonlinear
  differential equations}, J. Sci. Comp., 4 (1989), pp.~327--354.

\bibitem{GBP07}
{\sc S.~Gasner, P.~Blomgren, and A.~Palacios}, {\em Noise-induced intermittency
  in cellular pattern-forming systems}, Int. {J}. {B}if. {C}haos, 17 (2007),
  pp.~2765--2779.

\bibitem{GlWi03}
{\sc K.~B. Glaser and T.~P. Witelski}, {\em Coarsening dynamics of dewetting
  films}, Phys. Rev. E, 67 (2003), p.~016302.

\bibitem{GNDZ01}
{\sc A.~A. Golovin, A.~A. Nepomnyashchy, S.~H. Davis, and M.~A. Zaks}, {\em
  Convective {C}ahn-{H}illiard models: {F}rom coarsening to roughening}, Phys.
  Rev. Lett., 86 (2001), pp.~1550--1553.

\bibitem{Gruen03}
{\sc G.~Gr{\"u}n}, {\em On the convergence of entropy consistent schemes for
  lubrication type equations in multiple space dimensions}, Math. Comput., 72
  (2003), pp.~1251--79.

\bibitem{GrRu01}
{\sc G.~Gr\"un and M.~Rumpf}, {\em Simulation of singularities and
  instabilities arising in thin film flow}, Euro. Jnl of Applied Mathematics,
  12 (2001), pp.~293--320.

\bibitem{Ho1990}
{\sc D.~Ho}, {\em Tchebychev acceleration technique for large scale
  nonsymmetric matrices}, Numer. Math., 56 (1990), pp.~721--734.

\bibitem{HoLu97}
{\sc M.~Hochbruck and C.~Lubich}, {\em On krylov subspace approximations to the
  matrix exponential operator}, SIAM J. Numer. Anal., 34 (1997),
  pp.~1552--1574.

\bibitem{HLS98}
{\sc M.~Hochbruck, C.~Lubich, and H.~Selhofer}, {\em Exponential integrators
  for large systems of differential equations}, SIAM J. Sci. Comp., 19 (1998),
  pp.~1552--1574.

\bibitem{Isra92}
{\sc J.~N. Israelachvili}, {\em Intermolecular and Surface Forces}, Academic
  Press, London, 1992.

\bibitem{JeGr92}
{\sc O.~E. Jensen and J.~B. Grotberg}, {\em Insoluble surfactant spreading on a
  thin viscous film: {S}hock evolution and film rupture}, J. Fluid Mech., 240
  (1992), pp.~259--288.

\bibitem{JBT05}
{\sc K.~John, M.~B{\"a}r, and U.~Thiele}, {\em Self-propelled running droplets
  on solid substrates driven by chemical reactions}, Eur. Phys. J. E, 18
  (2005), pp.~183--199.

\bibitem{JHT08}
{\sc K.~John, P.~H{\"a}nggi, and U.~Thiele}, {\em Ratchet-driven fluid
  transport in bounded two-layer films of immiscible liquids}, Soft Matter, 4
  (2008), pp.~1183--1195.

\bibitem{Kall00}
{\sc S.~Kalliadasis}, {\em Nonlinear instability of a contact line driven by
  gravity}, J. Fluid Mech., 413 (2000), pp.~355--378.

\bibitem{KaTh07}
{\sc S.~Kalliadasis and U.~Thiele}, eds., {\em Thin Films of Soft Matter},
  Springer, Wien / New York, 2007.
\newblock {CISM~490}.

\bibitem{KaTr97}
{\sc D.~E. Kataoka and S.~M. Troian}, {\em A theoretical study of instabilities
  at the advancing front of thermally driven coating films}, J. Colloid
  Interface Sci., 192 (1997), pp.~350--362.

\bibitem{KNS90}
{\sc I.~G. Kevrekidis, B.~Nicolaenko, and J.~C. Scovel}, {\em Back in the
  saddle again - a computer-assisted study of the {K}uramoto-{S}ivashinsky
  equation}, SIAM J. Appl. Math., 50 (1990), pp.~760--790.

\bibitem{KSR00}
{\sc R.~Khanna, A.~Sharma, and G.~Reiter}, {\em The {ABC} of pattern evolution
  in self-destruction of thin polymer films}, EPJdirect, E2 (2000), pp.~1--9.

\bibitem{LMV01}
{\sc V.~Lapuerta, F.~J. Mancebo, and J.~M. Vega}, {\em Control of
  {R}ayleigh-{T}aylor instability by vertical vibration in large aspect ratio
  containers}, Phys. Rev. E, 64 (2001), p.~016318.

\bibitem{LDL05}
{\sc N.~Le~Grand, A.~Daerr, and L.~Limat}, {\em Shape and motion of drops
  sliding down an inclined plane}, J. Fluid Mech., 541 (2005), pp.~293--315.

\bibitem{LiTi04}
{\sc J.~Liesen and P.~Tich{\'y}}, {\em Convergence analysis of {K}rylov
  subspace methods}, GAMM Mitt. Ges. Angew. Math. Mech., 27 (2004),
  pp.~153--173.

\bibitem{Lin01}
{\sc Z.~Lin, T.~Kerle, S.~M. Baker, D.~A. Hoagland, E.~Sch{\"a}ffer,
  U.~Steiner, and T.~P. Russell}, {\em Electric field induced instabilities at
  liquid/liquid interfaces}, J. Chem. Phys., 114 (2001), pp.~2377--2381.

\bibitem{Manteuffel1977}
{\sc T.~A. Manteuffel}, {\em The {T}chebychev iteration for nonsymmetric linear
  systems}, Numer. Math., 28 (1977), pp.~307--327.

\bibitem{MaCr01}
{\sc O.~K. Matar and R.~V. Craster}, {\em Models for {M}arangoni drying}, Phys.
  Fluids, 13 (2001), pp.~1869--1883.

\bibitem{Meerbergen1996}
{\sc K.~Meerbergen and D.~Roose}, {\em Matrix transformations for computing
  rightmost eigenvalues of large sparse non-symmetric eigenvalue problems}, IMA
  J. Numer. Anal., 16 (1996), pp.~297--346.

\bibitem{MPBT05}
{\sc D.~Merkt, A.~Pototsky, M.~Bestehorn, and U.~Thiele}, {\em Long-wave theory
  of bounded two-layer films with a free liquid-liquid interface: {S}hort- and
  long-time evolution}, Phys. Fluids, 17 (2005), p.~064104.

\bibitem{Mitl93}
{\sc V.~S. Mitlin}, {\em Dewetting of solid surface: {A}nalogy with spinodal
  decomposition}, J. Colloid Interface Sci., 156 (1993), pp.~491--497.

\bibitem{Muen03}
{\sc A.~M{\"u}nch}, {\em Pinch-off transition in {M}arangoni-driven thin
  films}, Phys. Rev. Lett., 91 (2003), p.~016105.

\bibitem{Muench05}
\leavevmode\vrule height 2pt depth -1.6pt width 23pt, {\em Dewetting rates of
  thin liquid films}, J. Phys.: Condens. Matter, 17 (2005), pp.~S309--S318.

\bibitem{Oron00c}
{\sc A.~Oron}, {\em Nonlinear dynamics of three-dimensional long-wave
  {M}arangoni instability in thin liquid films}, Phys. Fluids, 12 (2000),
  pp.~1633--1645.

\bibitem{Oron00}
\leavevmode\vrule height 2pt depth -1.6pt width 23pt, {\em Three-dimensional
  nonlinear dynamics of thin liquid films}, Phys. Rev. Lett., 85 (2000),
  pp.~2108--2111.

\bibitem{ODB97}
{\sc A.~Oron, S.~H. Davis, and S.~G. Bankoff}, {\em Long-scale evolution of
  thin liquid films}, Rev. Mod. Phys., 69 (1997), pp.~931--980.

\bibitem{OrRo92}
{\sc A.~Oron and P.~Rosenau}, {\em Formation of patterns induced by
  thermocapillarity and gravity}, J. Physique II France, 2 (1992),
  pp.~131--146.

\bibitem{OrRo94}
\leavevmode\vrule height 2pt depth -1.6pt width 23pt, {\em On a nonlinear
  thermocapillary effect in thin liquid layers}, J. Fluid Mech., 273 (1994),
  pp.~361--374.

\bibitem{PTTK07b}
{\sc A.~Pereira, P.~M.~J. Trevelyan, U.~Thiele, and S.~Kalliadasis}, {\em
  Dynamics of a horizontal thin liquid film in the presence of reactive
  surfactants}, Phys. Fluids, 19 (2007), p.~112102.

\bibitem{PFL01}
{\sc T.~Podgorski, J.-M. Flesselles, and L.~Limat}, {\em Corners, cusps, and
  pearls in running drops}, Phys. Rev. Lett., 87 (2001), p.~036102.

\bibitem{PBMT04}
{\sc A.~Pototsky, M.~Bestehorn, D.~Merkt, and U.~Thiele}, {\em Alternative
  pathways of dewetting for a thin liquid two-layer film}, Phys. Rev. E, 70
  (2004), p.~025201(R).

\bibitem{PBMT05}
\leavevmode\vrule height 2pt depth -1.6pt width 23pt, {\em Morphology changes
  in the evolution of liquid two-layer films}, J. Chem. Phys., 122 (2005),
  p.~224711.

\bibitem{PBMT06}
{\sc A.~Pototsky, M.~Bestehorn, D.~Merkt, and U.~Thiele}, {\em 3d surface
  patterns in liquid two-layer films}, Europhys. Lett., 74 (2006),
  pp.~665--671.

\bibitem{Reit92}
{\sc G.~Reiter}, {\em Dewetting of thin polymer films}, Phys. Rev. Lett., 68
  (1992), pp.~75--78.

\bibitem{ReSh01}
{\sc G.~Reiter and A.~Sharma}, {\em Auto-optimization of dewetting rates by rim
  instabilities in slipping polymer films}, Phys. Rev. Lett., 87 (2001),
  p.~166103.

\bibitem{RuJa74}
{\sc E.~Ruckenstein and R.~K. Jain}, {\em Spontaneous rupture of thin liquid
  films}, J. Chem. Soc. Faraday Trans. II, 70 (1974), pp.~132--147.

\bibitem{Saad92}
{\sc Y.~Saad}, {\em Analysis of some krylov subspace approximations to the
  matrix exponential operator}, SIAM J. Numer. Anal., 29 (1992), pp.~209--228.

\bibitem{Sadkane1993}
{\sc M.~Sadkane}, {\em A block {A}rnoldi-{C}hebyshev method for computing the
  leading eigenpairs of large sparse unsymmetric matrices}, Numer. Math., 64
  (1993), pp.~181--193.

\bibitem{Seem05}
{\sc R.~Seemann, S.~Herminghaus, C.~Neto, S.~Schlagowski, D.~Podzimek,
  R.~Konrad, H.~Mantz, and K.~Jacobs}, {\em Dynamics and structure formation in
  thin polymer melt films}, J. Phys.-Condes. Matter, 17 (2005), pp.~S267--S290.

\bibitem{Seydel}
{\sc R.~Seydel}, {\em Practical Bifurcation and Stability Analysis},
  Springer-Verlag, New York, 1994.

\bibitem{Shar93}
{\sc A.~Sharma}, {\em Relationship of thin film stability and morphology to
  macroscopic parameters of wetting in the apolar and polar systems}, Langmuir,
  9 (1993), pp.~861--869.

\bibitem{ShKh98}
{\sc A.~Sharma and R.~Khanna}, {\em Pattern formation in unstable thin liquid
  films}, Phys. Rev. Lett., 81 (1998), pp.~3463--3466.

\bibitem{ShRe96}
{\sc A.~Sharma and G.~Reiter}, {\em Instability of thin polymer films on coated
  substrates: {R}upture, dewetting and drop formation}, J. Colloid Interface
  Sci., 178 (1996), pp.~383--399.

\bibitem{SRLL05}
{\sc J.~H. Snoeijer, E.~Rio, N.~Le~Grand, and L.~Limat}, {\em Self-similar flow
  and contact line geometry at the rear of cornered drops}, Phys. Fluids, 17
  (2005), p.~072101.

\bibitem{SpHo96}
{\sc M.~A. Spaid and G.~M. Homsy}, {\em Stability of {N}ewtonian and
  viscoelastic dynamic contact lines}, Phys. Fluids, 8 (1996), pp.~460--478.

\bibitem{SBB03}
{\sc J.~Sur, A.~L. Bertozzi, and R.~P. Behringer}, {\em Reverse
  undercompressive shock structures in driven thin film flow}, Phys. Rev.
  Lett., 90 (2003), p.~126105.

\bibitem{Thie03}
{\sc U.~Thiele}, {\em Open questions and promising new fields in dewetting},
  Eur. Phys. J. E, 12 (2003), pp.~409--416.

\bibitem{Thie07}
\leavevmode\vrule height 2pt depth -1.6pt width 23pt, {\em Structure formation
  in thin liquid films}, in Thin films of Soft Matter, S.~Kalliadasis and
  U.~Thiele, eds., Wien, 2007, Springer, pp.~25--93.

\bibitem{Thie10}
{\sc U.~Thiele}, {\em Thin film evolution equations from (evaporating)
  dewetting liquid layers to epitaxial growth}, J. Phys.-Cond. Mat., 22 (2010), p.~084019.
\newblock (at press).

\bibitem{TBBB03}
{\sc U.~Thiele, L.~Brusch, M.~Bestehorn, and M.~B{\"a}r}, {\em Modelling
  thin-film dewetting on structured substrates and templates: {B}ifurcation
  analysis and numerical simulations}, Eur. Phys. J. E, 11 (2003),
  pp.~255--271.

\bibitem{TGV09}
{\sc U.~Thiele, B.~Goyeau, and M.~G. Velarde}, {\em Film flow on a porous
  substrate}, Phys. Fluids, 21 (2009), p.~014103.

\bibitem{TJB04}
{\sc U.~Thiele, K.~John, and M.~B{\"a}r}, {\em Dynamical model for chemically
  driven running droplets}, Phys. Rev. Lett., 93 (2004), p.~027802.

\bibitem{ThKn03}
{\sc U.~Thiele and E.~Knobloch}, {\em Front and back instability of a liquid
  film on a slightly inclined plate}, Phys. Fluids, 15 (2003), pp.~892--907.

\bibitem{ThKn04}
{\sc U.~Thiele and E.~Knobloch}, {\em Thin liquid films on a slightly inclined
  heated plate}, Physica D, 190 (2004), pp.~213--248.

\bibitem{ThKn06b}
\leavevmode\vrule height 2pt depth -1.6pt width 23pt, {\em Driven drops on
  heterogeneous substrates: {O}nset of sliding motion}, Phys. Rev. Lett., 97
  (2006), p.~204501.

\bibitem{ThKn06}
\leavevmode\vrule height 2pt depth -1.6pt width 23pt, {\em On the depinning of
  a driven drop on a heterogeneous substrate}, New J. Phys., 8 (2006), pp.~313,
  1--37.

\bibitem{Thie09}
{\sc U.~Thiele, I.~Vancea, A.~J. Archer, M.~J. Robbins, L.~Frastia,
  A.~Stannard, E.~Pauliac-Vaujour, C.~P. Martin, M.~O. Blunt, and P.~J.
  Moriarty}, {\em Modelling approaches to the dewetting of evaporating thin
  films of nanoparticle suspensions}, J. Phys.-Cond. Mat., 21 (2009),
  p.~264016.

\bibitem{TVK06}
{\sc U.~Thiele, J.~M. Vega, and E.~Knobloch}, {\em Long-wave {M}arangoni
  instability with vibration}, J. Fluid Mech., 546 (2006), pp.~61--87.

\bibitem{TVN01}
{\sc U.~Thiele, M.~G. Velarde, and K.~Neuffer}, {\em Dewetting: {F}ilm rupture
  by nucleation in the spinodal regime}, Phys. Rev. Lett., 87 (2001),
  p.~016104.

\bibitem{Thie01}
{\sc U.~Thiele, M.~G. Velarde, K.~Neuffer, M.~Bestehorn, and Y.~Pomeau}, {\em
  Sliding drops in the diffuse interface model coupled to hydrodynamics}, Phys.
  Rev. E, 64 (2001), p.~061601.

\bibitem{TVNP01}
{\sc U.~Thiele, M.~G. Velarde, K.~Neuffer, and Y.~Pomeau}, {\em Film rupture in
  the diffuse interface model coupled to hydrodynamics}, Phys. Rev. E, 64
  (2001), p.~031602.

\bibitem{Tokman06}
{\sc M.~Tokman}, {\em Efficient integration of large stiff systems of odes with
  exponential propagation (epi) methods}, J. Comput. Phys., 213 (2006),
  pp.~748--776.

\bibitem{TB00}
{\sc L.~S.~Tuckerman and D.~Barkley}, {\em Bifurcation analysis for timesteppers},
  in Numerical Methods for Bifurcation Problems and Large-Scale Dynamical,
  L.~Doedel, E.~andTuckerman, ed., vol.~119 of IMA Volumes in Mathematics and
  its Applications, New-York, 2000, Springer, pp.~466--543.

\bibitem{VSKB05}
{\sc R.~Verma, A.~Sharma, K.~Kargupta, and J.~Bhaumik}, {\em Electric field
  induced instability and pattern formation in thin liquid films}, Langmuir, 21
  (2005), pp.~3710--3721.

\bibitem{WhBa06}
{\sc P.~Wheeler and D.~Barkley}, {\em Computation of spiral spectra}, SIAM J.
  Applied Dynamical Systems, 5 (2006), pp.~157--177.

\bibitem{WiBo03}
{\sc T.~Witelski and M.~Bowen}, {\em Adi schemes for higher-order nonlinear
  diffusion equations}, Applied Numerical Mathematics, 45 (2003), p.~331Ð351.

\bibitem{ZhBe00}
{\sc L.~Zhornitskaya and A.~L. Bertozzi}, {\em Positivity-preserving numerical
  schemes for lubrication-type equations}, SIAM J. Numer. Anal., 37 (2000),
  pp.~523--555.

\end{thebibliography}

\end{document}